\documentclass[aps,prx,reprint]{revtex4-2}

\usepackage{amsmath}
\usepackage{amssymb}
\usepackage{graphicx}
\usepackage{physics}
\usepackage{xcolor}
\usepackage{cancel}
\usepackage{titletoc}
\usepackage{hyperref}
\usepackage{bm}
\hypersetup{
  colorlinks=true,
  linkcolor=blue,
  citecolor=blue,
  urlcolor=blue,
  pdftitle={MF-SOR},
  pdfauthor={Blumenthal, Reddy}
}
\DeclareMathOperator*{\argmin}{arg\,min}

\DeclareMathOperator*{\E}{\operatorname{E}}

\newcommand{\ve}{\varepsilon}

\newcommand{\bx}{\bm{x}}
\newcommand{\by}{\bm{y}}
\newcommand{\bu}{\bm{u}}
\newcommand{\bze}{{\bm{\zeta}}}
\newcommand{\bvp}{{\bm{\varphi}}}
\newcommand{\bl}{{\bm{\lambda}}}

\newcommand{\bph}{{\bm{\phi}}}
\newcommand{\bb}{{\bm{\beta}}}
\newcommand{\be}{{\bm{\eta}}}
\newcommand{\bv}{{\bm{v}}}
\newcommand{\bch}{{\bm{\chi}}}

\newcommand{\bw}{{\bm{w}}}
\newcommand{\bW}{{\bm{W}}}
\newcommand{\bxi}{{\bm{\xi}}}

\newcommand{\bG}{{\bm{G}}}
\newcommand{\bC}{{\bm{C}}}

\newcommand{\bPi}{{\bm{\Pi}}}
\newcommand{\bSi}{{\bm{\Sigma}}}
\newcommand{\dS}{\mathsf{S}}
\newcommand{\dL}{\mathsf{L}}
\newcommand{\dC}{\mathsf{C}}
\newcommand{\dbC}{{\bm{\mathsf{C}}}}
\newcommand{\dG}{\mathsf{G}}
\newcommand{\dbG}{{\bm{\mathsf{G}}}}
\newcommand{\dQ}{\mathsf{Q}}
\newcommand{\dbQ}{{\bm{\mathsf{Q}}}}
\newcommand{\dtQ}{\widetilde{\mathsf{Q}}}
\newcommand{\dtbQ}{\widetilde{\bm{\mathsf{Q}}}}
\newcommand{\ds}{{s}}
\newcommand{\dhC}{\widehat{\mathsf{C}}}
\newcommand{\dhbC}{\widehat{\bm{\mathsf{C}}}}
\newcommand{\dhG}{\widehat{\mathsf{G}}}
\newcommand{\dhbG}{\widehat{\bm{\mathsf{G}}}}
\newcommand{\dK}{\mathsf{K}}
\newcommand{\dbK}{{\bm{\mathsf{K}}}}

\begin{document}

\title{Self-organized robustness in mean-field interacting systems}

\author{Emmy Blumenthal}
\email{eblu@princeton.edu}
\author{Gautam Reddy}
\email{greddy@princeton.edu}
\affiliation{
Joseph Henry Laboratories of Physics, Princeton University, Princeton, New Jersey 08544, USA
}

\begin{abstract}
Self-organization is a defining feature of living systems, with order often maintained through interactions between constituent units rather than centralized feedback. We introduce a tractable mean-field model of self-organized robustness, formulated as meta-optimization over the system's response to perturbations. The resulting interaction structure has an intuitive picture as a dynamically modulated landscape (``seascape'') whose shape is determined self-consistently to accelerate relaxation back to equilibrium. The collective dynamics follows an optimized Wasserstein gradient flow toward an attractor in the space of collective states. When communication is limited, interactions preferentially encode slowly relaxing modes and modes that are frequently perturbed. The model further shows that robust collective states are associated with flatter equilibrium landscapes and predicts a continuum of intermediate ``reservoir states'' in such systems. %
The model offers a perspective of self-organization as a hierarchical associative memory that operates on the scale of a collective of interacting computational units.
\end{abstract}

\maketitle

\section{Introduction}
\label{sec:intro}

A remarkable feature of multicellular systems is their ability to self-organize, the ability to achieve and maintain collective states through distributed signaling among constituent elements~\cite{li2019communication, Bruckner2024InformationContent,Tripathi2025CollectiveDifferent,Romeo2026InformationBounds,Smart2023EmergentProperties,Skinner2025PhysicalModelingEmbryonicTranscriptomes}. During animal development, for example, populations of cells establish robust spatial patterns of gene expression and cell identity through local cell-cell interactions, without necessarily leveraging a global positional blueprint~\cite{Schweisguth2019SelfOrganizationPatternFormation, collinet2021programmed}. %
In many tissues, homeostasis of tissue composition is mediated by collective mechanisms for feedback control~\cite{simons2011strategies, mesa2018homeostatic, kitadate2019competition, Klein2011UniversalPatternsStemCellFate, jorg2021stem, meizlish2021tissue, zhou2018circuit}. A common theme in these systems is that collective behavior is achieved through interactions among constituent units: cells of different types release, sense, and respond to a diverse collection of signals that report on the collective state, achieving a desired state through self-consistent feedback.

Associative memory~\cite{hopfield1982neural, amari1972learning} has served as a powerful conceptual framework for describing self-organization in biology and machine learning~\cite{krotov2023new, yampolskaya2025hopfield, Smart2023EmergentProperties, boukacem2024waddington, khona2022attractor}. In this framework, self-organization is a problem of designing attractors over collective states of the system. Here, we ask if a similar conceptual model can be constructed for describing the behavior of a population that achieves collective homeostasis using a collection of state-dependent signals. We describe a tractable model that satisfies this criterion, and discuss the nature of interactions that equip the system with desirable properties. Our framework can be viewed as a hierarchically structured associative memory that operates on the scale of a population of interacting computational units: each unit follows intrinsic gradient dynamics along a multistable attractor landscape and interacts with other units such that the collective system also follows a gradient flow toward a desired collective state. 

We consider interacting particle systems where each particle's state $\bx$ evolves stochastically according to $\dot{\bx} =  \bv(\bx, \bvp)+ \sqrt{\ve}\,\be$, where $\be$ is white noise and $\bx$ is a multi-dimensional state variable whose dynamics depend on $\bx$ and environmental signals $\bvp$ through $\bv$.
The density over single-particle states $\rho_t$ at time $t$ evolves under $\bv$. We work in the mean-field limit where the collective state is defined by this density $\rho_t$.
We are interested in scenarios where the control $\bv$ is set such that $\rho_t$ rapidly converges to a desired collective state $q$. The desired set point $q$ is determined by other external contextual factors that are not explicitly included in this model.

\begin{figure}[t]
  \includegraphics[width=\linewidth]{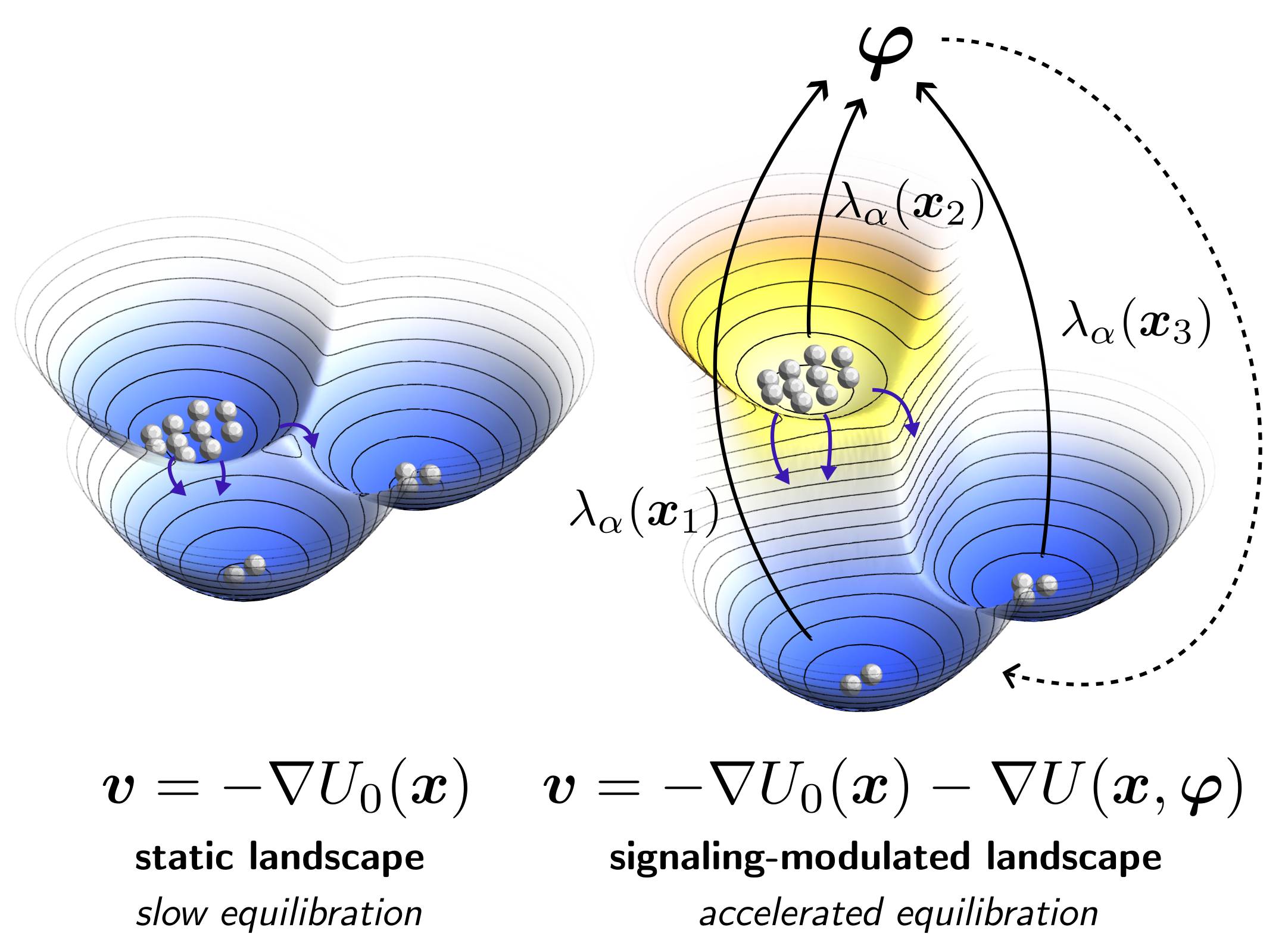}
  \caption{\label{fig:three-wells-cartoon}
    Signaling-modulated landscape accelerates self-organization.
    Left: a static potential landscape equilibrates to a desired state (equal distribution of populations in three wells) slowly due to the presence of high barriers.
    Right: a signaling-modulated dynamic landscape (``seascape''), where the barrier heights are modulated by the population distribution, achieves the desired state much faster.
    Particles release state-dependent signals $\bl(\bx)$ that are pooled to form a shared signal $\bvp$, which in turn modulates the landscape.
  }
\end{figure}

While this formulation is general, it is related to two data-driven perspectives for describing the physics of cell fate transitions in biological systems: (1) a dynamical systems perspective, where the environmental signals $\bvp$ `tilt' a potential landscape that reflects cellular decision-making~\cite{corson2017gene, rand2021geometry, raju2023geometrical, cislo2025reconstructing, Howe2025LearningGeometricModels}, and (2) an optimal transport perspective, which models cell fate transitions as the time-dependent transformation of single-cell densities over cellular states in different contexts~\cite{schiebinger2019optimal, bunne2023learning, bunne2024optimal, Reddy2024DynamicLandscapesGrowth}. In these scenarios, $\bx$ is a proxy for cellular state, for example, its gene expression profile, and $\bv$ represents gene regulatory control that modulates this profile based on the cell's current state and external signals. In this picture, our model corresponds to a scenario where the environment contains endogenous signals $\bvp$ released by cells which in turn shape heterogeneity by influencing changes in cellular state. The model is also conceptually related to nonequilibrium protocols for thermodynamic state preparation, stochastic optimal control and generative modeling frameworks in machine learning~\cite{amos2022meta,atanackovic2025meta, ikeda2025speed, klinger2025universal, orland2025schrodinger, Chen2021StochasticControlLiaisons, Reddy2024DynamicLandscapesGrowth, chen2016relation}. The key distinction is that $\bvp$, which is typically an externally set time-dependent annealing parameter, should instead generate a self-consistent annealing scheme that guides $\rho_t$ to $q$ regardless of the initial state.

How can such a system be designed? Suppose each particle's dynamics does not depend on the environment, say $\bv(\bx) = - \nabla U_0(\bx) \propto \nabla \log q(\bx)$. With non-zero fluctuations, the collective state will eventually converge to $q$, but convergence can be extremely slow if $U_0$ has minima separated by large barriers. More effective strategies are possible if the initial density (after a perturbation, say) is known. For example, one may anneal the landscape with a time-dependent signal $\bvp(t)$: $\bv(\bx) = - \nabla U_0(\bx) - \nabla U(\bx,\bvp(t))$, which, however, requires a centralized feedback mechanism. Now, suppose a particle with state $\bx$ conveys state-dependent signals, say $\bl(\bx)$, representing its contribution to a globally shared signal $\bvp$. This shared signal encodes partial information about the collective state, which can be used by every particle to control its state toward homeostasis (Fig.~\ref{fig:three-wells-cartoon}). %
This intuition would suggest that given a constraint on the number of shared signals, the choice of what to signal and how to respond should depend on the kinetics of equilibration and the typical perturbations experienced by the system.

In what follows, we introduce a minimal model where the design principles of such a system can be examined in detail. First, we show that feedback control in this model is the gradient of a dynamically modulated landscape (``seascape''), offering an intuitive picture of signaling-based modulation of system dynamics. Second, by introducing a geometric framework, we show that there is a direct relationship between signaling and control. That is, the finite modes along which the landscape can be modulated are determined by which signals are shared by the particles. Third, we use adjoint-based optimization to develop a differentiable numerical scheme to solve for the optimal feedback control using neural networks. Fourth, we formulate optimal signaling as a generalized eigenvalue problem, and show that optimal signaling modes reflect a tradeoff between conveying information about slowly relaxing modes and those that are frequently perturbed. Finally, we show that when perturbations to the system occur on fast timescales, the optimal design favors modifying the signaling-independent dynamics to introduce intermediate `reservoir' states that act as bridges between metastable regions.

\section{Self-organized robustness as an attractor on collective states}
\label{sec:model}

Consider a system with $N \gg 1$ particles whose collective state at time $t$ is represented by the distribution $
\rho_t(\bx)$ of the individual particle states $\{ \bx_i \}$ in the population.
The evolution of $\rho_t$ is given by the Fokker--Planck equation, which describes how $\rho_t$ evolves under the influence of a drift field $\bv$ and noise $\varepsilon$ (isotropic, for simplicity):
\begin{equation}
  \begin{gathered}
  \pdv{\rho_t}{t} = -\nabla \cdot  (\bv \rho_t) + \frac{\varepsilon}{2}\nabla^2\rho_t, 
  \\
  \bv(\bx) = \bu_0(\bx) + \bu(\bx;\rho_t),
\label{eq:fokker-planck}
\end{gathered}
\end{equation}
where $\bu_0 = -\nabla U_0$ specifies the feedback-independent passive dynamics of each particle and $\bu$ is a feedback-mediated control term that depends on the collective state $\rho_t$. For example, when $U_0$ has many minima, each minimum defines a single-particle attractor state that particles could occupy in the absence of signaling. Signaling can act through $\bu$ to force transitions between these attractor states, thereby shifting the occupancies of attractor states across the population.

The dependence on the current collective state $\rho_t$ is mediated by mean-field interactions between particles: a particle with state $\bx_i$ releases $M$ state-dependent signals $\lambda_{\alpha}(\bx_i)$ (where $\alpha = 1,2, \dots, M$). The signals from all particles are pooled together to form a shared signal,
\begin{align}
\varphi_{\alpha} = \frac{1}{N} \sum_{i=1}^N \lambda_{\alpha}(\bx_i) = \int \dd{\bx} \rho_t(\bx) \lambda_{\alpha}(\bx),
\label{eq:varphi}
\end{align}
where we are considering the $N\to\infty$ limit.
Since $M$ is finite, each particle has a partial estimate $\hat{\rho}_t$ of the collective state through $\bvp$. The ensemble of perturbations and the frequency of perturbations play an important role in determining the design of the system.
We assume that perturbations to the system take the form of a Poisson process with rate $\gamma$ in which the system is reset to a new collective initial condition $\rho_0 \sim p_{\rm pert}$ where $p_{\rm pert}$ is the distribution over possible initial conditions.

We consider a system that is designed to efficiently achieve a desired collective state $q$. Given passive dynamics $\bu_0$ and the choice of signals $\bl$, we first optimize over the feedback control $\bu$ that accelerates relaxation to $q$ given a signal $\bvp$. Then, we consider the optimal choice of $\bu_0$ and $\bl$ that enables the feedback to most effectively drive the system to $q$, averaged over the ensemble of perturbations.
In other words, this meta-optimization has two levels; an inner loop optimizes the feedback control $\bu$ given a fixed choice of $\bu_0$ and $\bl$ and defines a cost-to-go functional $S(\rho_t; \bu_0, \bl)$ which quantifies the expected cost incurred by the system given the current collective state $\rho_t$, and an outer loop optimizes over the choice of $\bu_0$ and $\bl$ to minimize the expected cost-to-go averaged over perturbations:
\begin{align}
\mathcal{S} &= \min_{\bu_0, \bl}\left\{ \left\langle S(\rho_0; \bu_0, \bl) \right\rangle_{\rho_0 \sim p_{\rm pert}} \right\}, \quad \text{where}  \label{eq:S1}\\
S(\rho_t; & \bu_0,\bl) = \min_{\bu} \left\{ \int_t^{\infty} \dd{t'} e^{-\gamma (t' - t)} \ell(\bu, \rho_{t'})\right\}.  \label{eq:S}
\end{align}
The loss (analogous to the Lagrangian) $\ell(\bu, \rho)$ penalizes the system for being far from $q$ and regularizes the strength of feedback control. We consider a particular tractable form that captures these two effects:
\begin{equation}
\begin{aligned}
  \ell(\bu,\rho_t) = D_{\rm KL}(\rho_t \Vert q) + \frac{r}{2} \int \dd{\bx} \rho_t(\bx) \, \norm{\bu(\bx; \rho_t)}^2,
\end{aligned}
\label{eq:ell-full}
\end{equation}
where $D_{\rm KL}(\rho_t \Vert q)$ is the Kullback-Leibler divergence from the target distribution $q$, and the second term is a quadratic cost on the strength of feedback control, which constrains how much the system can exploit signaled information to accelerate relaxation; the parameter $r$ sets the relative weight of the two terms.
The exponential weighting factor with rate $\gamma$ in Eq.~\eqref{eq:S} accounts for the probability that the system is perturbed to a new state. $S(\rho_t) $ measures the expected cost-to-go given the current collective state $\rho_t$ and is related to the principal function in Hamilton--Jacobi theory (the dependence on $\bu_0, \bl$ is dropped for convenience).

The optimal feedback control is the minimizer in Eq.~\eqref{eq:S}, but this is achievable only if each particle has perfect information about $\rho_t$. The field $\bvp$ instead provides an estimate $\hat{\rho}_t(\bvp)$ (the map from $\bvp$ to $\hat{\rho}_t$ is discussed further below). A local rule for the feedback control $\bu$ given $\hat{\rho}_t$ (equivalently, $\bvp$) is obtained by expressing $S(\hat{\rho}_t)$ in a recursive form:
\begin{equation}
\begin{aligned}
  S(\hat{\rho}_t) &=\min_{\bu}
 \left\{\ell(\bu,\hat{\rho}_t) \, \dd{t} + e^{-\gamma \,\dd{t}} S(\hat{\rho}_{t+\dd{t}}) \right\}. \label{eq:dp-estimated-state}
\end{aligned}
\end{equation}
Assuming that the \emph{estimated} $\hat{\rho}_t$ also follows Eq.~\eqref{eq:fokker-planck} (known as the certainty equivalence principle in control theory), integrating by parts and keeping only $O(\dd{t})$ terms gives the Hamilton--Jacobi--Bellman (HJB) equation:
\begin{equation}
\begin{aligned}
  0 = & \min_{\bu}
  \Bigg\{
    \ell(\bu,\hat{\rho}_t)
    +
    \int
    \dd{\bx}
    \hat{\rho}_t\,
    \qty[
      \bv
      \cdot
      \nabla
      \fdv{S}{\hat{\rho}_t}
      +
      \frac{\varepsilon}{2}
      \nabla^2
      \fdv{S}{\hat{\rho}_t}
      -
      \gamma S
    ]
  \Bigg\}. \label{eq:hjb-estimated-state}
\end{aligned}
\end{equation}
With the choice of loss in Eq.~\eqref{eq:ell-full}, the feedback control is given by
\begin{equation}
  \bu^\ast =  -\frac{1}{r}\nabla
  \fdv{S}{\hat{\rho}_t(\bx)} = -\nabla U(\bx; \hat{\rho}_t).  \label{eq:ustar}
\end{equation}
Eq.~\eqref{eq:ustar} shows that feedback control can be written as a gradient of a potential ${U}$, which is set self-consistently through $\bvp$. This result leads to a picture of the system's dynamics as being governed by a dynamically modulated landscape, which is a consequence of the quadratic transport cost.

Importantly, the transport equation Eq.~\eqref{eq:fokker-planck} under optimality has a precise mathematical interpretation as a gradient flow over a manifold of distributions equipped with the Wasserstein metric~\cite{Chewi2024StatisticalOptimalTransport}. Denote $q_0(\bx) \propto \exp(-2U_0(\bx)/\varepsilon)$. The collective state $\rho_t$ follows a Wasserstein gradient flow over the functional $\mathcal{F}(\rho) = (\varepsilon/2) D_{\rm KL}(\rho \Vert q_0) + r^{-1}S(\rho)$. That is, Eq.~\eqref{eq:fokker-planck} can be written as
\begin{align}
\pdv{\rho_t}{t} = \nabla\cdot \left( \rho_t  \nabla \fdv{\mathcal{F}}{\rho_t}\right),
\label{eq:wasserstein-gradient-flow}
\end{align}
and the dynamics ensure $\dv*{\mathcal{F}}{t} \le 0$. If $q_0 = q$, the global minimum of $\mathcal{F}$ is achieved when $\rho = q$. Throughout most of the paper, we will take $U_0(\bx) = -(\varepsilon/2) \log q(\bx)$ so that $q_0 = q$, and the role of the feedback term $\bu(\bx;\rho_t)$ is to accelerate relaxation to $q$. This assumption will be relaxed in Sec.~\ref{sec:preconditioning}, where we investigate alternative choices of $U_0$.

\begin{figure}[ht!]
  \includegraphics[width=\linewidth]{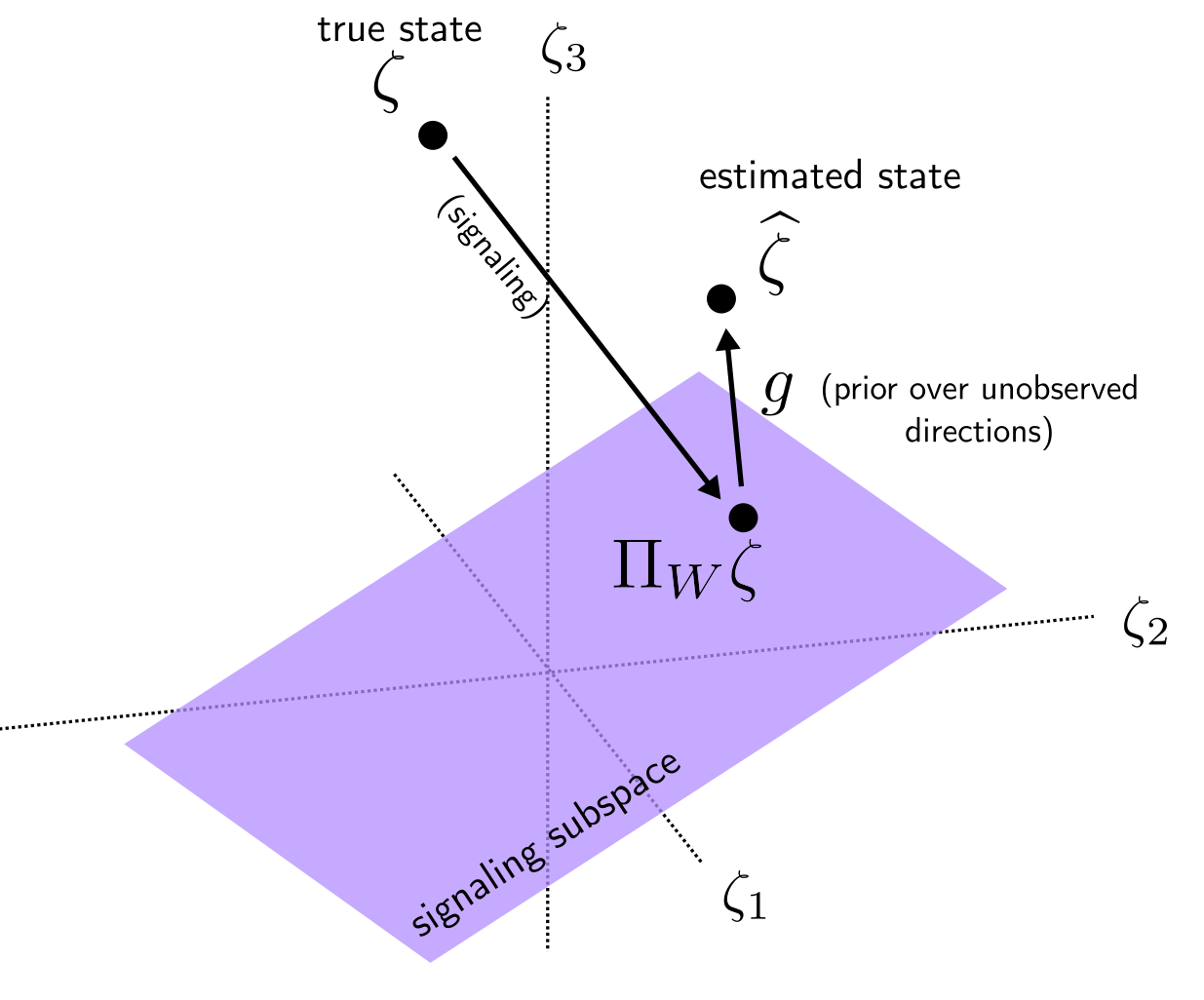}
  \caption{
    \label{fig:projection}
    The geometry of signaling-mediated inference of the collective state. The collective state of the system is represented using the dual variable $\bze$. Its estimate $\hat{\bze}$ from the shared signal $\bvp$ is obtained by projecting the true state onto the signaling subspace (defined by $\bl$) and adding a prior offset vector $\bm{g}$ to account for directions that are unobserved.
  }
\end{figure}

\section{Subspace selection and the connection between signaling and control}
In general, the principal function $S$ satisfies the functional HJB equation in Eq.~\eqref{eq:hjb-estimated-state}, which is a generally intractable nonlinear PDE. An alternative geometric representation makes it amenable to systematic approximation and leads to a relationship between signaling (determined by $\bl$) and the degrees of freedom along which each particle controls its state.

Consider the exponential representation:
\begin{align}
\rho_{\bze}(\bx) = q(\bx) \exp \qty(\sum_{\mu = 1}^\infty \zeta_\mu  \chi_\mu(\bx) - F(\bze)),
\end{align}
where $\{\chi_\mu(\bx)\}$ is a set of complete basis functions, $\bze$ are dual variables and $F$ is a log-partition which normalizes the density. In this representation, each distribution is represented by a unique infinite-dimensional vector $\bze$ and the magnitude of each component $\zeta_\mu$ quantifies the contribution of mode $\chi_\mu$ to the divergence from the target distribution $q$ (and thus $\zeta_\mu = 0$ for all $\mu$ when $\rho = q$).
The log-partition function $F(\bze)$ satisfies:
\begin{equation}
\begin{aligned}
  \pdv{F}{\zeta_\mu} =
  \int \dd{\bx} \rho_{\bze}(\bx) \, \chi_\mu(\bx)
  \equiv
  \xi_\mu
  . \label{eq:dual}
\end{aligned}
\end{equation}
That is, given the generalized moments $\bxi$ of $\rho$, the dual variables $\bze$ can be solved by imposing Eq.~\eqref{eq:dual}.

Now, consider the mean-field interaction scheme in Eq.~\eqref{eq:varphi}. If the state-dependent signal is expressed in terms of $\bch$ as $\bl(\bx) = \bW\bch(\bx)$, we have $\bvp = \bW\bxi$. The signaling field $\bvp$ can thus be interpreted as a finite set of $M$ generalized moments which provide partial information about $\rho$.

Given $\bvp$, an estimate $\hat{\rho}$ is given by:
\begin{align}
  \hat{\rho}(\bx) = \widetilde q(\bx) \exp\qty(
    \bb\cdot
    \bl(\bx) -
    \widetilde F(\bb)), \label{eq:hatrho}
\end{align}
where $\widetilde q$ is a prior (see SI~Sec.~\ref{si-sec:posterior-estimate} for a discussion), and $\beta_{\alpha}$ (for $\alpha = 1,2, \dots, M$) is obtained by enforcing consistency with the observed signal:
\begin{align}
  \pdv{\widetilde F}{\beta_\alpha} =  \int \dd{\bx} \hat{\rho}_t(\bx) \, \lambda_\alpha(\bx) = \varphi_\alpha.
  \label{eq:beta}
\end{align}
The coefficients $\bb$ thus serve as internal ``belief'' states that encode all information the particles know about the present collective state.
From Eq.~\eqref{eq:hatrho}, the estimated collective state at time $t$ can be written as:
\begin{equation}
\begin{aligned}
  \hat{\bze}_t =  \bW^T\bb_t + \bm{g},
\end{aligned}
\end{equation}
where $\bm{g}$ is a prior offset. Therefore, the effect of mean-field interactions is a projection of the true state $\bze$ to the ``signaling subspace'' defined by the state-dependent signals $\bl$. The prior offset $\bm{g}$ is the best estimate of $\hat{\rho}$ along directions that are not observed and is separately encoded in $\widetilde q$. This geometric picture is illustrated in Fig.~\ref{fig:projection}.

In this description, feedback control only has access to the information in $\bb$ about the collective state: $\bu = \bu(\bx;\bb)$. The physical effect of such feedback on the belief states is thus constrained by the choice of $\bl$. Concretely, given $\bb$, the signaling variables evolve according to
\begin{equation}
\begin{aligned}
  \dot \varphi_\alpha
  =
  \int \dd{\bx}
  \hat{\rho}(\bx)\,
  \bu(\bx;\bb)
  \cdot
  \nabla \lambda_\alpha(\bx)
  +
  \text{indep. of } \bu.
\end{aligned}
\end{equation}
So, the instantaneous effect of the control on the signaling field is only due to its components along the gradients of the state-dependent signals. We can therefore decompose the feedback as:
\begin{equation}
\begin{gathered}
  \bu = \bu_\parallel + \bu_\perp,
  \\
  \bu_\parallel \in \operatorname{span}\qty{\nabla \lambda_\alpha},
  \qquad
  \int \dd{\bx}\hat{\rho}\,
  \bu_\perp \cdot \nabla \lambda_\alpha = 0 .
\end{gathered}
\end{equation}
The component $\bu_\perp$ does not change the signaling field in the subsequent step to first order, and does not change the estimated collective state that determines subsequent feedback. It does, however, contribute to the quadratic cost of control. Consequently, within this closed belief-state description, the optimal feedback should have no orthogonal component $\bu_{\perp}$ and must take the form
\begin{equation}
\begin{aligned}
  \bu^\ast(\bx;\bb)
  =
  \sum_{\alpha = 1}^M
  c_\alpha(\bb)
  \nabla \lambda_\alpha(\bx).
\end{aligned}
\label{eq:optimal-feedback-limited-signaling-main}
\end{equation}
Eq.~\eqref{eq:optimal-feedback-limited-signaling-main} implies that the landscape is only modulated in directions along which information is available.
The overall choice of state-dependent signals selects a finite fixed set of modes (`tilts') along which the landscape is modulated, and at any particular instant the shared signal $\bvp$ determines the magnitude by which each mode tilts the landscape.
The coefficients $c_\alpha(\bb)$ are determined by solving the HJB equation expressed in terms of $\bb$ rather than $\hat{\rho}$.

What is a good choice for the basis $\{\chi_\mu\}$? In this exponential representation, the evolution of $\rho$ can be expressed as a system of ODEs:
\begin{equation}
\begin{aligned}
  \dot \zeta_\mu &= -
  \frac{\varepsilon}{2}
  C_{\mu\nu}^{-1}
  G_{\nu\omega}
  \zeta_\omega
  +
  \text{control term}
  \label{eq:zetadot}
\end{aligned}
\end{equation}
where
\begin{equation}
\begin{aligned}
  C_{\mu\nu}(\bze)
  &\equiv
  \int \dd{\bx} \rho_{\bze} \, \qty(
    \chi_\mu - \xi_\mu
  )
  \qty(
    \chi_\nu - \xi_\nu
   ),
  \\
  G_{\mu\nu}(\bze)
  &\equiv
  \int \dd{\bx} \rho_{\bze} \, \nabla \chi_\mu \cdot \nabla \chi_\nu
  .
\end{aligned}
\end{equation}
Eq.~\eqref{eq:zetadot} shows that the passive evolution of the system is described by the matrices $\bC$ and $\bG$. These matrices capture the geometry of the similarity between nearby distributions and the geometry of how costly it is to transport one distribution to another under the dynamics of the system, respectively; we discuss this connection in more detail in SI~Sec.~\ref{si-sec:info-geom}.

It is convenient to choose the basis so as to simultaneously diagonalize $\bC$ and $\bG$ at $\bze = 0$ (i.e., $\rho = q$), which can be achieved by choosing them to be the orthonormal eigenfunctions of the generalized eigenvalue problem:
\begin{equation}
\begin{aligned}
  -\nabla\cdot \qty( q \nabla\chi_\mu) = \kappa_\mu q \chi_\mu, \quad  \int \dd{\bx} q\,\chi_\mu  \chi_\nu = \delta_{\mu\nu}.
\end{aligned}
\end{equation}
In this basis, for small perturbations about $\rho = q$, the mode $\zeta_\mu$ relaxes back to the desired state at a rate $(\varepsilon/2) \kappa_\mu$ under the passive dynamics of the system.
$\bC$ is the identity matrix at $\bze = \bxi = 0$, and the generalized eigenvalues $\kappa_{\mu}$ thus reflect those of $\bG$.
In other words, the slow modes are more expensive to relax despite having the same contribution to the divergence from the target state, and this discrepancy is quantified by the spectrum $\{\kappa_{\mu} \}$.

\begin{figure*}[ht]
  \includegraphics[width=0.9\linewidth]{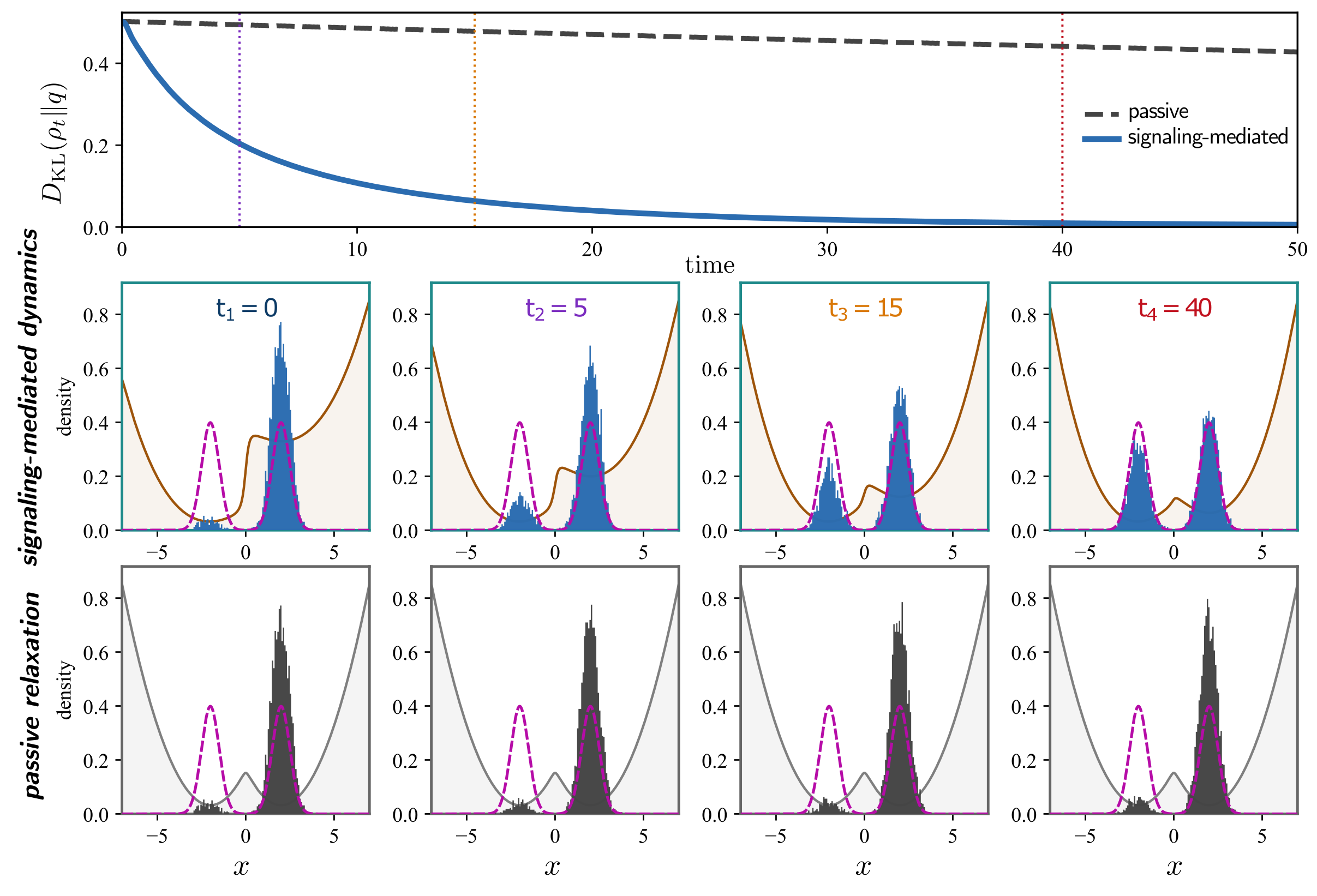}
  \caption{
    \label{fig:dynamics-gallery}
    A signaling-modulated landscape accelerates convergence to the desired state.
    Top row: the KL divergence from the desired state as a function of time for the system with a signaling-modulated landscape and a static landscape.
    Middle and bottom rows: snapshots of a swarm of $N = 4096$ particles evolving toward the desired state in the signaling-modulated (with $M = 1$ signal, the top mode in Fig.~\ref{fig:optimal-signaling-modes}) and static landscapes, respectively.
  }
\end{figure*}

\section{Adjoint-based numerical scheme}

Solving the HJB equation is challenging because it is infinite-dimensional and contains a nonlinearity in the gradient of $S$. The exponential representation of $\rho$ is particularly convenient for numerically solving this equation because it allows us to represent the dynamics of the system as a finite-dimensional system of ODEs after truncating Eq.~\eqref{eq:zetadot} to a finite number of modes~\cite{brigo2009projectingfokkerplanckequationfinite}.
Higher modes correspond to finer features of the distribution, and truncating these modes corresponds to approximating the distribution as a smoothed version of itself (similar to a Galerkin truncation~\cite{Grandclement2006spectral,gottlieb1977spectral}).

Even with this representation, the HJB equation is still a nonlinear PDE.
To solve for $S$ (and thus the optimal control), we substitute the expression for the control in terms of $S$ back into the objective Eq.~\eqref{eq:ell-full} written in terms of the truncated $\bze$-dynamics in order to evaluate a candidate solution.
We parameterize $S$ by a neural network and optimize the parameters by rolling out the dynamics under the candidate control with various initial conditions sampled from $p_{\rm pert}$ and using the adjoint method to compute gradients of the objective with respect to the parameters of $S$.
The adjoint method is a technique for computing gradients of an objective function with respect to the parameters of a system of ODEs by solving an auxiliary system of ODEs backward in time~\cite{Chen2018NeuralODE,Sinha2023OptimalControlActiveParticles,Krishnan2026HamiltonianBridge}.
In practice, we find that it is necessary to initialize $S$ to an approximate solution, which we obtain by solving the HJB equation in the weak-control limit where the quadratic term can be neglected.
We discuss further details of the numerical implementation in SI~Sec.~\ref{si-sec:numerics}.

The learned principal function $S$ is optimized using the $N \to \infty$ representation of the collective dynamics (expressed as ODEs), but the resulting optimal policy can be implemented on swarms of finite numbers of particles by calculating the observables $\varphi_\alpha = \frac{1}{N}\sum_{i=1}^N \lambda_\alpha(\bx_i)$, computing $\bb$ from $\bvp$ through Eq.~\eqref{eq:beta} using Newton iteration (as $\widetilde F$ is convex), and applying the feedback control $\bu(\bx;\bb)$ to each particle.
The resulting dynamics of the swarm and the signaling-dependent landscape $U(\bx;\bvp)$ under the optimal policy are shown in Fig.~\ref{fig:dynamics-gallery}.
The signaling-modulated landscape allows the system to accelerate convergence to the desired state, as demonstrated by the rapid decay of the KL divergence from the desired state compared to the static landscape case.
An animation of these dynamics and a two-dimensional example are available as SI~Movies 1 and 2, respectively.

\begin{figure*}[ht!]
  \includegraphics[width=\linewidth]{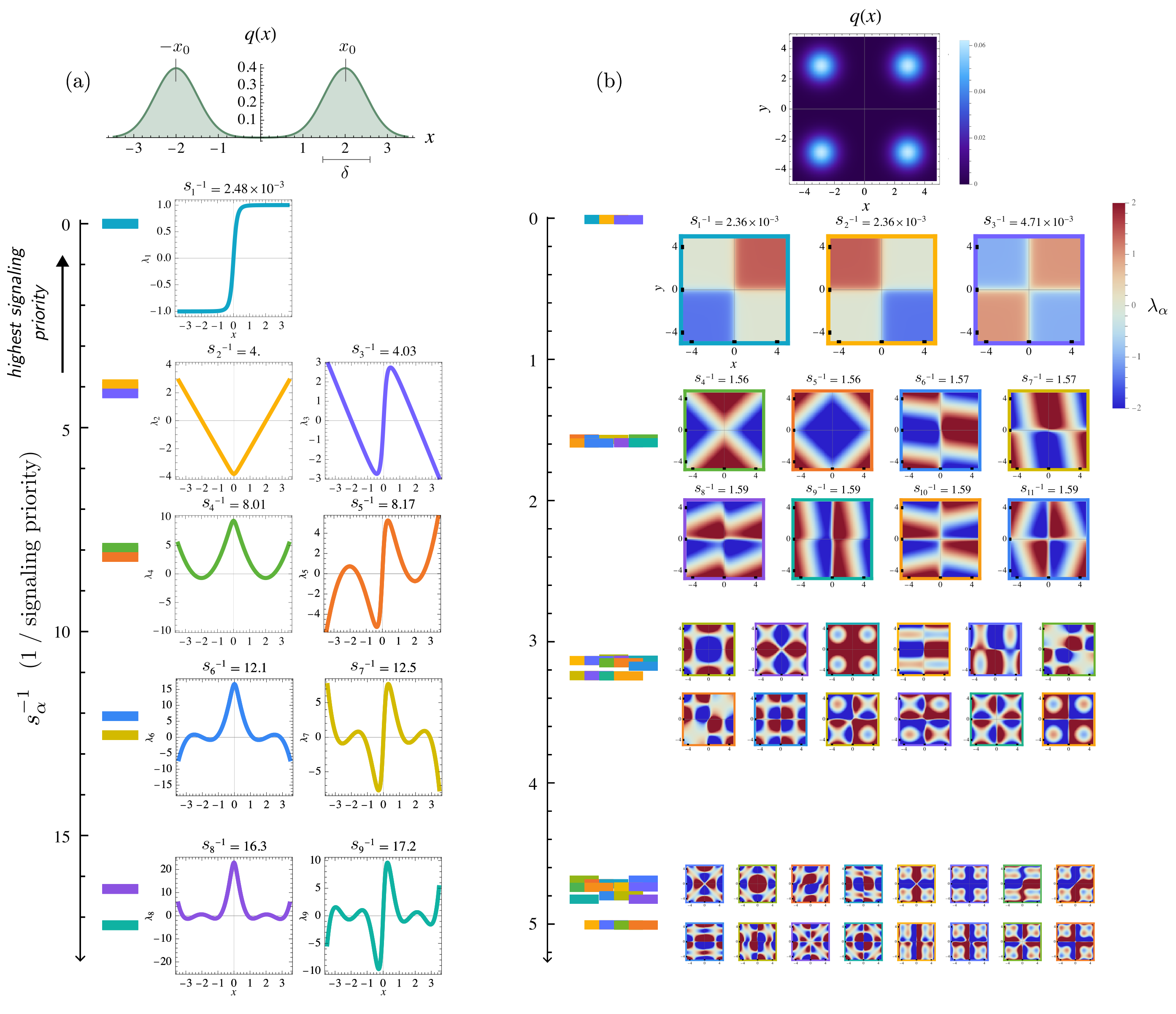}
  \caption{
    \label{fig:optimal-signaling-modes}
    Optimal signaling modes for example distributions experiencing perturbations along the passive modes ($\bSi = \bm{I}$).
    (a) For a two-peak Gaussian mixture distribution, the optimal signaling mode detects how unbalanced the distribution is between the two peaks and all other lower-priority modes detect features localized to individual peaks.
    The inverse signaling priorities $s_\alpha^{-1}$ of candidate modes are shown as a spectrum on the left and the plots of the corresponding modes are shown on the right, with the highest-priority mode at the top and lower-priority modes below.
    Because $\bSi = \bm{I}$ in this example, $s_\alpha^{-1} = \kappa_\alpha$ is the passive relaxation rate and the optimal signaling modes are the passive relaxation modes.
    (b) For a four-peak two-dimensional Gaussian mixture distribution arranged in a square, the highest-priority optimal signaling modes detect the peak balance of diagonally opposite peaks, the next highest-priority mode detects a checkerboard pattern, and the lower-priority modes detect features localized to individual peaks.
    Frame colors of the plots of the modes correspond to the colors of the spectrum on the left.
  }
\end{figure*}

\section{Signaling}

\subsection{Signaling architecture as optimal subspace selection}

Given a communication bottleneck $M$, the system should select appropriate signaling modes that allow for rapid convergence to homeostasis. Formally, the signaling modes are found by minimizing the outer loop over $\bl$ (equivalently, $\bW$) in Eq.~\eqref{eq:S1}. To do this, one would substitute the optimal feedback control and the resulting transport law into the objective and minimize the resulting cost $\mathcal{L}_{\bW} = \langle S(\rho_0)\rangle_{\rho_0 \sim p_{\rm pert}}$ with respect to $\bW$. This is a difficult problem in general because $S$ cannot be solved exactly.

Some intuition can be gained from a reformulation of the signaling problem as a projection $\bPi$ from the full state $\bze$ to some lower-dimensional subspace. Generally, such projections from the full state to the state estimated through signaling are non-linear, and the projection may depend on the local information available to each particle. When projections can be treated as linear, the theory of Grassmann manifolds provides a natural framework for the optimal subspace selection problem (see SI~Sec.~\ref{si-sec:grassmann} for a quick review)~\cite{Bendokat2024GrassmannHandbook}.
A central result of Grassmann-manifold optimization is that the stationarity condition for optimizing a function $\mathcal{L}(\bPi)$ with respect to a projector $\bPi$ takes the form of a commutator relation:
\begin{equation}
\begin{aligned}
  \qty[
    \bPi,
    \frac{1}{2}
    \qty(
      \nabla_{\bPi}
      \mathcal{L}
      +
      \nabla_{\bPi}^T
      \mathcal{L}
    )
  ]
  =
  0,
\end{aligned}
\end{equation}
where $\bPi$ is the projection operator onto the signaling subspace, $\nabla_{\bPi} \mathcal{L}$ is the Euclidean gradient of some cost $\mathcal{L}$ that quantifies the quality of the signaling architecture with respect to $\bPi$, and $[A,B] = AB - BA$ is the commutator.
This condition states that the optimal signaling subspace is the span of some subset of the eigenvectors of a particular operator defined by the cost function, and thus the optimal signaling architecture can be found by diagonalizing this operator. This picture implies that the choice of signaling modes can be cast as an eigenvalue problem and thus introduces a concept of a ``signaling priority'' spectrum that quantifies the relative value of signaling different features of the collective state. One would intuitively expect the signaling priority of any particular mode to depend on how often it is perturbed and how slowly it relaxes. We now consider a limit where this intuition can be made concrete.

\subsection{Linearized dynamics}

The signaling priority spectrum can be calculated in the limit where the dynamics are linear around the desired state $\bze= 0$.
In SI~Sec.~\ref{si-sec:dimensionless-parameters}, we identify two relevant dimensionless quantities:
\begin{equation}
\begin{aligned}
  \nu_\mu
  =
  \frac{\varepsilon \kappa_\mu}{\gamma},
  \qquad
  \Lambda = \frac{1}{r \varepsilon \gamma}
  .
\end{aligned}
\end{equation}
The former quantifies the passive relaxation rate of mode $\mu$ relative to the perturbation rate, and the latter quantifies the strength of the feedback relative to the perturbation rate.
Inspired by the linear-quadratic-regulator approximation in control theory, we consider the ansatz that the principal function is quadratic in $\bb$: $S(\bb) = \frac{1}{2} \bb^T\bm{Q} \bb$.
With this linearization, we evaluate all metric matrices at $\bze = 0$. We work perturbatively in the weak-control limit (i.e., $\Lambda \gg 1$) and further assume perturbations are drawn from a Gaussian distribution in $\bze$-space,
\begin{equation}
  p_{\rm pert}(\bze) \propto
  \exp\!\qty(-\bze^T
  \bSi^{-1}\bze/2),
\end{equation}
which together with linearized dynamics admits an exact solution for $\mathcal{L}_{\bW}$.

We will begin by considering the case where the distribution of perturbations is aligned with the passive relaxation modes of the system, so $\Sigma_{\mu\nu} = \sigma_\mu^2 \, \delta_{\mu\nu}$.
In SI~Sec.~\ref{si-sec:limited-signaling}, we show that:
\begin{equation}
\begin{aligned}
  \mathcal{L}_{\bW}
  &=\frac{1}{2\gamma}\tr\qty[\frac{ \bSi }{\bm{I} + \varepsilon \bG/\gamma}]
  \\
  &\qquad
  - \frac{1}{2\gamma \Lambda}\tr\qty[\bSi
  \frac{ \varepsilon \bG/\gamma}{
    \qty(\bm{I} + \varepsilon \bG/\gamma)^3
    }
    \bPi_{\bW}
  ]
  +
  O(\Lambda^{-2}),
\end{aligned}
\end{equation}
where $\bPi_{\bW} = \bW^T(\bW \bW^T)^{-1}\bW$ is the projection operator onto the signaling subspace defined by $\bW$.
Because $[\bSi,\bG] = 0$ in this case, the optimal signaling modes are simply some subset of the passive relaxation modes of the system, and the optimal subset depends on $\kappa_\alpha$. We show in the SI~Sec.~\ref{si-sec:diagonal_perturbations} that the optimal signaling modes are those that have the largest signaling priority $\ds_{\alpha}$, where
\begin{align}
  \ds_\alpha^2 = \sigma_\alpha^2 \frac{
  \nu_{\alpha}}{
   (1 +  \nu_{\alpha})^3}
  .
  \label{eq:diagonal-priority}
\end{align}
The optimal signaling strategy is to choose the $M$ modes with the highest signaling priority.
Selecting modes $\alpha = 1,\ldots,M$ with the largest $\ds_\alpha$ decreases the expected accumulated KL divergence from the desired state as:
\begin{equation}
\begin{aligned}
  \int \dd{t}
  \gamma
  &
  e^{-\gamma t}
  \ev{
    D_{\rm KL}(\rho_t \Vert q)
  }
  \\
  &
  =
  \frac{1}{2}
  \sum_{\mu = 1}^\infty
  \frac{
    \sigma_{\mu}^2
  }{
    1 + \nu_\mu
  }
  -
  \frac{1}{\Lambda}
  \sum_{\alpha = 1}^M
  \ds_\alpha^2
  +
  O\qty(
    \Lambda^{-2},
    \zeta^3
  )
  .
\end{aligned}
\end{equation}
This form of $\ds_\alpha$ reveals a crossover: the term that depends on $\kappa_{\alpha}$ in Eq.~\eqref{eq:diagonal-priority} is maximized at $\nu_{\alpha}^{\star} = 1/2$ (i.e., $\kappa_{\alpha}^{\star} = \gamma/(2\varepsilon)$).
To understand why there is an optimum, note first that under passive dynamics $(\dv*{t})D_{\rm KL}(\rho_t\Vert q) \approx -\frac{\varepsilon}{2} \kappa_\mu \zeta_\mu^2 $ for a perturbation along mode $\mu$.
Thus, $\ve\kappa_\mu$ is the contribution of that mode to the passive decay rate of the KL divergence. Moreover, the same eigenvalue also controls the cost of actively transporting the distribution along that mode. That is, a control of the form $\bu=a_\mu\nabla\chi_\mu$ has transport cost $\frac{r}{2}\int \dd{\bx}q(\bx)\norm{\bu(\bx)}^2=\frac{r}{2}\kappa_\mu a_\mu^2,$ while it generates dynamics $\dot\zeta_\mu=\kappa_\mu a_\mu$ to leading order near $q$. Generating a fixed velocity $\dot\zeta_\mu=b_\mu$ costs $\frac{r}{2}\frac{b_\mu^2}{\kappa_\mu}$. Thus, large-$\kappa_\mu$ modes are easy to transport but also relax quickly on their own, whereas small-$\kappa_\mu$ modes are both expensive to transport and persist for a long time if left uncontrolled. An intermediate scale $\kappa_{\alpha}^{\star}$ is selected because the fast modes $\nu_{\alpha} \gg 1$ relax much faster relative to the rate of perturbations. The slow modes $\nu_{\alpha} \ll 1$, on the other hand, relax slowly relative to this rate, and it is preferable to wait for a perturbation to reset this mode rather than pay a significant cost to rectify it.

In Fig.~\ref{fig:optimal-signaling-modes}, we plot the optimal signaling modes for two example distributions when the perturbations are aligned along the passive modes (i.e., $\bSi$ is diagonal) and perturbations are infrequent ($\gamma \to 0$). For the two-peak Gaussian mixture distribution target, the optimal signaling mode detects how unbalanced the distribution is between the two peaks, and all other lower-priority modes detect features localized to individual peaks. This preference is because rebalancing mass across the peaks is very slow as particles must be transported across the barrier, whereas features localized to individual peaks relax much faster. For the four-peak two-dimensional Gaussian mixture distribution arranged in a square, the highest-priority optimal signaling modes detect the peak balance of diagonally opposite peaks, the next highest-priority mode detects a checkerboard pattern, and the lower-priority modes detect features localized to individual peaks.
Generally, when there are separated peaks, there is a spectral gap in the passive relaxation rates in which the very slow modes correspond to the peak (im)balance and the faster modes correspond to features localized to individual peaks; in these configurations, the slowest relaxation rates are analogous to Kramers escape rates and are determined not just by barrier height but also by the width of the barrier and the curvature of the peaks~\cite{hanggi1990reaction}.
In SI~Sec.~\ref{si-sec:additional-optimal-signaling-modes}, we show optimal signaling modes for additional example distributions.

In the more general case where the distribution of perturbations is not aligned with the passive relaxation modes (i.e., $[\bSi,\bG] \neq 0$), we investigate the infrequent perturbation limit (i.e., $\nu_1 \gg 1$).
In this limit, we show that the signaling modes should be chosen to be the top $M$ generalized eigenvectors of the problem:
\begin{equation}
  \qty(
    \int_0^\infty\dd{t}
    e^{-\frac{\varepsilon}{2}\bG t}
    \bSi\,
    e^{-\frac{\varepsilon}{2}\bG t}
  )\bw_\alpha
  =
  \frac{\varepsilon}{\gamma^2} \ds_\alpha^2
  \,
  \bG\,\bw_\alpha,
  \label{eq:generalized-eigenvalue-problem}
\end{equation}
where $\bw_{\alpha}$ is the $\alpha$th row of $\bW$. The left-hand side is the time-integrated perturbation covariance evolved under the passive dynamics; it measures how much a perturbation in direction $\bw_\alpha$ persists before being passively corrected.
The right-hand side encodes how expensive it is to control in direction $\bw_\alpha$.
The optimal signaling modes therefore balance the accumulated cost of uncorrected perturbations against the geometric cost of transport in the space of distributions as measured by $\bG$.

\begin{figure}[t]
  \includegraphics{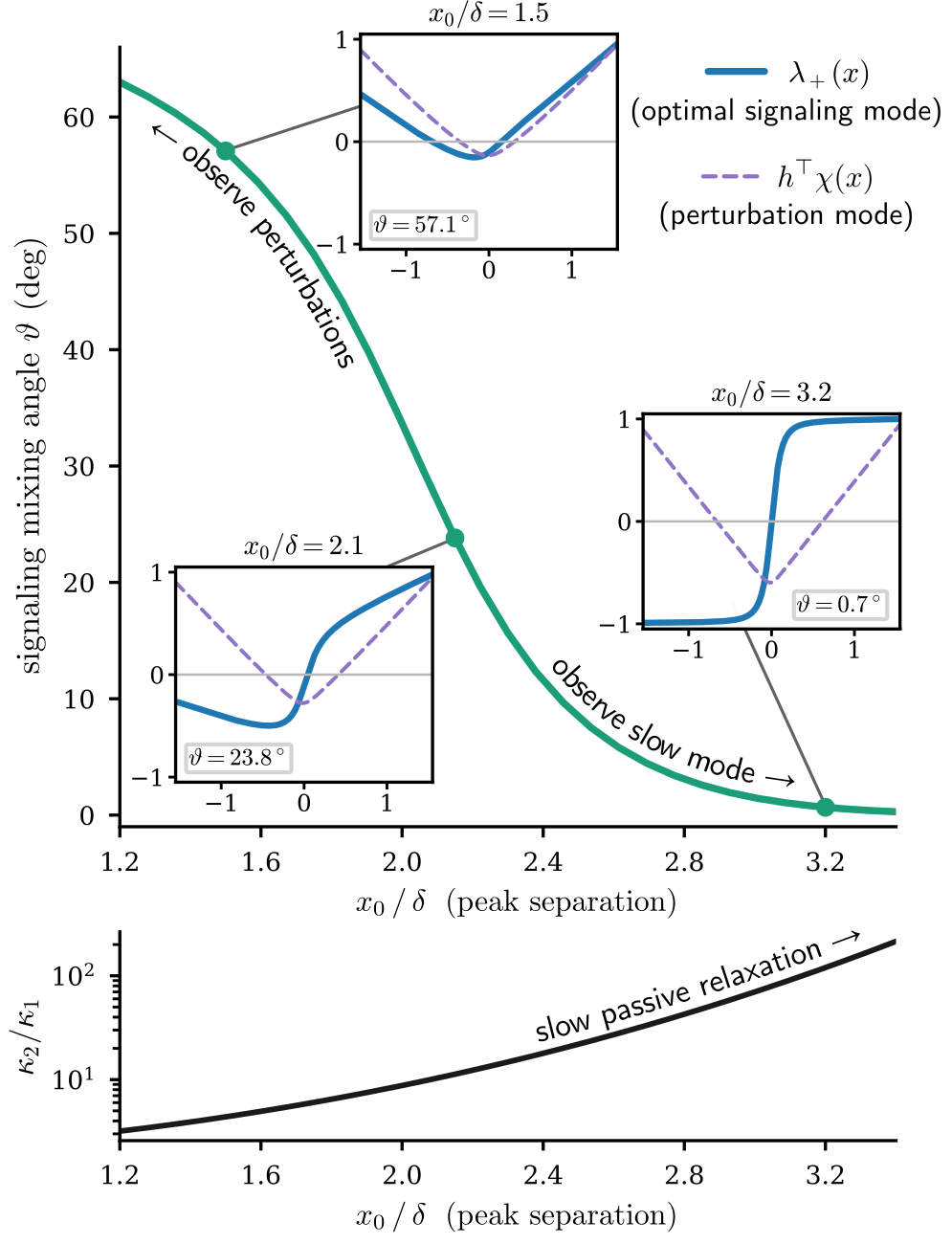}
  \caption{\label{fig:signaling-mixing}
  Optimal signaling trades off between signaling the most perturbed mode and signaling the slowest mode.
  Top: the mixing angle $\vartheta$ between signaling the direction of perturbations $\bm{h}^T\bch$ and the slowest mode $\chi_1$ biases toward signaling the slow mode as the peak separation $x_0$ increases. The insets show comparisons between the perturbation direction and the optimal signaling mode $\lambda_+$ for different values of $x_0$.
  Bottom: the ratio of the passive decay rates between the second-slowest and slowest modes $\kappa_2/\kappa_1$ becomes very large as peak separation $x_0$ increases, so the slowest mode becomes much slower than the other modes and thus more important to signal.
  SI~Fig.~\ref{si-fig:signaling-mixing} serves as a companion figure to this one, visualizing the perturbations and optimal signaling modes in more detail.
  }
\end{figure}
\subsection{An example of optimal signaling modes for a two-peak Gaussian mixture distribution}
\label{sec:two-peak-example}

\begin{figure*}[ht!]
  \includegraphics[width=\linewidth]{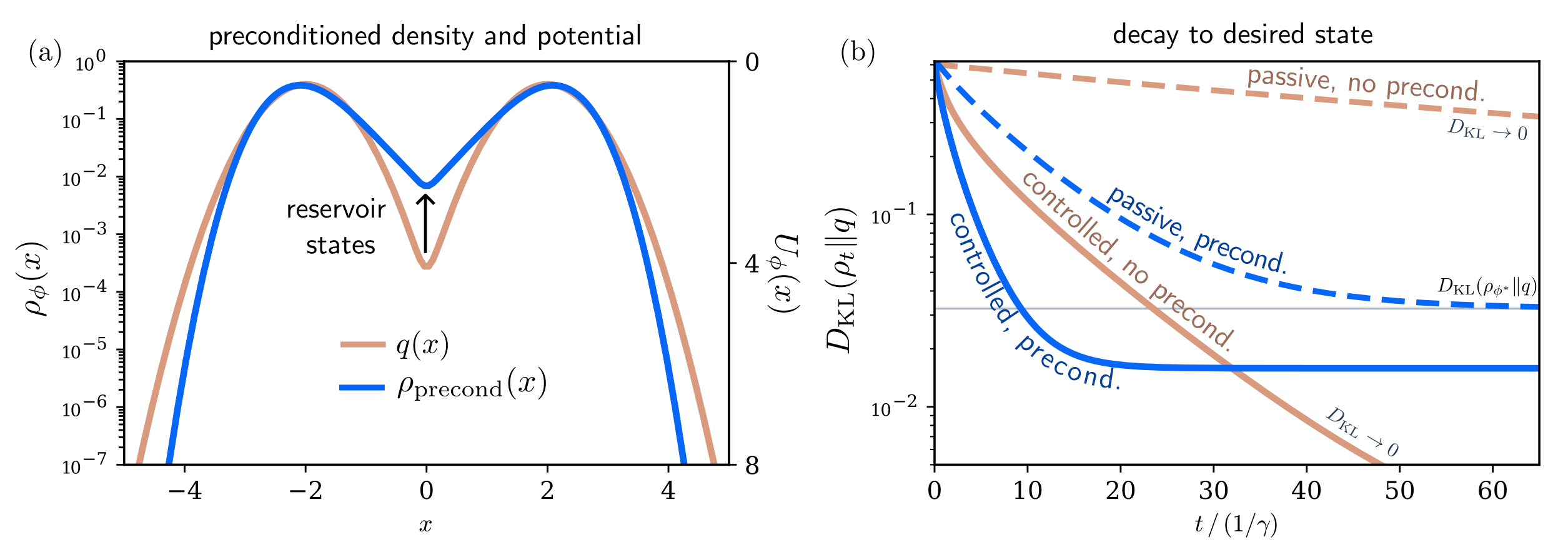}
  \caption{
    \label{fig:preconditioning}
    Preconditioning introduces reservoir states that accelerate convergence at short times at the cost of long-time accuracy.
    (a) The preconditioned stationary distribution $\rho_{\rm precond}$ (blue, dark) develops elevated probability density between the peaks of the target $q$ (orange, light), creating \textit{reservoir states} that lower the inter-peak barrier in the preconditioned potential $U_{\bph^\ast}$ relative to $U_0$.
    (b) $D_{\rm KL}(\rho_t \Vert q)$ as a function of time for passive and feedback-mediated dynamics with ($\bph = \bph^\ast$) and without ($\bph = \bm{0}$) preconditioning, in response to perturbations along the first (inter-peak) mode only ($\bSi \propto \bm{e}_1 \bm{e}_1^T$).
    Without preconditioning, both passive and feedback-mediated dynamics converge to $q$ but slowly, limited by the large inter-peak barrier.
    Preconditioning dramatically accelerates convergence at short times for both passive and feedback-mediated dynamics, but the system plateaus at
    $D_{\rm KL}(\rho_{\bph^\ast} \Vert q) > 0$ rather than converging to $q$, since the preconditioned passive dynamics drive the system toward $\rho_{\bph^\ast}$ rather than $q$.
  }
\end{figure*}

To demonstrate how this generalized eigenvalue problem encodes the tradeoff, we now consider a simple example where the desired state is a one-dimensional two-peak Gaussian mixture distribution:
\begin{equation}
\begin{aligned}
  q(x)
  &=
  \frac{1}{2\sqrt{2 \pi \delta^2}}
  \qty(
    e^{-(x-x_0)^2/2 \delta^2}
    +
    e^{-(x+x_0)^2/2 \delta^2}
  )
\end{aligned}
\end{equation}
with $x_0 \gg \delta$.
Asymptotically, we show in the SI that the passive relaxation rates are $\kappa_1 \approx \frac{4 x_0}{\pi\sqrt{2\pi} \delta^3} e^{-x_0^2/2\delta^2}$ for the first mode and $\kappa_2, \kappa_3 \approx 1/\delta^2$, $\kappa_4, \kappa_5 \approx 2/\delta^2$, etc. for the higher modes.
The first basis mode takes the form $\chi_1(x) \approx \frac{4}{\pi} \arctan \tanh(x x_0/2 \delta^2)$ which is a sigmoid switching mode, and the higher modes are Hermite polynomials localized to each peak, symmetrized and antisymmetrized across the two peaks.
These modes and their corresponding signaling priorities (in the case where perturbations are rare and are aligned with the passive relaxation modes, so that $\ds_\alpha \sim \gamma/(\varepsilon\kappa_\alpha)$) are plotted in Fig.~\ref{fig:optimal-signaling-modes}(a).

With these closed-form expressions for the basis functions and passive relaxation rates, we can evaluate the optimal signaling modes for different choices of the perturbation covariance $\bSi$.
For example, let us consider rank-one perturbations that only excite the first three modes of the form:
\begin{equation}
\begin{aligned}
  \bSi \propto
  \bm{h}\bm{h}^T,
  \qquad
  \bm{h}
  =
  \begin{bmatrix}
    \cos \alpha \\
    \eta \sin \alpha \cos \beta \\
    \eta \sin \alpha \sin \beta \\
    \vdots
  \end{bmatrix}
  .
\end{aligned}
\end{equation}
Write the perturbation vector as $\bm{h} = \cos \alpha \, \bm{e}_1 + \eta \sin \alpha \,\bm{e}_\parallel$ where $\bm{e}_\parallel \equiv \cos \beta \, \bm{e}_2 + \sin \beta \, \bm{e}_3$ is the unit vector aligned to the perturbation but orthogonal to the first mode, and let $\bm{e}_\perp \equiv - \sin \beta \, \bm{e}_2 + \cos \beta \, \bm{e}_3$ be the unit vector orthogonal to the perturbation and orthogonal to the first mode.
In SI~Sec.~\ref{si-sec:signaling-priority-mixing-angle-example}, we show that the signaling modes and their signaling priorities are given by:
\begin{equation}
\begin{aligned}
  &
  \bm{w}_+
  \propto
  \frac{
    \cos \vartheta
  }{
    \sqrt{\kappa_1}
  }
  \bm{e}_1
  +
  \frac{
    \sin \vartheta
  }{
    \sqrt{\kappa_2}
  }
 \bm{e}_\parallel,
  \quad
  &&
  \ds_+
  \approx
  \sigma
  \frac{
    \abs{\cos \alpha}
  }{
    \gamma
    \varepsilon
    \kappa_1
  }
  \\
  &
  \bm{w}_-
  \propto
  -
  \frac{
    \sin \vartheta
  }{
    \sqrt{\kappa_1}
  }
  \bm{e}_1
  +
  \frac{
    \cos \vartheta
  }{
    \sqrt{\kappa_2}
  }
  \bm{e}_\parallel,
  \quad
  &&
  \ds_-
  \approx
  \sigma
  \eta
  \frac{\abs{ \sin \alpha}
  }{
    \gamma
    \varepsilon
    \kappa_2
  }
  \\
  &
  \bm{w}_\perp
  \propto
  \bm{e}_\perp,
  \quad
  &&
  \ds_\perp
  =
  0
\end{aligned}
\end{equation}
where
\begin{equation}
\begin{aligned}
  \tan 2 \vartheta
  &=
  \frac{
    4\eta \tan \alpha
    \frac{
      \kappa_1^{3/2}
    }{
      \sqrt{\kappa_2}
      (\kappa_1 + \kappa_2)
    }
  }{
    1
    -
    \qty(
      \frac{\kappa_1}{\kappa_2}
      \eta
      \tan \alpha
    )^2
  }
  .
\end{aligned}
\end{equation}
We can see there is nontrivial mixing and a crossover; when $\eta\abs{\tan \alpha} \ll \frac{\kappa_2}{\kappa_1}$, $\vartheta \sim 0$ and the optimal signaling mode is aligned with the first passive relaxation mode, but when $\eta\abs{\tan \alpha} \gg \frac{\kappa_2}{\kappa_1}$, $\vartheta \sim \pi/2$ and the optimal signaling mode is aligned with $\bm{e}_\parallel$ which is the direction of the perturbation but orthogonal to the first passive relaxation mode.
Because $\kappa_1$ is exponentially smaller than $\kappa_2$, generically, it is often better to signal the slowest passive relaxation mode.
In Fig.~\ref{fig:signaling-mixing}, we plot how the mixing angle $\vartheta$ changes as a function of $x_0/\delta$ (i.e., peak separation) which affects the ratio $\kappa_2/\kappa_1$ and the crossover point.
In the figure, we demonstrate how the optimal signaling mode changes from being aligned with the perturbation $\bm{h}^T\bch$ to being aligned with the slowest passive relaxation mode $\chi_1$ as the peak separation increases and $\kappa_1$ becomes smaller relative to $\kappa_2$.

\section{Homeostatic equilibrium and reservoir states}
\label{sec:preconditioning}

So far, we have assumed that the intrinsic dynamics of each particle is $\bu_0 = - \nabla U_0 = (\varepsilon/2)\nabla \log q$. This choice is intuitive as it ensures that the system will still converge to $q$ in the absence of feedback, while the role of feedback is essentially to accelerate convergence. This intuition turns out to be incorrect, especially when the system is frequently perturbed. To see why this is the case, consider the two-peak Gaussian mixture distribution discussed in Sec.~\ref{sec:two-peak-example}. The passive dynamics $\bu_0 = - \nabla U_0$ has a large barrier between the two peaks of the distribution. Any imbalance in the mass on each side of the barrier requires the system to significantly lower the barrier. In other words, the feedback control has to act against the system's passive dynamics, thereby incurring a significant cost. If such imbalances occur frequently, the system might as well choose a modified landscape for the passive dynamics, even though this means the system will not asymptotically converge to $q$.

We conceive of this modification as a form of \emph{preconditioning} or design of a \emph{homeostatic equilibrium} that prepares the system to respond more efficiently to expected perturbations. To numerically optimize for this modified passive dynamics, we represent it as:
\begin{equation}
\begin{aligned}
  \bu_0(\bx)
  =
  \frac{\varepsilon}{2}\nabla\log\qty[q(\bx)\,\exp\qty(\sum_\mu\phi_\mu\chi_\mu(\bx))],
\end{aligned}
\end{equation}
where $\{\phi_\mu\}$ are the $\bze$-space coefficients of the preconditioned passive stationary distribution, $\rho_{\bph}$. This representation of $\bu_0$ makes the preconditioning problem compatible with the exponential representation we introduced previously and makes the problem numerically tractable. Recall that this preconditioning takes the form of an outer-loop meta-optimization over the passive dynamics in Eq.~\eqref{eq:S},
where the inner optimization involves solving an identical optimal control problem but with modified passive dynamics.

In Fig.~\ref{fig:preconditioning}, we show the result of this meta-optimization for the two-peak Gaussian mixture distribution with perturbations along the first (inter-peak) mode only.
We see that preconditioning introduces \emph{reservoir states}, regions of elevated probability density between the metastable peaks of $q$, which lower the inter-peak barrier. When the collective state is imbalanced, the particles occupying the reservoir states quickly shift to the appropriate side to restore homeostasis. While this accelerates convergence to the desired state at short times, the system plateaus at $D_{\rm KL}(\rho_{\bph^\ast} \Vert q) > 0$ rather than converging to $q$.

There is a natural analogy of the preconditioned stationary distribution $\rho_{\bph}$ with the concept of an entropy-regularized Wasserstein barycenter of the distribution of perturbations~\cite{chizat2025doubly, Chewi2024StatisticalOptimalTransport}, which is the distribution that minimizes the expected Wasserstein distance to the distribution of perturbations plus an entropy regularization term, as discussed in SI~Sec.~\ref{si-sec:preconditioning-wasserstein-barycenter}.

\begin{figure}[t]
  \includegraphics[width=\linewidth]{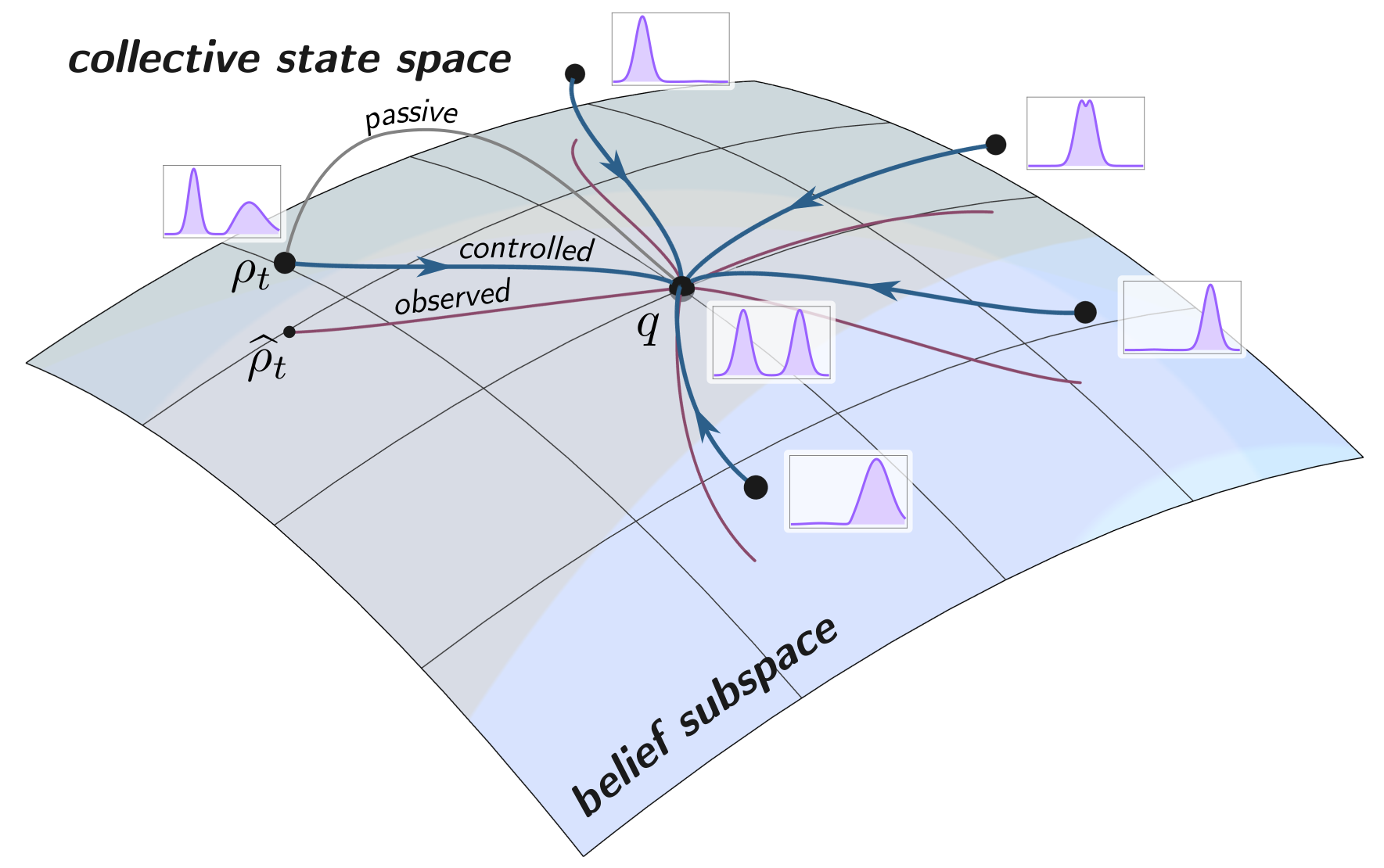}
  \caption{
    \label{fig:trajecs-schem}
    Self-organized robustness can be conceived of as an attractor in the space of distributions which is designed to be more robust, more efficient, and faster to converge than a passive non-feedback system.
    The location of the system's current state cannot be directly observed, so control must be determined based on an estimated state which lives in a lower-dimensional belief subspace. The system's state is influenced by both passive dynamics and active control, which together determine the trajectory of the system in the space of distributions.
  }
\end{figure}

\section{Discussion}

Self-organization occurs at different scales in biology~\cite{kirschner2000molecular}. %
In our model, we sought to capture a notion of self-organization in a system of interacting units where each unit has the capacity to process relatively sophisticated information. This feature is especially striking in tissue-scale biology where interactions between cells of different types with complex intracellular regulatory pathways lead to emergent functional properties~\cite{meizlish2021tissue, Smart2023EmergentProperties}. 
Here, we focus on a model where each unit's state follows recurrent dynamics on a multistable landscape, and units interact such that the collective state also follows an optimized gradient flow toward a desired collective state. 
This gradient-flow picture can be made precise: under the quadratic transport cost, the closed-loop Fokker--Planck dynamics are a Wasserstein gradient flow on the space of probability distributions (Eq.~\eqref{eq:wasserstein-gradient-flow}). Fig.~\ref{fig:trajecs-schem} illustrates this distribution-level view of self-organized robustness.

The model offers a preliminary theoretical framework for describing design principles underlying such systems. In general, there are at least four properties that define a self-organizing system with the aforementioned features: (1) which signals each unit conveys, $\bl(\bx)$, (2) how signals mix, $\bl(\bx) \to \bvp$, (3) how each unit modulates its state based on signals, $\bu(\bx,\bvp)$, and (4) the signaling-independent passive dynamics of each unit's state, $\bu_0(\bx)$. The closed loop between signaling and response leads to nontrivial population dynamics.

We consider a scenario where signals are well-mixed and thus convey information about the marginal density, $\rho(\bx)$, which characterizes heterogeneity in particle states. Implicit in our model is the assumption that the desired collective state, $q$, is set by (here, unspecified) exogenous signals that determine which collective state this system should maintain, and the endogenous signals $\bvp$ enable the system to achieve this desired state.

The mean-field assumption simplifies the analysis: each unit receives identical information and the signals $\bvp$ can be interpreted as generalized moments of the collective state $\rho$. Each unit's estimate of the collective state can then be viewed through the lens of moment-matching. Specifically, since each unit knows what signals it should expect to sense when the system is at $q$, the instantaneous signals $\bvp$ convey the extent to which this constraint is not satisfied. It is natural, then, to work in the dual space of Lagrange multipliers $\bb$, which capture both the extent and directions along which these constraints are broken. The dual variables can thus be interpreted as internal belief states that encode each unit's estimate of the collective state. This internal belief state allows each unit to efficiently modulate its state to satisfy appropriate constraints.

Since there can only be a finite set of signals (say, $M$), the variations in $\bvp$ (and $\bb$) lie on a restricted manifold. When we work in a linearized regime close to homeostasis, this manifold is linear, allowing us to describe the evolution of the internal belief variables (and thus each unit's estimate of the collective state) within an $M$-dimensional subspace. The choice of what signals to convey determines which subspace the dynamics will span, and thus selecting an interaction structure involves selecting an appropriate subspace, or equivalently, optimizing for the right projection given some loss.

There are possibly many modeling choices for the loss that are reasonable; our choice is based on two desiderata: (1) it contains two terms, one of which penalizes deviations from homeostasis while the other regularizes the rate at which each unit can modulate its state, i.e., the feedback-based control, and (2) the regularization term on control is quadratic. A term that regularizes the control is necessary to fully specify the problem. The quadratic choice in particular leads to an intuitive description of the underlying dynamics as a dynamically modulated potential landscape and makes the analysis tractable.

Given a choice of loss, the model offers insights on which signals each unit must employ and how each unit must respond. First, since signals convey information along a restricted manifold, the cost of changing each unit's state forces each unit to modulate its state only in directions along which it can sense. That is, changes in state along other directions cannot be measured and should instead be accounted for in each unit's passive dynamics. This leads to a connection between signaling and control: the modes along which each unit senses should be the ones along which feedback control modulates the landscape. Second, signals should convey information about the modes that the passive dynamics cannot quickly relax to equilibrium by themselves. However, this is not a strict rule as it is possible that these slow modes are infrequently perturbed. Thus, the optimal choice of signaling modes should take into account both the rate at which passive dynamics relax those modes and the rate at which those modes are perturbed. In the language of self-assembly, these results quantify the intuition that a robust self-organizing system should favor mechanisms that account for frequently encountered kinetic traps~\cite{Hagan2021EquilibriumMechanisms, Pelc2025CouplingDifferentialAdhesion}.

We further show that the optimal choice of passive dynamics is one where the passive dynamics do \emph{not} eventually relax the system to the desired collective state. Rather, it is preferable to converge to a flatter equilibrium landscape, or, in other words, it is optimal to maintain a reservoir of otherwise spurious intermediate states that help the system respond rapidly to perturbations.

While inspired by biological self-organization in multicellular systems, the model should be viewed as a conceptual device that helps isolate key features of a self-organizing system. Perhaps the most closely related biological system is the control of adult stem cell populations in many tissues, including testes, hair follicles and intestinal crypts~\cite{mesa2018homeostatic, simons2011strategies, Klein2011UniversalPatternsStemCellFate, jorg2021stem, kitadate2019competition}. Here, emerging evidence points toward collective mechanisms for feedback control rather than an individual-level mechanism that requires stem cells to achieve a fine balance between self-renewal and differentiation. Other examples include global feedback regulation during hematopoiesis~\cite{Wu2024IL17APrimesErythropoieticFeedback} and communication-mediated homeostasis of macrophage-fibroblast ratios in tissues~\cite{zhou2018circuit}. One should note, however, that local cell-cell contacts and spatial structure play a role in these examples, and interactions may not necessarily be described in mean-field terms. Nevertheless, reservoir states emerge naturally from an optimization principle, and offer an explanation for the existence of hybrid cell types in developmental systems~\cite{velten2017human, farrell2018single, soldatov2019spatiotemporal, byrnes2018lineage, weng2024deciphering} as cell states that enable rapid response to perturbations.

Our framework connects several threads in the mathematical physics literature, including optimal transport and stochastic control~\cite{Chen2021StochasticControlLiaisons}.
The optimal control strategy takes the form of a Wasserstein gradient flow on the exponential family manifold~\cite{Ambrosio2008GradientFlows,Santambrogio2015OptimalTransportApplied}, and the belief-state reduction has a natural interpretation in terms of information geometry---the signaling projection is an oblique projection in the Fisher metric~\cite{Amari2016InformationGeometryApplications}.
The structure of the value function and its relationship to the generalized eigenvalue problem bears a close resemblance to successor features in reinforcement learning, where a similar decomposition of future cost into geometry and policy arises~\cite{Ollivier2025WhichFeaturesBestSuccessorFeatures}.
Our work is also related to the literature on mean-field games and mean-field control in which a large population of agents compete or cooperate to optimize individual or collective rewards in an environment where inter-agent interactions are mediated through the population distribution rather than through pairwise interactions~\cite{bertucci2026meanfieldgamesincomplete,sen2019mean,huang2006large,cardaliaguet2025mfg,yong2013linear}.
Because our objective emphasizes repeated recovery from perturbations rather than convergence from a single initial condition, it is also conceptually adjacent to stochastic resetting and restart processes~\cite{evans2020stochastic}.
More broadly, this perspective is close in spirit to particle- and flow-based approaches to generative modeling and variational inference, where distributions are transformed by learned or geometry-driven dynamics rather than represented only through static densities~\cite{Liu2016SteinVariationalGradientDescent,Lipman2024FlowMatchingGuide,Haviv2024WassersteinFlowMatching}.

Our results suggest that the geometry of the collective state space, the statistics of perturbations, and the architecture of signaling are not independent features of a robust system but are aspects of a single co-design problem.
A particularly promising direction to explore this design space is to generalize our formulation to multi-agent reinforcement learning (MARL) settings~\cite{Deshpande2025EngineeringMorphogenesis,Nimonkar2025SelfSupervisedGoalReaching}.
A MARL perspective allows for the incorporation of more complex signaling architectures, including spatially structured and time-delayed signaling.
Proliferation, death, and the co-evolution of spatial structure and internal states allow for the emergence of collective behaviors, and a MARL formulation can potentially accommodate these features to explore more biologically plausible models of self-organization.
A machine learning perspective also allows for investigation of high-dimensional dynamics whose study is not tractable with the numerical methods we have employed here.
We believe this optimization-driven path opens the door for the discovery of new emergent phenomena in such systems.

\section*{Code availability}

Code demonstrating the main results of this paper is available at~\cite{code}.

\begin{acknowledgments}
  We thank Fran\c{c}ois Bourassa, Nikolas Claussen, Catherine Ji, Pankaj Mehta, Matthew Smart, and members of the Reddy lab for helpful discussions and comments on the manuscript.
  E.B. acknowledges support from the Fannie and John Hertz Foundation Fellowship.
  This material is based upon work supported by the National Science Foundation Graduate Research Fellowship Program under Grant No. DGE-2444107. Any opinions, findings, and conclusions or recommendations expressed in this material are those of the authors and do not necessarily reflect the views of the National Science Foundation.
  G.R. was partially supported by a joint research agreement between NTT Research Inc and Princeton University, and a grant from Coefficient Giving.
  The simulations presented in this article were performed on computational resources managed and supported by Princeton Research Computing, a consortium of groups including the Princeton Institute for Computational Science and Engineering (PICSciE) and Research Computing at Princeton University.
\end{acknowledgments}

\bibliography{references}

\onecolumngrid
\clearpage

\begin{center}
  \textbf{\large Supplementary Information: Self-organized robustness in mean-field interacting systems}
\end{center}
\vspace{2ex}

\setcounter{equation}{0}
\setcounter{figure}{0}
\setcounter{table}{0}
\setcounter{page}{1}
\setcounter{section}{0}
\renewcommand{\theequation}{S\arabic{equation}}
\renewcommand{\thefigure}{S\arabic{figure}}
\renewcommand{\thepage}{S\arabic{page}}
\renewcommand{\thesection}{S\arabic{section}}
\providecommand{\theHequation}{}
\providecommand{\theHfigure}{}
\providecommand{\theHtable}{}
\providecommand{\theHsection}{}
\providecommand{\theHsubsection}{}
\providecommand{\theHsubsubsection}{}
\renewcommand{\theHequation}{SI.\arabic{equation}}
\renewcommand{\theHfigure}{SI.\arabic{figure}}
\renewcommand{\theHtable}{SI.\arabic{table}}
\renewcommand{\theHsection}{SI.\arabic{section}}
\renewcommand{\theHsubsection}{SI.\arabic{section}.\arabic{subsection}}
\renewcommand{\theHsubsubsection}{SI.\arabic{section}.\arabic{subsection}.\arabic{subsubsection}}

\startcontents[SI]
\printcontents[SI]{}{1}{%
  \setcounter{tocdepth}{3}%
  \titlecontents{section}[2.5em]{\addvspace{0.7em}\bfseries}%
    {\contentslabel{2.5em}}{\hspace*{-2.5em}}%
    {\titlerule*[0.5pc]{.}\bfseries\contentspage}%
  \titlecontents{subsection}[5em]{}%
    {\contentslabel{2.5em}}{\hspace*{-2.5em}}%
    {\titlerule*[0.5pc]{.}\contentspage}%
  \titlecontents{subsubsection}[7.5em]{}%
    {\contentslabel{2.5em}}{\hspace*{-2.5em}}%
    {\titlerule*[0.5pc]{.}\contentspage}%
}
\vspace{2ex}

\clearpage

\section{Collective optimal control with full-state information and the exponential family representation}
\label{si-sec:full-state-exponential-family}

\subsection{Optimal control of a distribution}

In the main text, we have argued that self-organization can be conceived of as a problem of designing attractors in the space of distributions.
In particular, we have considered a system of particles $\{\bx_i\}$ with dynamics:
\begin{equation}
\begin{aligned}
  \dot{\bx}_i
  =
  \bu_0(\bx_i) + \bu(\bx_i; \rho_t)
  +
  \sqrt{\varepsilon}
  \,\be_i(t)
\end{aligned}
\end{equation}
where $\bu_0$ is the drift due to passive dynamics, $\bu$ is the drift due to feedback control which can depend on the collective state $\rho_t$ of the system, and $\be_i(t)$ are independent white noise processes. The collective state $\rho_t$ evolves according to the Fokker-Planck equation:
\begin{equation}
\begin{aligned}
  \pdv{\rho_t}{t}
  &=
  -
  \nabla
  \cdot
  \qty(
    \rho_t \, (\bu_0 + \bu)
  )
  +
  \frac{\varepsilon}{2}
  \nabla^2 \rho_t
  .
\end{aligned}
\end{equation}
We are interested in finding the control $\bu(\bx;\rho)$ that minimizes the long-term cost:
\begin{equation}
\begin{aligned}
  \mathcal{L}(\bu)
  &=
  \ev{
    \int_0^\infty
    \dd{t}
    e^{-\gamma t}
    \qty(
      D_{\rm KL}(\rho_t \Vert q)
      +
      \frac{r}{2}
      \int \dd{\bx} \rho_t(\bx) \norm{\bu(\bx;\rho_t)}^2
    )
  }
\end{aligned}
\end{equation}
where $q$ is the desired state, $\gamma$ is the discount factor, and $r$ is the control cost coefficient.
The expectation is taken over the stochastic trajectories and also the distribution of perturbations which set the initial condition $\rho_0$ for each trajectory.
We can solve this problem by introducing the value function
\begin{equation}
\begin{aligned}
  S(\rho_t)
  =
  \min_{\bu,\,(t'>t)} \,
  \ev{
    \int_t^\infty
    \dd{t'}
    e^{-\gamma (t'-t)}
    \qty(
      D_{\rm KL}(\rho_{t'} \Vert q)
      +
      \frac{r}{2}
      \int \dd{\bx} \rho_{t'}(\bx) \norm{\bu(\bx;\rho_{t'})}^2
    )
  }
\end{aligned}
\end{equation}
which is the optimal cost-to-go from any state $\rho_t$.
The dynamic programming principle states that the expected future cost starting from $\rho_t$ and following the optimal policy must be equal to the immediate cost at time $t$ plus the expected future cost starting from $\rho_{t+\dd{t}}$ and following the optimal policy, which leads to the Hamilton-Jacobi-Bellman equation~\cite{Bensoussan2004StochasticControlPartialObservation}:
\begin{equation}
\begin{aligned}
  S(\rho_t)
  =
  \min_{\bu} \,
  \ev{
    \qty(
      D_{\rm KL}(\rho_t \Vert q)
      +
      \frac{r}{2}
      \int \dd{\bx} \rho_t(\bx) \norm{\bu(\bx;\rho_t)}^2
    )
    \dd{t}
    +
    e^{-\gamma \dd{t}} S(\rho_{t+\dd{t}})
  }
\end{aligned}
\end{equation}
\begin{equation}
\begin{aligned}
  &
  S(\rho_{t+\dd{t}})
  =
  S(\rho_t)
  +
  \dd{t}
  \int \dd{\bx}
  \fdv{S}{\rho_t}
  \pdv{\rho_t}{t}
  \\
  &\quad=
  S(\rho_t)
  +
  \dd{t}
  \int \dd{\bx}
  \rho_t
  \Big[
    \bv
    \cdot
    \nabla
    \fdv{S}{\rho_t}
    +
    \frac{\varepsilon}{2}
    \nabla^2
    \fdv{S}{\rho_t}
  \Big]
  ,
\end{aligned}
\end{equation}
Keeping only $O(\dd{t})$ terms and rearranging gives the Hamilton--Jacobi--Bellman (HJB) equation:
\begin{equation}
\begin{aligned}
  0
  =
  &
  \min_{\bu}
  \Bigg\{
    D_{\rm KL}(\rho_t \Vert q)
    +
    \int
    \dd{\bx}
    \rho_t\,
    \qty[
      \frac{r}{2} \norm{\bu}^2
      +
      \bv
      \cdot
      \nabla
      \fdv{S}{\rho_t}
      +
      \frac{\varepsilon}{2}
      \nabla^2
      \fdv{S}{\rho_t}
      -
      \gamma S
    ]
  \Bigg\}
\end{aligned}
\end{equation}
\begin{equation}
\begin{gathered}
  \bv
  =
  \bu_0
  +
  \bu^\ast
  =
  -\nabla U_0(\bx)
  -
  \nabla U(\bx; \rho_t),
  \\
  U(\bx;\rho_t)
  =
  \frac{1}{r}
  \fdv{S}{\rho_t(\bx)}
  .
\end{gathered}
\end{equation}
The cost-to-go function $S$ then is determined by the functional HJB equation:
\begin{equation}
\begin{aligned}
  \gamma \,S
  &=
  D_{\rm KL}(\rho \Vert q)
  +
  \int \dd{\bx} \rho
  \qty[
    \bu_0\cdot\nabla\fdv{S}{\rho}
    -
    \frac{1}{2r}\abs{\nabla\fdv{S}{\rho}}^2
    +
    \frac{\varepsilon}{2}\nabla^2\fdv{S}{\rho}
  ]
\end{aligned}
\label{eq:full-state-hjb}
\end{equation}
While this equation is generally intractable, the exponential family representation introduced in the following section renders it amenable to systematic approximation.

\subsection{Exponential family representation}

We represent a collective state $\rho$ by expanding it in terms of basis functions:
\begin{equation}
\begin{aligned}
  \rho_{\bze}(\bx)
  &=
  q(\bx)
  \exp\qty(
    \zeta_\mu \, \chi_\mu(\bx) - F(\bze)
  )
\end{aligned}
\end{equation}
where we use Einstein summation convention; $F(\bze) = \log \int \dd{\bx} q(\bx) \, e^{\zeta_\mu \chi_\mu(\bx)}$ is the log-partition function; and $\{\chi_\mu\}$ are the basis functions.
If $\rho$ is sufficiently generic and the basis functions $\{\chi_\mu\}$ are sufficiently rich, then there is a one-to-one mapping between $\rho$ and $\bze$.
We introduce the generalized moments $\bxi$ which are the expectation values of the basis functions under $\rho$:
\begin{equation}
\begin{aligned}
  \xi_\mu
  &=
  \int \dd{\bx} \rho(\bx) \, \chi_\mu(\bx)
  =
  \pdv{F}{\zeta_\mu}
  .
\end{aligned}
\end{equation}
We choose the basis functions so that the desired state $q$ corresponds to $\bze = 0$ and thus $\bxi = 0$.
Let us now consider the geometry of this representation.
The KL divergence from the desired state can be written as:
\begin{equation}
\begin{aligned}
  D_{\rm KL}(\rho_{\bze} \Vert q)
  &=
  \int \dd{\bx} \rho(\bx) \, \log\qty(\frac{\rho(\bx)}{q(\bx)})
  =
  \zeta_\mu \, \xi_\mu - F(\bze)
  =
  F^\ast(\bxi)
\end{aligned}
\end{equation}
where $F^\ast(\bxi) = \sup_{\bze} \qty{\zeta_\mu \, \xi_\mu - F(\bze)}$ is the Legendre dual of $F$.
A useful quantity to define is the Hessian of $F$:
\begin{equation}
\begin{aligned}
  C_{\mu\nu}(\bze)
  &=
  \pdv[2]{F}{\zeta_\mu}{\zeta_\nu}
  =
  \int \dd{\bx} \rho_{\bze}(\bx) \, \qty(\chi_\mu(\bx) - \xi_\mu) \qty(\chi_\nu(\bx) - \xi_\nu)
   ,
\end{aligned}
\end{equation}
which is also the Fisher information metric and the covariance of the basis functions under $\rho$.
We also introduce the matrix
\begin{equation}
\begin{aligned}
  G_{\mu\nu}(\bze)
  &=
  \int \dd{\bx} \rho_{\bze}(\bx) \, \nabla  \chi_\mu(\bx) \cdot \nabla \chi_\nu(\bx),
\end{aligned}
\end{equation}
which is an inner product on the space of basis functions that encodes the geometry of optimal transport and is related to the Wasserstein metric as discussed in SI~Sec.~\ref{si-sec:info-geom}.
This is particularly nice for analyzing the evolution of the system under the dynamics:
\begin{equation}
\begin{aligned}
  \dot{\bx}_i
  &=
  \frac{\varepsilon}{2} \nabla \log q(\bx_i)
  +
  a_\mu
  \nabla \chi_\mu(\bx_i)
  +
  \sqrt{\varepsilon} \, \be_i(t),
\end{aligned}
\end{equation}
where $a_\mu$ are some $\bze$-dependent coefficients, later to be determined by the control strategy. We have assumed $U_0 = -(\ve/2)\log q$.
Using the Fokker--Planck equation, we can derive the dynamics of $\bze$:
\begin{equation}
\begin{aligned}
  \dot \zeta_\mu
  &
  =
  C_{\mu\nu}^{-1}
  \dot \xi_\nu
  =
  C_{\mu\nu}^{-1}
  \int \dd{\bx} \pdv{\rho}{t} \,\chi_\nu(\bx)
  \\
  &
  =
  C_{\mu\nu}^{-1}
  \int \dd{\bx}
  \chi_\nu
  \qty[
    -
    \nabla \cdot \qty(
      \rho\qty(
        \frac{\varepsilon}{2} \nabla \log q
        +
        a_\omega
        \nabla \chi_\omega
      )
    )
    +
    \frac{\varepsilon}{2} \nabla^2 \rho
  ]
  \\
  &
  =
  C_{\mu\nu}^{-1}
  \qty[
    \frac{\varepsilon}{2}
    \int
    \dd{\bx}
    \rho\,
    \nabla
    \chi_\nu
    \cdot
    \nabla \log q
    +
    a_\omega
    \int \dd{\bx}
    \rho\,
    \nabla \chi_\omega
    \cdot
    \nabla \chi_\nu
    -
    \frac{\varepsilon}{2}
    \int \dd{\bx}
    \rho\,
    \nabla
    \log \rho
    \cdot
    \nabla
    \chi_\nu
  ]
  \\
  &
  =
  C_{\mu\nu}^{-1}
  \qty[
    a_\omega
    \int \dd{\bx}
    \rho\,
    \nabla \chi_\omega
    \cdot
    \nabla \chi_\nu
    -
    \frac{\varepsilon}{2}
    \zeta_\omega
    \int \dd{\bx}
    \rho\,
    \nabla \chi_\omega
    \cdot
    \nabla
    \chi_\nu
  ]
  \\
  &
  =
  C_{\mu\nu}^{-1}
  G_{\nu\omega}
  \qty(
    a_\omega
    -
    \frac{\varepsilon}{2}
    \zeta_\omega
  )
  .
\end{aligned}
\label{eq:full-zeta-dynamics}
\end{equation}
where we have used $\pdv*{\zeta_\mu}{\xi_\nu} = C_{\mu\nu}^{-1}$, the Fokker--Planck equation, integrated by parts, and used the definition of $C_{\mu\nu}$ and $G_{\mu\nu}$.
This equation shows that without control (i.e., $a_\omega = 0$), the system relaxes toward the desired state at a rate set by $\varepsilon/2$ and the geometry encoded in $G_{\mu\nu}$ and $C_{\mu\nu}$.

A particularly nice choice of basis functions which diagonalizes Eq.~\eqref{eq:full-zeta-dynamics} close to $\bze = 0$ is to choose $\{\chi_\mu\}$ to solve the generalized eigenvalue problem:
\begin{equation}
\begin{aligned}
  -\nabla \cdot (q(\bx) \nabla \chi_\mu(\bx))
  =
  \kappa_\mu \, q(\bx) \, \chi_\mu(\bx)
\end{aligned}
\end{equation}
where $\{\kappa_\mu\}$ are generalized eigenvalues, and we choose the orthonormalization condition:
\begin{equation}
\begin{aligned}
  \int \dd{\bx} q(\bx) \, \chi_\mu(\bx) \, \chi_\nu(\bx)
  &=
  \delta_{\mu\nu}
   .
\end{aligned}
\end{equation}
This choice of basis functions as generalized eigenfunctions is special because it diagonalizes both of these metrics at the origin of $\bze$-space:
\begin{equation}
\begin{aligned}
  C_{\mu\nu}(\bze = 0) = \delta_{\mu\nu},
  \qquad
  G_{\mu\nu}(\bze = 0) = \kappa_\mu \, \delta_{\mu\nu}.
\end{aligned}
\end{equation}
The Fisher metric, $\bC$, being diagonal means that the basis functions are uncorrelated under the desired distribution.
$\bG$ being diagonal means that the basis functions represent orthogonal directions in the space of distributions with respect to the geometry of optimal transport, and $1/\kappa_\mu$ represents the local cost of moving along direction $\mu$ in this space as we will discuss in SI~Sec.~\ref{si-sec:info-geom}.
If we linearize Eq.~\eqref{eq:full-zeta-dynamics} around $\bze = 0$ in the $a_\omega = 0$ case, this becomes:
\begin{equation}
\begin{aligned}
  \dot \zeta_\mu
  =
  -\frac{\varepsilon}{2} \kappa_\mu \delta_{\mu\nu} \zeta_\nu
  +
  O(\zeta^2)
  ,
\end{aligned}
\end{equation}
which means that in this basis, $\{\chi_\mu\}$ represent the passive relaxation modes of the system, and $\{\tfrac{\varepsilon}{2}\kappa_\mu\}$ represent the passive relaxation rates of these modes.
Using that $\xi_\mu \approx \zeta_\mu$ to leading order, we can write the time derivative of the KL divergence as:
\begin{equation}
\begin{aligned}
  \dv{t}
  D_{\rm KL}(\rho_{\bze} \Vert q)
  &
  =
  \dot\zeta_\mu
  \zeta_\mu
  +
  O(\zeta^3)
  =
  -\frac{\varepsilon}{2} \kappa_\mu \delta_{\mu\nu} \zeta_\nu \zeta_\mu
  +
  O(\zeta^3)
\end{aligned}
\end{equation}
which means the contribution of mode $\mu$ to the divergence from the desired state decays at a rate set by $\kappa_\mu$ under the passive dynamics.
This means that $\kappa_\mu$ sets both the natural decay rate of the KL divergence and the (inverse) cost of moving along mode $\mu$ in the space of distributions.

\section{Signaling as projection, belief dynamics, and optimal control with limited signaling}
\label{si-sec:limited-signaling}

\subsection{Posterior estimate of collective state}
\label{si-sec:posterior-estimate}
In our setup, all information about the state is available only through the moments of the signaling modes $\{\lambda_\alpha\}$:
\begin{equation}
\begin{aligned}
  \varphi_\alpha(t)
  =
  \frac{1}{N}
  \sum_{i=1}^N
  \lambda_\alpha(\bx_i(t)),
  \qquad
  \alpha = 1, \ldots, M.
\end{aligned}
\end{equation}
With this data, Sanov's theorem~\cite{touchette2009large} implies that in the $N \to \infty$ case, the most likely collective state of the system is given by the distribution $\hat{\rho}$ that matches the observed moments and minimizes the KL divergence from some prior state $\widetilde q(\bx)$:
\begin{equation}
\begin{aligned}
  \argmin_{\hat{\rho}}
  \;
  &
  D_{\rm KL}(\hat{\rho} \Vert \widetilde q)
  \\
  \text{s.t.,}
  \;&
  \int \dd{\bx} \hat{\rho}(\bx) \, \lambda_\alpha(\bx)
  =
  \varphi_\alpha(t)
  \qquad
  \alpha = 1, \ldots, M
  .
\end{aligned}
\end{equation}
What is the appropriate choice of the prior $\widetilde q$? We provide a more mechanistic Bayesian perspective on estimating the state from the signaling data. A prior over $\bze$ should take into account all values of $\bze$ encountered on the path toward equilibrium from a perturbation $\bze_0$, weighted by the probability $p_{\rm pert}(\bze_0)$ for this perturbation and the time spent before the system is reset to another perturbation. This prior over $\bze$ can be formally written as:
\begin{equation}
\begin{aligned}
  p_{\rm prior}(\bze)
  &=
  \int_0^\infty
  \dd{t}
  \gamma
  e^{-\gamma t}
  \int \dd{\bze_0}
  p_{\rm pert}(\bze_0)
  \delta\qty(
    \bze - \bze(0 \to t)
  )
\end{aligned}
\end{equation}
where $\bze(0 \to t)$ is the trajectory of $\bze$ under the controlled dynamics starting from $\bze_0$.
This is a self-consistent prior and cannot be solved exactly (we will not ever have to), and the posterior is given by:
\begin{equation}
\begin{aligned}
  p_{\rm post}(\bze\vert \{\varphi_\alpha\})
  &
  \propto
  p_{\rm prior}(\bze)
  \prod_{\alpha=1}^M
  \delta\qty(
    \int \dd{\bx} \rho_{\bze}(\bx) \, \lambda_\alpha(\bx)
    -
    \varphi_\alpha
  )
\end{aligned}
\end{equation}
The posterior mean state estimate is:
\begin{equation}
\begin{aligned}
  \hat{\rho}(\bx)
  =
  \int \dd{\bze} p_{\rm post}(\bze\vert \{\varphi_\alpha\})
  q(\bx)
  \exp\qty(
    \zeta_\mu \, \chi_\mu(\bx) - F(\bze)
  )
  =
  \widetilde q(\bx)
  \exp\qty(
    \beta_\alpha \lambda_\alpha(\bx) - \widetilde F(\bb)
  )
\end{aligned}
\end{equation}
where
\begin{equation}
\begin{aligned}
  \widetilde q(\bx)
  &
  =
  \int \dd{\bze} p_{\rm prior}(\bze)
  q(\bx)
  \exp\qty(
    \zeta_\mu \, \chi_\mu(\bx) - F(\bze)
  )
  \\
  \widetilde F
  &
  =
  \log \int \dd{\bx} \widetilde q(\bx) \exp\qty(
    \beta_\alpha \lambda_\alpha(\bx)
  )
\end{aligned}
\end{equation}
with $\beta_\alpha$ being the Lagrange multipliers for the moment-matching constraints:
\begin{equation}
\begin{aligned}
  \pdv{\widetilde F}{\beta_\alpha}
  =
  \int \dd{\bx} \hat{\rho}(\bx) \, \lambda_\alpha(\bx)
  =
  \varphi_\alpha
  .
\end{aligned}
\end{equation}
If we decompose the signaling functions in our basis:
\begin{equation}
\begin{aligned}
  \lambda_\alpha(\bx)
  &=
  W_{\alpha\mu} \, \chi_\mu(\bx),
\end{aligned}
\end{equation}
then we can write the state estimate as:
\begin{equation}
\begin{aligned}
  \hat{\rho}(\bx)
  =
  q(\bx)
  \exp\qty(
    \hat{\zeta}_\mu
    \chi_\mu(\bx) - F(\hat{\bze})
  ),
  \qquad
  \hat{\zeta}_\mu
  =
  \beta_\alpha
  W_{\alpha\mu}
  +
  g_\mu
\end{aligned}
\end{equation}
where $g_\mu$ is an offset to the state estimate that arises from the prior and is independent of the signaling data.
Thus, the prior over $\bze$, which considers self-consistent trajectories over paths in $\bze$-space, is absorbed into this offset $\bm{g}$. For the purpose of numerically solving the problem, this offset can be optimized in the outer-loop of the meta-optimization problem.
We will choose $W_{\alpha\mu} g_\mu = 0$ so that the prior offset does not affect the moment-matching constraints.

The introduction of the prior offset $g_\mu$ allowed us to write $\hat{\rho}$ in the same exponential family form as $\rho_{\bze}$ (i.e., with base measure $q(\bx)$ instead of $\widetilde q(\bx)$), which is a convenient choice for analyzing the dynamics of the system under the control strategy.
The coefficients $\beta_\alpha$ are determined by the moment-matching constraints which means that the state estimate must match the observed signaling data:
\begin{equation}
\begin{aligned}
  \varphi_\alpha
  &=
  \int \dd{\bx} \hat{\rho}(\bx) \, \lambda_\alpha(\bx)
  =
  W_{\alpha\mu}
  \int \dd{\bx} \hat{\rho}(\bx) \, \chi_\mu(\bx)
  =
  W_{\alpha\mu}
  \eval{\xi_\mu}_{\bze = \hat{\bze}}
  =
  W_{\alpha\mu}
  \eval{\pdv{F}{\zeta_\mu}}_{\bze = \hat{\bze}}
\end{aligned}
\end{equation}
This means that in $\bxi$-space, the signaling data is a linear projection of the state estimate, but in $\bze$-space, the signaling data is a nonlinear projection of the state estimate.

\subsection{The limited-signaling optimal control problem and certainty equivalence}
\label{si-sec:limited-signaling-ocp}

In order to make progress with the limited-signaling optimal control problem, we must make an assumption about the relationship between the true state $\rho$ and the posterior state estimate.
Formally, the optimal control problem with limited signaling is the same as the full-state case, but we conceive of it as being solved by a controller that only has access to the signals $\bvp$.
We can write this as:
\begin{equation}
\begin{aligned}
  \argmin_{\bu(\cdot ; \bvp)}
  \;
  \int_0^\infty
  \dd{t}
  e^{-\gamma t}\,
  \E\left[
    D_{\rm KL}(\rho_t \Vert q)
    +
    \frac{r}{2}
    \int \dd{\bx} \rho_t(\bx) \, \norm{\bu(\bx)}^2
    \middle\vert
    \bvp
  \right]
\end{aligned}
\end{equation}
where the expectation is taken over the posterior distribution of the full state given the signaling data, and the dynamics are the same as before.
It is important to note that this is distinct from just replacing $\rho_t$ with $\hat{\rho}_t$ in the full-state optimal control problem because $\hat{\rho}_t$ is the posterior mean state estimate.
For the control regularization term, this is not an issue because $\rho$ appears linearly, so we can simply pass the expectation through and replace it with $\hat{\rho}$.
However, this is not the case for the state cost term:
\begin{equation}
\begin{aligned}
  \E\left[
    D_{\rm KL}(\rho \Vert q)
  \middle\vert
    \{\varphi_\alpha\}
  \right]
  =
  \E\left[
    \int \dd{\bx} \rho(\bx) \, \log\frac{\rho(\bx)}{q(\bx)}
  \middle\vert
    \{\varphi_\alpha\}
  \right]
  \neq
  \int \dd{\bx} \hat{\rho}(\bx) \, \log\frac{\hat{\rho}(\bx)}{q(\bx)}
\end{aligned}
\end{equation}
However, in order to make the problem tractable, we will make the approximation that the controller uses the state estimate $\hat{\rho}$ in place of the true state $\rho$ when evaluating the cost of being far from the desired state.
This assumption allows us to write the trajectory of the estimated state $\hat{\rho}$ as autonomous and thus write a dynamic programming principle for the value function in terms of $\hat{\rho}$.
This is a nontrivial approximation, but it is not uncommon in partially observed optimal control problems, and it is a useful starting point for analyzing the problem; it is often called certainty equivalence in the control literature~\cite{Bensoussan2004StochasticControlPartialObservation}.
The certainty equivalence assumption is valid when the posterior distribution of the true state given the signaling data is sufficiently concentrated around the state estimate $\hat{\rho}$ so that the expected cost of being far from the desired state is well-approximated by the cost evaluated at $\hat{\rho}$.
This is the case when the signaling modes are well-chosen to capture typical values of $\bze$ that the system encounters during its trajectory, which is implicitly the goal of the optimal signaling mode problem.
With these assumptions, the limited-signaling optimal control problem reduces to the same form as the full-state case, but with the state estimate $\hat{\rho}$ in place of the true state $\rho$.

\subsection{Solving the limited-signaling optimal control problem}

The certainty equivalence assumption allows us to write the limited-signaling optimal control problem solely in terms of the belief variables $\bb(t)$ that parameterize the state estimate $\hat{\rho}$.
To solve this problem, we can follow the same steps as in the full-state case, but with the value function now being a function of $\bb$ rather than $\bze$:
\begin{equation}
\begin{aligned}
  S(\bb(t))
  =
  \min_{
    \{a_\mu(\bb(t'))\}\, t' > t
  }
  \ev{
    \int_t^\infty
    \dd{t'}
    e^{-\gamma (t'-t)}\,
    \widehat\ell
  },
  \qquad
  \widehat\ell =
  D_{\rm KL}(\hat{\rho}_{t'} \Vert q)
  +
  \frac{r}{2}
  a_\mu(\bb(t'))
  G_{\mu\nu}
   a_\nu(\bb(t')).
\end{aligned}
\end{equation}
The dynamic programming principle is the same as before:
\begin{equation}
\begin{aligned}
  S(\bb(t))
  =
  \min_{\{a_\mu(\bb(t))\}}
  \qty{
    \qty(
      D_{\rm KL}(\hat{\rho}_t \Vert q)
      +
      \frac{r}{2}
      a_\mu
      G_{\mu\nu}
      a_\nu
    )
    \dd{t}
    +
    (1 - \gamma \dd{t})
    \ev{
      S(\bb(t + \dd{t}))
    }
    +
    \text{const.}
  }.
\end{aligned}
\end{equation}
The time evolution of the cost-to-go function can be determined by the Fokker--Planck equation:
\begin{equation}
\begin{aligned}
  \dv{t}
  S(\bb)
  &
  =
  \pdv{S}{\beta_\alpha} \dot \beta_\alpha
  =
  \pdv{S}{\beta_\alpha}
  \pdv{\beta_\alpha}{\varphi_\beta}
  \dot \varphi_\beta
  \\
  &=
  \pdv{S}{\beta_\alpha}
  \widehat C_{\alpha\beta}^{-1}
  \int
  \dd{\bx}
  \pdv{\hat{\rho}}{t} \,\lambda_\beta(\bx)
  =
  \pdv{S}{\beta_\alpha}
  \widehat C_{\alpha\beta}^{-1}
  W_{\beta\mu}
  \int
  \dd{\bx}
  \pdv{\hat{\rho}}{t} \chi_\mu(\bx)
  \\
  &=
  \pdv{S}{\beta_\alpha}
  \widehat C_{\alpha\beta}^{-1}
  W_{\beta\mu}
  \int
  \dd{\bx}
  \qty[
    -
    \nabla \cdot
    \qty(
      \hat{\rho}
      \qty(
        \frac{\varepsilon}{2} \nabla \log q
        +
        a_\omega
        \nabla \chi_\omega
      )
    )
    +
    \frac{\varepsilon}{2}
    \nabla^2 \hat{\rho}
  ]
  \chi_\mu(\bx)
  \\
  &=
  \pdv{S}{\beta_\alpha}
  \widehat C_{\alpha\beta}^{-1}
  W_{\beta\mu}
  \int
  \dd{\bx}
  \hat{\rho}
  \qty[
    \frac{\varepsilon}{2} \nabla \log q
    +
    a_\nu
    \nabla \chi_\nu
    -
    \frac{\varepsilon}{2}
    \nabla \log\hat{\rho}
  ]
  \cdot
  \nabla
  \chi_\mu(\bx)
  \\
  &=
  \pdv{S}{\beta_\alpha}
  \widehat C_{\alpha\beta}^{-1}
  W_{\beta\mu}
  \int
  \dd{\bx}
  \hat{\rho}
  \qty[
    a_\nu
    \nabla \chi_\nu
    -
    \frac{\varepsilon}{2}
    \beta_\gamma
    \nabla \lambda_\gamma
  ]
  \cdot
  \nabla
  \chi_\mu(\bx)
  \\
  &=
  \pdv{S}{\beta_\alpha}
  \widehat C_{\alpha\beta}^{-1}
  W_{\beta\mu}
  G_{\mu\nu}
  \qty[
    a_\nu
    -
    \frac{\varepsilon}{2}
    W_{\gamma\nu}
    \beta_\gamma
  ],
\end{aligned}
\end{equation}
where $\widehat{C}_{\alpha\beta} = \pdv*{\varphi_\alpha}{\beta_\beta} = \int \dd{\bx} \hat{\rho}(\bx) (\lambda_\alpha(\bx) - \varphi_\alpha)(\lambda_\beta(\bx) - \varphi_\beta)$ is the Fisher information matrix in the space of signaling modes (also the covariance of the signaling modes under $\hat{\rho}$), and we have used the Fokker--Planck equation, integrated by parts, and used the definition of $G_{\mu\nu}$.
The dynamic programming principle reduces to:
\begin{equation}
\begin{aligned}
  0
  &=
  \min_{\{a_\mu\}}
  \qty{
    D_{\rm KL}(\hat{\rho} \Vert q)
    +
    \frac{r}{2}
    a_\mu
    G_{\mu\nu}
    a_\nu
    +
    \pdv{S}{\beta_\alpha}
    \widehat C_{\alpha\beta}^{-1}
    W_{\beta\mu}
    G_{\mu\nu}
    \qty[
      a_\nu
      -
      \frac{\varepsilon}{2}
      W_{\gamma\nu}
      \beta_\gamma
    ]
    -
    \gamma S
  }.
\end{aligned}
\end{equation}
Varying gives:
\begin{equation}
\begin{aligned}
  a_\mu
  =
  -
  \frac{1}{r}
  W_{\alpha\mu}
  \widehat C_{\alpha\beta}^{-1}
  \pdv{S}{\beta_\beta}.
\end{aligned}
\end{equation}
This means the real-space control is:
\begin{equation}
\begin{aligned}
  \bu(\bx; \bb)
  =
  -
  \frac{1}{r}
  \widehat C_{\alpha\beta}^{-1}
  \pdv{S}{\beta_\beta}
  \nabla \lambda_\alpha(\bx)
  =
  -
  \frac{1}{r}
  \pdv{S}{\varphi_\alpha}
  \nabla \lambda_\alpha(\bx)
  .
\end{aligned}
\label{eq:optimal-feedback-limited-sensing}
\end{equation}
Substituting this back into the dynamic programming relation:
\begin{equation}
\begin{aligned}
  0
  &=
  D_{\rm KL}(\hat{\rho} \Vert q)
  +
  \frac{1}{2r}
  \pdv{S}{\beta_\beta}
  \widehat C_{\alpha\beta}^{-1}
  W_{\alpha\mu}
  G_{\mu\nu}
  W_{\gamma\nu}
  \widehat C_{\gamma\delta}^{-1}
  \pdv{S}{\beta_\delta}
  +
  \pdv{S}{\beta_\alpha}
  \widehat C_{\alpha\beta}^{-1}
  W_{\beta\mu}
  G_{\mu\nu}
  \qty[
    -
    \frac{1}{r}
    W_{\gamma\nu}
    \widehat C_{\gamma\delta}^{-1}
    \pdv{S}{\beta_\delta}
    -
    \frac{\varepsilon}{2}
    W_{\gamma\nu}
    \beta_\gamma
  ]
  -
  \gamma S
\end{aligned}
\end{equation}
The HJB equation is:
\begin{equation}
\begin{aligned}
    \gamma S
    =
    D_{\rm KL}(\hat{\rho} \Vert q)
    -
    \frac{1}{2r}
    \pdv{S}{\beta_\beta}
    \widehat C_{\alpha\beta}^{-1}
    \widehat G_{\alpha\gamma}
    \widehat C_{\gamma\delta}^{-1}
    \pdv{S}{\beta_\delta}
    -
    \frac{\varepsilon}{2}
    \pdv{S}{\beta_\alpha}
    \widehat C_{\alpha\beta}^{-1}
    \widehat G_{\beta\gamma}
    \beta_\gamma
    .
\end{aligned}
\label{eq:hjb-limited-signaling}
\end{equation}
where $\widehat G_{\alpha\beta} = W_{\alpha\mu} G_{\mu\nu} W_{\beta\nu} = \int \dd{\bx} \hat{\rho}(\bx) \nabla \lambda_\alpha(\bx) \cdot \nabla \lambda_\beta(\bx)$ is the metric in the space of signaling modes induced by the control cost.

If we replace $\beta_\alpha$ with $\zeta_\mu$, we can immediately obtain the full-state exponential-family HJB equation:
\begin{equation}
\begin{aligned}
  \gamma S =
  D_{\rm KL}(\rho \Vert q)
  -
  \frac{1}{2r}
  \pdv{S}{\zeta_\mu}
  C_{\mu\nu}^{-1}
  G_{\nu\omega}
  C_{\omega\gamma}^{-1}
  \pdv{S}{\zeta_\gamma}
  -
  \frac{\varepsilon}{2}
  \pdv{S}{\zeta_\mu}
  C_{\mu\nu}^{-1}
  G_{\nu\omega}
  \zeta_\omega
  .
\end{aligned}
\end{equation}
This is equivalent to Eq.~\eqref{eq:full-state-hjb}.

\subsection{Weak-control limit, dimensionless parameters, and linearization of the HJB equation}
\label{si-sec:dimensionless-parameters}
\label{si-sec:weak-control-solution}

\subsubsection{The weak-control limit}

We can write the HJB in dimensionless form by rescaling $S, \widehat G, \widehat C$:
\begin{equation}
\begin{aligned}
  \dS(\bb)
  \equiv
  \gamma S(\bb),
  \qquad
  \dhC_{\alpha\beta}(\bb)
  \equiv
  \widehat C_{\alpha\beta}(\bb),
  \qquad
  \dhG_{\alpha\beta}(\bb)
  \equiv
  \frac{\varepsilon}{\gamma}
  \widehat G_{\alpha\beta}(\bb).
\end{aligned}
\end{equation}
The HJB equation becomes:
\begin{equation}
\begin{aligned}
  \dS
  =
  D_{\rm KL}(\hat{\rho} \Vert q)
  -
  \frac{1}{2\Lambda}
  \pdv{\dS}{\beta_\beta}
  \dhC_{\alpha\beta}^{-1}
  \dhG_{\alpha\gamma}
  \dhC_{\gamma\delta}^{-1}
  \pdv{\dS}{\beta_\delta}
  -
  \frac{1}{2}
  \pdv{\dS}{\beta_\alpha}
  \dhC_{\alpha\beta}^{-1}
  \dhG_{\beta\gamma}
  \beta_\gamma
  ,
\end{aligned}
\label{eq:hjb-limited-signaling-dimensionless}
\end{equation}
where we have introduced the dimensionless parameter:
\begin{equation}
\begin{aligned}
  \Lambda
  =
  r
  \varepsilon
  \gamma,
\end{aligned}
\end{equation}
which represents the relative strength of control.
We will proceed to analyze the weak-control limit $\Lambda \gg 1$ in which the control is weak relative to the passive dynamics:
\begin{equation}
\begin{aligned}
  \dS
  =
  \dS_0
  +
  \frac{1}{\Lambda}
  \dS_1
  +
  O(\Lambda^{-2})
  ,
\end{aligned}
\end{equation}
At successive orders in $\Lambda^{-1}$, Eq.~\eqref{eq:hjb-limited-signaling-dimensionless} becomes:
\begin{align}
  \label{eq:weak-control-hjb-S0}
  \dS_0
  &=
  D_{\rm KL}(\hat{\rho} \Vert q)
  -
  \frac{1}{2}
  \pdv{\dS_0}{\beta_\alpha}
  \dhC_{\alpha\beta}^{-1}
  \dhG_{\beta\gamma}
  \beta_\gamma
  \\
  \dS_1
  &=
  -
  \frac{1}{2}
  \pdv{\dS_0}{\beta_\beta}
  \dhC_{\alpha\beta}^{-1}
  \dhG_{\alpha\gamma}
  \dhC_{\gamma\delta}^{-1}
  \pdv{\dS_0}{\beta_\delta}
  -
  \frac{1}{2}
  \pdv{\dS_1}{\beta_\alpha}
  \dhC_{\alpha\beta}^{-1}
  \dhG_{\beta\gamma}
  \beta_\gamma
  .
\end{align}
These are linear transport equations for $\dS_0$ and $\dS_1$ along the passive dynamics in $\bb$-space, and they can be solved by the method of characteristics.
Let $\tau = \gamma t$ be the dimensionless time, and let $\bb^{(0)}(\tau ; \bb)$ denote the passive trajectory:
\begin{equation}
\begin{aligned}
  \dv{\bb^{(0)}}{\tau}
  =
  -
  \frac{1}{2}
  \dhbC^{-1}(\bb^{(0)})
  \dhbG(\bb^{(0)})
  \bb^{(0)},
  \qquad
  \bb^{(0)}(0 ; \bb) = \bb.
\end{aligned}
\end{equation}
Then, any equation of the form:
\begin{equation}
\begin{aligned}
  U + \frac{1}{2}
  \pdv{U}{\beta_\alpha}
  \dhC_{\alpha\beta}^{-1}
  \dhG_{\beta\gamma}
  \beta_\gamma
  =
  F
\end{aligned}
\label{eq:beta-transport-equation}
\end{equation}
satisfies:
\begin{equation}
\begin{aligned}
  \dv{\tau}
  \qty[
    e^{-\tau}
    U(\bb^{(0)}(\tau ; \bb))
  ]
  =
  e^{-\tau}
  F(\bb^{(0)}(\tau ; \bb))
  .
\end{aligned}
\end{equation}
We can then obtain the solutions for $\dS_0$ and $\dS_1$:
\begin{align}
  \label{eq:weak-control-value-function-S0}
  \dS_0(\bb)
  &=
  \int_0^\infty
  \dd{\tau}
  e^{-\tau}
  D_{\rm KL}(\hat{\rho}_{\bb^{(0)}(\tau ; \bb)} \Vert q)
  \\
  \label{eq:weak-control-value-function-S1}
  \dS_1(\bb)
  &=
  -
  \frac{1}{2}
  \int_0^\infty
  \dd{\tau}
  e^{-\tau}
  \qty[
    \pdv{\dS_0}{\beta_\beta}
    \dhC_{\alpha\beta}^{-1}
    \dhG_{\alpha\gamma}
    \dhC_{\gamma\delta}^{-1}
    \pdv{\dS_0}{\beta_\delta}
  ]_{\bb^{(0)}(\tau ; \bb)}.
\end{align}
This, to first order in $\Lambda^{-1}$, is the value function of the discounted KL cost accumulated under passive relaxation, corrected by the transport norm of the zeroth-order value-function gradient along the same passive trajectory.

\subsubsection{Linearization of the HJB equation}

We will now consider the linearization of the HJB equation around the desired state $\bb = 0$.
In the linearization, we will evaluate all interaction matrices at $\zeta = 0$ so that $C_{\mu\nu} = \delta_{\mu\nu}$ and $G_{\mu\nu} = \kappa_\mu \delta_{\mu\nu}$.
This means $\widehat C_{\alpha\beta} = W_{\alpha\mu} W_{\beta\mu}$ and $\widehat G_{\alpha\beta} = W_{\alpha\mu} \kappa_\mu W_{\beta\mu}$.
The KL divergence is expanded as $D_{\rm KL}(\hat{\rho} \Vert q) = \frac{1}{2} \beta_\alpha \widehat C_{\alpha\beta} \beta_\beta + \text{const.} + O(\beta^3)$ where the constant term comes from the prior offset $g_\mu$ and does not affect the optimal control.
With this linearization, we can explicitly solve Eq.~\eqref{eq:beta-transport-equation}:
\begin{equation}
\begin{aligned}
  \bb^{(0)}(\tau ; \bb)
  =
  e^{-\frac{1}{2} \dhbC^{-1} \dhbG \tau} \, \bb,
\end{aligned}
\end{equation}
where $e^{\bm{X}}$ is the matrix exponential.
Substituting this ansatz and linearization into the weak-control expression in Eq.~\eqref{eq:weak-control-value-function-S0} gives:
\begin{equation}
\begin{aligned}
  \dS_0(\bb)
  &
  =
  \frac{1}{2}
  \bb^{\rm T}
  \qty[
    \int_0^\infty
    \dd{\tau}
    e^{-\tau}
    e^{-\frac{1}{2} \dhbC^{-1} \dhbG \tau}
    \dhbC
    e^{-\frac{1}{2} \dhbC^{-1} \dhbG \tau}
  ]
  \bb
  +
  O(\bb^3)
\end{aligned}
\end{equation}
These sorts of expressions will appear frequently in this analysis, so we will define the following matrix:
\begin{equation}
\begin{aligned}
  L(\bm{A},\bm{X},\gamma)
  =
  \int_0^\infty
  \dd{t}
  e^{-\gamma t}
  e^{-\bm{A} t}
  \bm{X}
  e^{-\bm{A}^{\rm T} t}
  .
\end{aligned}
\end{equation}
These matrices are solutions to the Lyapunov equation:
\begin{equation}
\begin{aligned}
  \bm{A} L + L \bm{A}^{\rm T} + \gamma L = \bm{X}
\end{aligned}
\end{equation}
These matrices have the following useful properties:
\begin{equation}
\begin{gathered}
  L(\bm{A}, \bm{B}\bm{X}, \gamma) = \bm{B} L(\bm{A},\bm{X},\gamma), \qquad
  \text{if $[\bm{A}^T,\bm{B}] =[\bm{A},\bm{B}]= 0$}
  \\
  L(\bm{A},\bm{X},\gamma)^T
  =
  L(\bm{A}^T,\bm{X}^T,\gamma)
  \\
  \tr\qty[
    L(\bm{A},\bm{X},\gamma) \bm{Y}
  ]
  =
  \tr\qty[
    \bm{X}^T
    L(\bm{A}^T,\bm{Y}^T,\gamma)
  ]
  \\
  \pdv{\gamma}
  L(\bm{A},\bm{X},\gamma)
  =
  -
  L(\bm{A},
    L(\bm{A},\bm{X},\gamma),
  \gamma
  )
  \\
  L(\bm{A}, a \bm{X} + b \bm{Y}, \gamma)
  =
  a L(\bm{A},\bm{X},\gamma)
  +
  b L(\bm{A},\bm{Y},\gamma)
\end{gathered}
\end{equation}
With the introduction of this notation, we can write:
\begin{equation}
\begin{aligned}
  \dS_0(\bb)
  &=
  \frac{1}{2}
  \bb^{\rm T}
  L\qty(
    \frac{1}{2} \dhbC^{-1} \dhbG,
    \dhbC,
    1
  )
  \bb
  +
  O(\bb^3)
\end{aligned}
\label{eq:weak-control-value-function-S0-linearized}
\end{equation}
which allows for a cleaner expression for $\dS_1$:
\begin{equation}
\begin{aligned}
  \dS_1(\bb)
  &=
  -
  \frac{1}{2}
  \int_0^\infty
  \dd{\tau}
  e^{-\tau}
  \bb^{\rm T}
  e^{-\frac{1}{2} \dhbC^{-1} \dhbG \tau}
  L\qty(
    \frac{1}{2} \dhbC^{-1} \dhbG,
    \dhbC,
    1
  )
  e^{-\frac{1}{2} \dhbC^{-1} \dhbG \tau}
  \bb
  +
  O(\bb^3)
  \\
  &=
  -
  \frac{1}{2}
  \bb^{\rm T}
  L\qty(
    \frac{1}{2} \dhbC^{-1} \dhbG,
    L\qty(
      \frac{1}{2} \dhbC^{-1} \dhbG,
      \dhbC,
      1
    ),
    1
  )
  \bb
  +
  O(\bb^3)
\end{aligned}
\label{eq:weak-control-value-function-S1-linearized}
\end{equation}
For simplicity, we will define the following matrix:
\begin{equation}
\begin{aligned}
  \dbQ
  \equiv
  L\qty(
    \frac{1}{2}
    \dhbC^{-1} \dhbG,
    \dhbC,
    1
  )
  -
  \frac{1}{\Lambda}
  L\qty(
    \frac{1}{2}
    \dhbC^{-1} \dhbG,
    L\qty(
      \frac{1}{2}
      \dhbC^{-1} \dhbG,
      \dhbC,
      1
    ),
    1
  )
  +
  O(\Lambda^{-2})
  ,
\end{aligned}
\label{eq:weak-control-value-function-Q-matrix}
\end{equation}
so that $\dS(\bb) = \frac{1}{2} \bb^{\rm T} \dbQ \bb + O(\Lambda^{-2}, \bb^3)$.

\subsection{Linearization in the full-state case}

In the case where the full state is available (i.e., $\bm{W} = \bm{I}$), the linearization is simpler because $\dC_{\mu\nu} = \delta_{\mu\nu}$ and $\dG_{\mu\nu} = \nu_\mu \delta_{\mu\nu}$ is diagonal where
\begin{equation}
\begin{aligned}
  \nu_\mu
  =
  \frac{
    \varepsilon
    \kappa_\mu
  }{
    \gamma
  }
\end{aligned}
\end{equation}
is the dimensionless parameter that measures the relaxation rate of the $\mu$th mode relative to the perturbation rate.
The integrals in the weak-control limit can be evaluated explicitly to find:
\begin{equation}
\begin{aligned}
  \dS(\bze)
  &=
  \frac{1}{2}
  \sum_\mu
  \qty[
    \frac{1}{1+\nu_\mu}
    -
    \frac{1}{\Lambda}
    \frac{
      \nu_\mu
    }{
      (1 + \nu_\mu)^3
    }
  ]
  \zeta_\mu^2
  +
  O\qty(
    \Lambda^{-2},
    \zeta^3
  )
  .
\end{aligned}
\end{equation}

\section{Derivation of the optimal signaling modes}
\label{si-sec:optimal-signaling-modes}

In order to compare different choices of signaling modes, we want to compare a constrained-signaling optimal control policy to how well it would perform if it had access to the full state.
To do this, we can take a constrained-signaling optimal control policy and evaluate the cost of this policy under the full-state cost function.
This means we take policies solved with the objective $\widehat \ell$ and evaluate them under the objective $\ell$.
In particular, we would like to solve for the dynamics of the full state $\bze$ under the constrained-signaling optimal control policy and substitute this and the constrained-signaling control into the full-state cost function and take the expectation.
This is only tractable analytically in the weak-control, linearized regime.

\subsection{Evolution of the full state under constrained-signaling optimal control}

Substituting the limited-signaling optimal feedback we derived (Eq.~\eqref{eq:optimal-feedback-limited-sensing}) into the dynamics for $\bze$ (Eq.~\eqref{eq:full-zeta-dynamics}) we obtain:
\begin{equation}
\begin{aligned}
  \dot \zeta_\mu
  =
  -
  C_{\mu\nu}^{-1}
  G_{\nu\omega}
  \qty(
    \frac{\varepsilon}{2}
    \zeta_\omega
    +
    \frac{1}{r}
    W_{\alpha\omega}
    \widehat C_{\alpha\beta}^{-1}
    \pdv{S}{\beta_\beta}
  )
  ,
\end{aligned}
\end{equation}
or in dimensionless form:
\begin{equation}
\begin{aligned}
  \dv{\zeta_\mu}{\tau}
  =
  -
  \dC_{\mu\nu}^{-1}
  \dG_{\nu\omega}
  \qty(
    \frac{1}{2}
    \zeta_\omega
    +
    \frac{1}{\Lambda}
    W_{\alpha\omega}
    \dhC_{\alpha\beta}^{-1}
    \pdv{\dS}{\beta_\beta}
  )
\end{aligned}
\end{equation}
In order to evaluate the cost of this policy under the full-state cost function, we will need to obtain a closed expression for the dynamics of $\bze$ in terms of $\bze$ itself, and the above equation is not closed because the value function is a function of $\bb$ rather than $\bze$.
The relationship between $\bb$ and $\bze$ is fixed by the moment-matching condition $\varphi_\alpha = \int \dd{\bx} \hat{\rho}_{\bb}(\bx) \lambda_\alpha(\bx) = \int \dd{\bx} \rho_{\bze}(\bx) \lambda_\alpha(\bx)$, which is a nonlinear relationship between $\bb$ and $\bze$.
Note that at the desired state $\bze = 0$, we have $\bb = 0$, so in order to linearize the relationship between $\bb$ and $\bze$, we must determine the matrix $\pdv*{\zeta_\mu}{\beta_\alpha}$ at $\bze = 0$.
We can exploit the chain rule to write:
\begin{equation}
\begin{aligned}
  \pdv{\varphi_\alpha}{\beta_\beta}
  =
  \pdv{\varphi_\alpha}{\xi_\mu}
  \pdv{\xi_\mu}{\zeta_\nu}
  \pdv{\zeta_\nu}{\beta_\beta}
  =
  W_{\alpha\mu}
  \dC_{\mu\nu}
  \pdv{\zeta_\nu}{\beta_\beta}
\end{aligned}
\end{equation}
We additionally know that $\pdv*{\varphi_\alpha}{\beta_\beta} = \dhC_{\alpha\beta} = W_{\alpha\mu}\dC_{\mu\nu}W_{\beta\nu}$, which gives the relationship:
\begin{equation}
\begin{aligned}
  W_{\alpha\mu}
  \dC_{\mu\nu}
  \pdv{\zeta_\nu}{\beta_\beta}
  =
  W_{\alpha\mu}
  \dC_{\mu\nu}
  W_{\beta\nu}.
\end{aligned}
\end{equation}
This means that the linearized relationship between $\bb$ and $\bze$ is:
\begin{equation}
\begin{aligned}
  \bb =
    \underbrace{
      \qty(
        \bW
        \dbC
        \bW^T
      )^{-1}
      \bW
      \dbC
    }_{
      \equiv \bW^\dagger
    }
  \bze
  +
  O(\bze^2)
  .
\end{aligned}
\end{equation}
The estimate of the full state $\hat{\bze} = \bW^T \bb + \bm{g}$ is then a projection of the full state onto the signaling modes:
\begin{equation}
\begin{aligned}
  \hat{\bze} = \bPi_\bW \bze + \bm{g} + O(\bze^2),
  \qquad
  \bPi_\bW
  \equiv
    \bW^T
    \qty(
      \bW
      \dbC
      \bW^T
    )^{-1}
    \bW
    \dbC
\end{aligned}
\end{equation}
where $\bPi_\bW$ is an oblique projection matrix (orthogonal in the Fisher metric).
With this in place, we can write the linearized dynamics of the full state under constrained-signaling optimal control:
\begin{equation}
\begin{aligned}
  \dv{\zeta_\mu}{\tau}
  =
  -
  \dC_{\mu\nu}^{-1}
  \dG_{\nu\omega}
  \qty(
    \frac{1}{2}
    \zeta_\omega
    +
    \frac{1}{\Lambda}
    W_{\alpha\omega}
    \dhC_{\alpha\beta}^{-1}
    \dQ_{\beta\gamma}
    W^\dagger_{\gamma\pi}
    \zeta_\pi
  )
\end{aligned}
\end{equation}
If we define the matrix $\dtbQ = (\bW^\dagger)^T \dbQ \bW^\dagger$ (which is the extension of the matrix $\dbQ$ to the full state space),
then the dynamics become:
  \begin{equation}
  \begin{aligned}
    \dv{\zeta_\mu}{\tau}
    =
    -
    \dC_{\mu\nu}^{-1}
    \dG_{\nu\omega}
    \qty(
      \frac{1}{2}
      \zeta_\omega
      +
      \frac{1}{\Lambda}
      \dC_{\omega\pi}^{-1}
      \dtQ_{\pi\rho}
      \zeta_\rho
    )
\end{aligned}
\end{equation}
so the projection onto the signaling modes is encoded in the matrix $\dtbQ$.
In the linearized regime, this becomes:
  \begin{equation}
  \begin{aligned}
    \dv{\zeta_\mu}{\tau}
    =
    -
    \dG_{\mu\omega}
    \qty(
      \frac{1}{2}
      \zeta_\omega
      +
      \frac{1}{\Lambda}
      \dtQ_{\omega\nu}
      \zeta_\nu
    )
\end{aligned}
\end{equation}
where $\dG_{\mu\nu} = (\varepsilon \kappa_\mu / \gamma) \delta_{\mu\nu}$.
We will now work exclusively in the linearized regime (i.e., $\dC$ and $\dG$ are diagonal and $\bPi_\bW = \bW^T (\bW \bW^T)^{-1} \bW$ is a constant, orthogonal, projection matrix).

\subsection{Evolution of the initial state statistics}

  Now that we have an expression for how the full state evolves under constrained-signaling optimal control, we will use it to understand how the initial perturbation ensemble evolves under this control policy.
  As a reminder, the initial perturbation ensemble is drawn from a distribution $p_{\rm pert}(\bze)$ with mean $\ev{\bze_0} = \bm{0}$ and covariance $\ev{\bze_0 \bze_0^T} = \bm{\Sigma}_{\rm pert}$.
  We now want to analyze how the statistics of $\bze$ evolve under the optimal control strategy.
  In particular, we will be interested in the mean $m_\mu(\tau) = \ev{\zeta_\mu(\tau)}$ and the covariance $\Sigma_{\mu\nu}(\tau) = \ev{(\zeta_\mu(\tau) - m_\mu(\tau))(\zeta_\nu(\tau) - m_\nu(\tau))}$.
  The mean evolves according to:
  \begin{equation}
  \begin{aligned}
    \dv{m_\mu}{\tau}
    &=
    -
    \dG_{\mu\nu}
    \qty(
      \frac{1}{2}
      m_\nu
      +
      \frac{1}{\Lambda}
      \dtQ_{\nu\omega}
      m_\omega
    )
\end{aligned}
\end{equation}
  and the covariance evolves according to:
  \begin{equation}
  \begin{aligned}
    \dv{\Sigma_{\mu\nu}}{\tau}
    &=
    -
    \dG_{\mu\omega}
    \qty(
      \frac{1}{2}
      \Sigma_{\omega\nu}
      +
      \frac{1}{\Lambda}
      \dtQ_{\omega\pi}
      \Sigma_{\pi\nu}
    )
    -
    \Sigma_{\mu\omega}
    \qty(
      \frac{1}{2}
      \dG_{\omega\nu}
      +
      \frac{1}{\Lambda}
      \dtQ_{\omega\pi}
      \dG_{\pi\nu}
    )
  \end{aligned}
  \end{equation}
  Now, we can solve perturbatively in $\Lambda^{-1}$ to get the leading-order behavior of the mean and covariance.
  Write:
  \begin{equation}
  \begin{aligned}
    m_\mu(\tau)
    =
    m_\mu^{(0)}(\tau)
    +
    \frac{1}{\Lambda}
    m_\mu^{(1)}(\tau)
    +
    O\qty(
      \Lambda^{-2}
    ),
    \qquad
    \Sigma_{\mu\nu}(\tau)
    =
    \Sigma_{\mu\nu}^{(0)}(\tau)
    +
    \frac{1}{\Lambda}
    \Sigma_{\mu\nu}^{(1)}(\tau)
    +
    O\qty(
      \Lambda^{-2}
    )
    .
  \end{aligned}
\end{equation}
  The zeroth-order equations are:
  \begin{equation}
  \begin{aligned}
    \dv{m_\mu^{(0)}}{\tau}
    &=
    -
    \frac{1}{2}
    \dG_{\mu\nu}
    m_\nu^{(0)}
    \\
    \dv{\Sigma_{\mu\nu}^{(0)}}{\tau}
    &=
    -
    \frac{1}{2}
    \qty(
      \dG_{\mu\omega}
      \Sigma_{\omega\nu}^{(0)}
      +
      \Sigma_{\mu\omega}^{(0)}
      \dG_{\omega\nu}
    ).
  \end{aligned}
  \end{equation}
The solutions are:
\begin{equation}
\begin{aligned}
    m^{(0)}(\tau)
    =
    e^{-\frac{1}{2} \dbG \tau}
    m(0)
    =
    0
    ,
    \qquad
    \bSi^{(0)}(\tau)
    =
    e^{
      -
      \frac{1}{2}
      \dbG
      \tau
    }
    \bSi(0)
    e^{
      -
      \frac{1}{2}
      \dbG
      \tau
    }
    .
\end{aligned}
\end{equation}
  The first-order equations are:
  \begin{equation}
  \begin{aligned}
    \dv{m_\mu^{(1)}}{\tau}
    &
    =
    -
    \frac{1}{2}
    \dG_{\mu\nu}
    m_\nu^{(1)}
    -
    \dG_{\mu\nu}
    \dtQ_{\nu\omega}
    m_\omega^{(0)}
    ,
    \\
    \dv{\Sigma_{\mu\nu}^{(1)}}{\tau}
    &=
    -
    \qty(
      \frac{1}{2}
      \dG_{\mu\omega}
      \Sigma_{\omega\nu}^{(1)}
      +
      \dG_{\mu\omega}
      \dtQ_{\omega\pi}
      \Sigma_{\pi\nu}^{(0)}
    )
    -
    \qty(
      \frac{1}{2}
      \Sigma_{\mu\omega}^{(1)}
      \dG_{\omega\nu}
      +
      \Sigma_{\mu\omega}^{(0)}
      \dtQ_{\omega\pi}
      \dG_{\pi\nu}
    )
  \end{aligned}
  \end{equation}
  The solutions are:
  \begin{equation}
  \begin{aligned}
    m_\mu^{(1)}(\tau) &= 0
    \\
    \bSi^{(1)}(\tau)
    &
    =
    -
    \int_0^\tau
    \dd{\tau'}
    e^{
      -
      \frac{1}{2}
      \dbG
      (\tau-\tau')
    }
    \qty(
      \dbG
      \dtbQ
      \bSi^{(0)}(\tau')
      +
      \bSi^{(0)}(\tau')
      \dtbQ
      \dbG
    )
    e^{
      -
      \frac{1}{2}
      \dbG
      (\tau-\tau')
    }
     .
  \end{aligned}
\end{equation}
  For our calculation, the objects of interest are really the discounted integrals of the covariance:
  \begin{equation}
  \begin{aligned}
    \dK_{\mu\nu}
    &=
    \int_0^\infty
    \dd{\tau}
    e^{-\tau}
    \Sigma_{\mu\nu}(\tau)
  \end{aligned}
  \end{equation}
  To obtain this, multiply the covariance evolution ODEs by $e^{-\tau}$ and integrate the LHS by parts to get:
  \begin{equation}
  \begin{aligned}
    \dK_{\mu\nu}^{(0)}
    -
    \Sigma_{\mu\nu}(0)
    &=
    -
    \frac{1}{2}
    \qty(
      \dG_{\mu\omega}
      \dK_{\omega\nu}^{(0)}
      +
      \dK_{\mu\omega}^{(0)}
      \dG_{\omega\nu}
    )
    \\
    \dK_{\mu\nu}^{(1)}
    &=
    -
    \qty(
      \frac{1}{2}
      \dG_{\mu\omega}
      \dK_{\omega\nu}^{(1)}
      +
      \dG_{\mu\omega}
      \dtQ_{\omega\pi}
      \dK_{\pi\nu}^{(0)}
    )
    -
    \qty(
      \frac{1}{2}
      \dK_{\mu\omega}^{(1)}
      \dG_{\omega\nu}
      +
      \dK_{\mu\omega}^{(0)}
      \dtQ_{\omega\pi}
      \dG_{\pi\nu}
    )
  \end{aligned}
  \end{equation}
  These are Lyapunov equations, so we can write the solutions as:
  \begin{equation}
  \begin{aligned}
    \dbK^{(0)}
    &=
    L\qty(
      \frac{1}{2} \dbG,
      \bSi(0),
      1
    )
    ,
    \\
    \dbK^{(1)}
    &=
    -
    L\qty(
      \frac{1}{2} \dbG,
      \dbG \dtbQ \dbK^{(0)} + \dbK^{(0)} \dtbQ \dbG,
      1
    )
    .
  \end{aligned}
\end{equation}

\subsection{Evaluation of the cost function}

Now that we have the evolution of the mean and covariance, we can evaluate the full-state cost function under the constrained-signaling optimal control policy.
\begin{equation}
\begin{aligned}
  \dL_{\bW}
  \equiv
  \gamma \mathcal{L}_{\bW}
  &=
  \int_0^\infty
  \dd{\tau}
  e^{-\tau}\,
  \ev{
    D_{\rm KL}(\rho_\tau \Vert q)
    +
    \frac{1}{2\Lambda}
    \zeta_\omega
    \dtQ_{\omega\mu}
    \dG_{\mu\nu}
    \dtQ_{\nu\pi}
    \zeta_\pi
  }
  \\
  &=
  \int_0^\infty
  \dd{\tau}
  e^{-\tau}\,
  \ev{
    \frac{1}{2}
    \zeta_\mu
    \delta_{\mu\nu}
    \zeta_\nu
    +
    \frac{1}{2\Lambda}
    \zeta_\omega
    \dtQ_{\omega\mu}
    \dG_{\mu\nu}
    \dtQ_{\nu\pi}
    \zeta_\pi
  }
  +
  O\qty(
    \Lambda^{-2}
    ,
    \zeta^3
  )
\end{aligned}
\end{equation}
Using the formula:
\begin{equation}
\begin{aligned}
  \int_0^\infty
  \dd{\tau}
  e^{-\tau}
  \ev{
    \bze^T
    \bm{A}
    \bze
  }
  =
  \tr\qty[
    \bm{A}
    \dbK
  ],
\end{aligned}
\end{equation}
we can write the cost as:
\begin{equation}
\begin{aligned}
  \dL_{\bW}
  &=
  \frac{1}{2}
  \tr\qty[
    \dbK^{(0)}
  ]
  +
  \frac{1}{2\Lambda}
  \tr\qty[
    \dbK^{(1)}
  ]
  +
  \frac{1}{2\Lambda}
  \tr\qty[
    \dtbQ
    \dbG
    \dtbQ
    \dbK^{(0)}
  ]
  +
  O\qty(
    \Lambda^{-2}
    ,
    \zeta^3
  )
  .
\end{aligned}
\end{equation}
The first term is the state cost of the uncontrolled dynamics, the second term is the state-cost reduction due to the control, and the third term is the control cost.
We can now use properties of the Lyapunov equation solution to manipulate the second term:
\begin{equation}
\begin{aligned}
  \tr\qty[
    \dbK^{(1)}
  ]
  &=
  -
  \tr\qty[
    L\qty(
      \frac{1}{2} \dbG,
      \dbG \dtbQ \dbK^{(0)} + \dbK^{(0)} \dtbQ \dbG,
      1
    )
  ]
  \\
  &=
  -
  \tr\qty[
    \dbK^{(0)}
    \dtbQ
    \dbG
    L\qty(
      \frac{1}{2} \dbG,
      \bm{I},
      1
    )
  ]
  -
  \tr\qty[
    \dbG
    \dtbQ
    \dbK^{(0)}
    L\qty(
      \frac{1}{2} \dbG,
      \bm{I},
      1
    )
  ]
  \\
  &=
  -
  2
  \tr\qty[
    \dbG
    \operatorname{sym}
    \qty(
      \dtbQ
      \dbK^{(0)}
      \qty(
        \bm{I} + \dbG
      )^{-1}
    )
  ]
\end{aligned}
\end{equation}
where $\operatorname{sym}(\bm{X}) = \frac{1}{2}(\bm{X} + \bm{X}^T)$ is the symmetric part of $\bm{X}$.
Additionally:
\begin{equation}
\begin{aligned}
  \tr\qty[
    \dbK^{(0)}
  ]
  =
  \tr\qty[
    \bSi(0)
    L\qty(
      \frac{1}{2} \dbG,
      \bm{I},
      1
    )
  ]
  =
  \tr\qty[
    \bSi(0)
    \qty(
      \bm{I} + \dbG
    )^{-1}
  ]
\end{aligned}
\end{equation}
This gives us the most general form of $\dL_{\bW}$ that we will work with:
\begin{equation}
\begin{gathered}
  \dL_{\bW}
  =
  \frac{1}{2}
  \tr\qty[
    \bSi(0)
    \qty(
      \bm{I} + \dbG
    )^{-1}
  ]
  -
  \frac{1}{\Lambda}
  \tr\qty[
    \dbG
    \operatorname{sym}
    \qty(
      \dtbQ
      \dbK^{(0)}
      \qty(
        \bm{I} + \dbG
      )^{-1}
    )
  ]
  +
  \frac{1}{2\Lambda}
  \tr\qty[
    \dtbQ
    \dbG
    \dtbQ
    \dbK^{(0)}
  ]
  +
  O\qty(
    \Lambda^{-2}
    ,
    \zeta^3
  )
  \\
  \text{where }\quad
  \dbK^{(0)}
  =
  L\qty(
    \frac{1}{2} \dbG,
    \bSi(0),
    1
  )
\end{gathered}
\label{eq:general_L_W}
\end{equation}
This means that all dependence on the choice of signaling modes appears through $\dtbQ$.
In order to calculate up to $O(\Lambda^{-1})$, we only need the $O(\Lambda^0)$ term in the expansion for $\dtbQ$ (in Eq.~\eqref{eq:weak-control-value-function-Q-matrix}).
With our expression for $\dtbQ = (\bW^\dagger)^T \dbQ \bW^\dagger$, we have the identity:
\begin{equation}
\begin{aligned}
  \bPi_\bW
  \dtbQ
  =
  \dtbQ
  \bPi_\bW
  =
  \dtbQ,
\end{aligned}
\end{equation}
meaning that $\dtbQ$ implicitly contains both the LQR solution and the projection onto the signaling modes.
This can be made manifest in the expression for $\dtbQ$ by writing:
\begin{equation}
\begin{aligned}
  \dtbQ
  =
  L\qty(
    \frac{1}{2}
    \dbG
    \bPi_\bW,
    \bPi_\bW,
    1
  )
  +
  O(\Lambda^{-1}).
\end{aligned}
\label{eq:dtQ_projection}
\end{equation}
This means that in Eq.~\eqref{eq:general_L_W}, all dependence on the choice of signaling modes appears through the projection matrix $\bPi_\bW$.

\subsection{Grassmann manifold optimization of signaling modes}
\label{si-sec:grassmann}

We will now take a detour to discuss the mathematical structure of optimizing a scalar function over projectors.
Grassmann manifold optimization is a technique for optimizing over subspaces of a vector space~\cite{Absil2008OptimizationMatrixManifolds,Bendokat2024GrassmannHandbook}.
We will recap and derive some basic results about Grassmann manifolds that we will need for our analysis.
A Grassmann manifold $\operatorname{Gr}(k, n)$ is the space of $k$-dimensional subspaces of $\mathbb{R}^n$.
This can be written in terms of orthogonal projectors as:
\begin{equation}
\begin{aligned}
  \operatorname{Gr}(k,n)
  =
  \qty{
    \bm{P} \in \mathbb{R}^{n\times n} :
    \bm{P}^2 = \bm{P},
    \bm{P}^T = \bm{P},
    \tr[\bm{P}] = k
  },
\end{aligned}
\end{equation}
meaning that every point in the Grassmannian can be represented by a rank-$k$ orthogonal projector onto the corresponding subspace.
Now, consider a path through this space $\bm{P}(t)$.
It must satisfy $\bm{P}(t)^2 = \bm{P}(t)$ for all $t$, so we can differentiate this to get:
\begin{equation}
\begin{aligned}
  \dot{\bm{P}}(t)
  \bm{P}(t)
  +
  \bm{P}(t)
  \dot{\bm{P}}(t)
  =
  \dot{\bm{P}}(t)
\end{aligned}
\end{equation}
Projecting $\bm{P}(t)$ onto the left and right gives:
\begin{equation}
\begin{aligned}
  \bm{P}(t)
  \dot{\bm{P}}(t)
  \bm{P}(t)
  =
  0
  ,
  \quad
  (\bm{I}-\bm{P}(t))
  \dot{\bm{P}}(t)
  (\bm{I}-\bm{P}(t))
  =
  0.
\end{aligned}
\end{equation}
From $\bm{P}(t)^T = \bm{P}(t)$, we also have $\dot{\bm{P}}(t)^T = \dot{\bm{P}}(t)$.
This means that the tangent space at a point $\bm{P}$ is given by:
\begin{equation}
\begin{aligned}
  T_{\bm{P}} \operatorname{Gr}(k,n)
  =
  \qty{
    \bm{X} \in \mathbb{R}^{n\times n} :
    \bm{X}^T = \bm{X},
    \bm{P} \bm{X} \bm{P} = 0,
    (\bm{I}-\bm{P}) \bm{X} (\bm{I}-\bm{P}) = 0
  }
\end{aligned}
\end{equation}
which makes sense because the only way to perturb a subspace is to mix it with its orthogonal complement.
Now, if we have a function $f : \operatorname{Gr}(k,n) \to \mathbb{R}$, we can compute its gradient by first computing the Euclidean gradient $\nabla f(\bm{P})$ and then projecting it onto the tangent space. We need to move it along a path $\bm{P}(t)$ such that $\dot{\bm{P}}(0)$ is in the tangent space:
\begin{equation}
\begin{aligned}
  \pdv{t}
  \eval{f(\bm{P}(t))}_{t=0}
  =
  \tr\qty[
  \dot{\bm{P}}(0)
  \nabla f(\bm{P})
  ]
\end{aligned}
\end{equation}
We found that $\dot{\bm{P}}(0)$ must satisfy $\bm{P} \dot{\bm{P}}(0) \bm{P} = 0$ and $(\bm{I}-\bm{P}) \dot{\bm{P}}(0) (\bm{I}-\bm{P}) = 0$, so we can write $\dot{\bm{P}}(0) = \bm{P} \dot{\bm{P}}(0) (\bm{I}-\bm{P}) + (\bm{I}-\bm{P}) \dot{\bm{P}}(0) \bm{P}$.
Then, we have:
\begin{equation}
\begin{aligned}
  \pdv{t}
  &
  \eval{f(\bm{P}(t))}_{t=0}
  =
  \tr\qty[
    \bm{P} \dot{\bm{P}} (\bm{I}-\bm{P})
    \nabla f(\bm{P})
  ]
  +
  \tr\qty[
    (\bm{I}-\bm{P}) \dot{\bm{P}} \bm{P}
    \nabla f(\bm{P})
  ]
  \\
  &=
  \tr\qty[
    \dot{\bm{P}}
    (\bm{I}-\bm{P})
    \nabla f(\bm{P})
    \bm{P}
  ]
  +
  \tr\qty[
    \dot{\bm{P}}
    \bm{P}
    \nabla f(\bm{P})
    (\bm{I}-\bm{P})
  ]
  \\
  &=
  \tr
  \qty[
    \dot{\bm{P}}
    \qty(
    \bm{P}
    \,
    \frac{1}{2}
    \qty(
      \nabla f(\bm{P})
      +
      \nabla f(\bm{P})^T
    )
    \,
    (\bm{I}-\bm{P})
    +
    (\bm{I}-\bm{P})
    \,
    \frac{1}{2}
    \qty(
      \nabla f(\bm{P})
      +
      \nabla f(\bm{P})^T
    )
    \,
    \bm{P}
    )
  ]
  \\
  &=
  \tr
  \qty[
    \dot{\bm{P}}\,
    \nabla_{\rm Gr} f(\bm{P})
  ]
\end{aligned}
\end{equation}
This expression is useful because it explicitly shows how any directional derivative must be computed by projecting the Euclidean gradient onto the tangent space of the Grassmannian.
The Grassmannian gradient can also be written as:
\begin{equation}
\begin{aligned}
  \nabla_{\rm Gr} f(\bm{P})
  =
  \qty[
    \bm{P},
    \qty[
      \bm{P},
      \tfrac{1}{2}
      \qty(
        \nabla f(\bm{P})
        +
        \nabla f(\bm{P})^T
      )
    ]
  ]
\end{aligned}
\end{equation}
where $[\bm{A},\bm{B}] = \bm{A}\bm{B} - \bm{B}\bm{A}$ is the commutator.

Now, let us consider a property of commutators of orthogonal projectors.
We can write:
\begin{equation}
\begin{aligned}
  \qty[\bm{P}, \bm{S}]
  &
  =
  \bm{P} \bm{S} - \bm{S} \bm{P}
  =
  (
    \bm{P} \bm{S} \bm{P}
    +
    \bm{P} \bm{S} (\bm{I}-\bm{P})
  )
  -
  (
    \bm{P} \bm{S} \bm{P}
    +
    (\bm{I}-\bm{P}) \bm{S} \bm{P}
  )
  \\
  &
  =
  \bm{P} \bm{S} (\bm{I}-\bm{P})
  -
  (\bm{I}-\bm{P}) \bm{S} \bm{P}
  =
  \bm{P} \bm{S}(\bm{I}-\bm{P})
  -
  \qty( \bm{P} \bm{S} (\bm{I}-\bm{P}) )^T
\end{aligned}
\end{equation}
One can do the same for $[\bm{P}, [\bm{P}, \bm{S}]]$ to find:
\begin{equation}
\begin{aligned}
  \qty[\bm{P}, [\bm{P}, \bm{S}]]
  &=
  \bm{P}\bm{S} - 2 \bm{P} \bm{S} \bm{P} + \bm{S} \bm{P}
  =
  (
    \bm{P} \bm{S} \bm{P}
    +
    \bm{P} \bm{S} (\bm{I}-\bm{P})
  )
  -
  2 \bm{P} \bm{S} \bm{P}
  +
  (
    \bm{P} \bm{S} \bm{P}
    +
    (\bm{I}-\bm{P}) \bm{S} \bm{P}
  )
  \\
  &=
  \bm{P} \bm{S} (\bm{I}-\bm{P})
  +
  (\bm{I}-\bm{P}) \bm{S} \bm{P}
  =
  \bm{P} \bm{S} (\bm{I}-\bm{P})
  +
  \qty( \bm{P} \bm{S} (\bm{I}-\bm{P}) )^T
\end{aligned}
\end{equation}
Call $\bm{\Omega} = \bm{P} \bm{S} (\bm{I}-\bm{P})$.
If we have $[\bm{P},\bm{S}] = 0$, then $\bm{\Omega}$ must be symmetric, so $\bm{\Omega} = \bm{\Omega}^T$.
Notice $\bm{P} \bm{\Omega} = \bm{\Omega}$ and $\bm{\Omega} \bm{P} = 0$.
Because $\bm{\Omega}$ is symmetric, then $\bm{P} \bm{\Omega} = (\bm{\Omega} \bm{P})^T = 0$ but since $\bm{P} \bm{\Omega} = \bm{\Omega}$, we have $\bm{\Omega} = 0$ which means $[\bm{P},[\bm{P},\bm{S}]] = 0$ as well.
Conversely, if we have $[\bm{P},[\bm{P},\bm{S}]] = 0$, then $\bm{\Omega}$ must be anti-symmetric, so $\bm{\Omega} = -\bm{\Omega}^T$.
Just like before, $\bm{P}\bm{\Omega} = (\bm{\Omega} \bm{P})^T = 0$ but since $\bm{P} \bm{\Omega} = \bm{\Omega}$, we have $\bm{\Omega} = 0$ which means $[\bm{P},\bm{S}] = 0$ as well.
This ultimately means:
\begin{equation}
\begin{aligned}
  \qty[\bm{P},\bm{S}] = 0
  \iff
  [\bm{P},[\bm{P},\bm{S}]] = 0
  ,\qquad
  \qty(
    \bm{P}^2 = \bm{P},
    \;
    \bm{P}^T = \bm{P},
    \;
    \bm{S}^T = \bm{S}
  )
\end{aligned}
\end{equation}
This is useful because if we are at a stationary point of the Grassmannian gradient, then we have:
\begin{equation}
\begin{aligned}
  \nabla_{\rm Gr} f(\bm{P}) = 0
  \iff
  \qty[
    \bm{P},
    \tfrac{1}{2}
    \qty(
      \nabla f(\bm{P})
      +
      \nabla f(\bm{P})^T
    )
  ]
  =
  0
\end{aligned}
\end{equation}
which can be a bit neater to work with.
The essential conceptual point is that an optimal subspace will be one that is an eigenspace of an operator related to the Euclidean gradient of the cost function.
Therefore, optimizing over subspaces is essentially an eigenvalue problem.
All this means that the optimal signaling modes are those that satisfy a commutation relation of the form:
\begin{equation}
\begin{aligned}
  \qty[
    \bPi_{\bW},
    \frac{1}{2}
    \qty(
      \nabla_{\bPi_{\bW}} \dL_{\bW}
      +
      \qty(\nabla_{\bPi_{\bW}} \dL_{\bW})^T
    )
  ]
  =
  0.
\end{aligned}
\end{equation}
We will now proceed to derive the eigenvalue problem for our specific cost function $\dL_{\bW}$ in tractable limits.

\subsection{Diagonal perturbations}
\label{si-sec:diagonal_perturbations}

We will make progress in analyzing $\dL_{\bW}$ by specializing to the case where the initial perturbation covariance $\bSi(0)$ is diagonal in the same basis as $\dbG$ so that $[\dbG, \bSi(0)] = 0$.
This physically means that perturbations along the different passive relaxation modes are uncorrelated.
In this case, there is no possible basis available other than the diagonal one, so the optimal signaling modes must be a subset of the diagonal basis vectors: $[\bPi_{\bW}, \dbG] = 0$.
This means that our expressions for $\dbK^{(0)}$ and $\dtbQ$ can be simplified to:
\begin{equation}
\begin{aligned}
  \dbK^{(0)}
  =
  \bSi(0)
  \qty(
    \bm{I} + \dbG
  )^{-1}
  ,
  \qquad
  \dtbQ
  =
  \qty(
    \bm{I} + \dbG
  )^{-1}
  \bPi_{\bW}
\end{aligned}
\end{equation}
This means the cost function takes the form:
\begin{equation}
\begin{aligned}
  \dL_{\bW}
  &=
  \frac{1}{2}
  \tr\qty[
    \frac{\bSi(0)}{\bm{I} + \dbG}
  ]
  -
  \frac{1}{\Lambda}
  \tr\qty[
    \frac{\dbG \bSi(0)}
    {
      (\bm{I} + \dbG)^3
    }
    \bPi_{\bW}
  ]
  +
  \frac{1}{2\Lambda}
  \tr\qty[
    \frac{\dbG \bSi(0)}
    {
      (\bm{I} + \dbG)^3
    }
    \bPi_{\bW}
  ]
  +
  O\qty(
    \Lambda^{-2},
    \zeta^3
  )
  \\
  &=
  \frac{1}{2}
  \sum_{\mu = 1}^\infty
  \frac{
    \sigma_{\mu}^2
  }{
    1 + \nu_\mu
  }
  -
  \frac{1}{2\Lambda}
  \sum_{\mu \in W}
  \frac{
    \sigma_{\mu}^2
    \nu_\mu
  }{
    (1 + \nu_\mu)^3
  }
  +
  O\qty(
    \Lambda^{-2},
    \zeta^3
  ),
  \qquad
  \qty([\bSi(0), \dbG] = 0).
\end{aligned}
\end{equation}
where $\sigma_\mu^2$ are the diagonal entries of $\bSi(0)$.
This means that to minimize $\dL_{\bW}$, we should choose the signaling modes to be the $M$ modes so that the last term is as negative as possible.
We can define the signaling priority $\ds_\alpha$ for each mode $\alpha$ as:
\begin{equation}
\begin{aligned}
  \ds_\alpha^2
  \equiv
  \frac{
    \sigma_\alpha^2
    \nu_\alpha
  }{
    (1 + \nu_\alpha)^3
  }
  \qquad
  \qty(
    \nu_\alpha
    =
    \frac{
      \varepsilon
      \kappa_\alpha
    }{
      \gamma
    }
  )
\end{aligned}
\end{equation}
which appears in the cost function as:
\begin{equation}
\begin{aligned}
  \dL_{\bW}
  &=
  \frac{1}{2}
  \sum_{\mu = 1}^\infty
  \frac{
    \sigma_{\mu}^2
  }{
    1 + \nu_\mu
  }
  -
  \frac{1}{2\Lambda}
  \sum_{\alpha = 1}^M
  \ds_\alpha^2
  +
  O\qty(
    \Lambda^{-2},
    \zeta^3
  ).
\end{aligned}
\end{equation}
The signaling priority captures a tradeoff in which modes that are expensive to control (small $\kappa_\mu$) are less desirable to signal when perturbations are frequent ($\gamma \gg \kappa_\mu \varepsilon$) but more desirable to signal when perturbations are rare because they will persist for a long time and cause significant damage if not controlled ($\gamma \ll \kappa_\mu \varepsilon$).

Note that in the above, we could have left out the control cost term and we would get the result:
\begin{equation}
\begin{aligned}
  \int \dd{t}
  \gamma
  e^{-\gamma t}
  \ev{
    D_{\rm KL}(\rho_t \Vert q)
  }
  =
  \frac{1}{2}
  \sum_{\mu = 1}^\infty
  \frac{
    \sigma_{\mu}^2
  }{
    1 + \nu_\mu
  }
  -
  \frac{1}{\Lambda}
  \sum_{\alpha = 1}^M
  \ds_\alpha^2
  +
  O\qty(
    \Lambda^{-2},
    \zeta^3
  )
  .
\end{aligned}
\end{equation}

\subsection{Non-diagonal perturbations in the rare perturbation limit}
\label{si-sec:non-diagonal-rare-perturbation}

In order to analyze the general case where $[\bSi(0), \dbG] \neq 0$, we will specialize to the limit of rare perturbations $\gamma \ll \varepsilon \kappa_1$ (i.e., $\nu_\mu \gg 1$ for all $\mu$).
First, we must expand $\dtbQ$ to leading order in $\nu_\mu^{-1}$.
When we wrote $\dtbQ$ in terms of the Lyapunov equation solution (Eq.~\eqref{eq:dtQ_projection}), we asserted it satisfied the Lyapunov equation:
\begin{equation}
\begin{aligned}
  \frac{1}{2}
  \bPi_\bW
  \dbG
  \bPi_\bW
  \dtbQ
  +
  \frac{1}{2}
  \dtbQ
  \bPi_\bW
  \dbG
  \bPi_\bW
  +
  \dtbQ
  =
  \bPi_\bW
\end{aligned}
\end{equation}
We will now expand this in the limit $\nu_\mu \gg 1$.
We can immediately see that $\dtbQ$ cannot have a $O(1)$ term because then the first two terms would be $O(\nu_\mu)$ and the last term would be $O(1)$, which cannot sum to zero.
Assume $\dtbQ = O(\nu_\mu^{-1})$ to expand the Lyapunov equation to leading order in $\nu_\mu^{-1}$:
\begin{equation}
\begin{aligned}
  \frac{1}{2}
  \bPi_\bW
  \dbG
  \bPi_\bW
  \dtbQ
  +
  \frac{1}{2}
  \dtbQ
  \bPi_\bW
  \dbG
  \bPi_\bW
  +
  O(\nu_1^{-1})
  =
  \bPi_\bW
  .
\end{aligned}
\end{equation}
This has the solution:
\begin{equation}
\begin{aligned}
  \dtbQ
  =
  \qty(
    \bPi_\bW
    \dbG
    \bPi_\bW
  )^{+}
  +
  O(\nu_1^{-2})
\end{aligned}
\end{equation}
where $\qty(\, \cdot \,)^+$ is the Moore--Penrose pseudoinverse.
From this we can immediately obtain:
\begin{equation}
\begin{aligned}
  \dtbQ
  \dbG
  \dtbQ
  =
  \qty(
    \bPi_\bW
    \dbG
    \bPi_\bW
  )^{+}
  +
  O(\nu_1^{-2}).
\end{aligned}
\end{equation}
Recall that $\dbK^{(0)}$ is defined as the solution to the Lyapunov equation:
\begin{equation}
\begin{aligned}
  \frac{1}{2}
  \dbG
  \dbK^{(0)}
  +
  \frac{1}{2}
  \dbK^{(0)}
  \dbG
  +
  \bm{I}
  =
  \bSi(0)
\end{aligned}
\end{equation}
Like before, we know that $\dbK^{(0)}$ cannot have a $O(1)$ term, so if we assume $\dbK^{(0)} = O(\nu_\mu^{-1})$, we can expand the Lyapunov equation to leading order in $\nu_\mu^{-1}$:
\begin{equation}
\begin{aligned}
  \frac{1}{2}
  \dbG
  \dbK^{(0)}
  +
  \frac{1}{2}
  \dbK^{(0)}
  \dbG
  +
  O(\nu_1^{-1})
  =
  \bSi(0)
  \implies
  \dbK^{(0)}
  =
  L\qty(
    \frac{1}{2} \dbG,
    \bSi(0),
    0
  )
  .
\end{aligned}
\end{equation}
Using this, the cost function $\dL_{\bW}$ (Eq.~\eqref{eq:general_L_W}) can be written as:
\begin{equation}
\begin{gathered}
  \begin{aligned}
    \dL_{\bW}
    &
    =
    \frac{1}{2}
    \tr\qty[
      \bSi(0)
      \dbG^{-1}
    ]
    -
    \frac{1}{\Lambda}
    \tr\qty[
      \dbG
      \operatorname{sym}
      \qty(
        (
          \bPi_\bW
          \dbG
          \bPi_\bW
        )^{+}
        \dbK^{(0)}
        \dbG^{-1}
      )
    ]
    +
    \frac{1}{2\Lambda}
    \tr\qty[
    \qty(
        \bPi_\bW
        \dbG
        \bPi_\bW
      )^{+}
      \dbK^{(0)}
    ]
    +
    O\qty(
      \Lambda^{-2}
      ,
      \zeta^3
      ,
      \nu_1^{-2}
    )
    \\
    &
    =
    \frac{1}{2}
    \tr\qty[
      \bSi(0)
      \dbG^{-1}
    ]
    -
    \frac{1}{2\Lambda}
    \tr\qty[
      \operatorname{sym}
      \qty(
        (
          \bPi_\bW
          \dbG
          \bPi_\bW
        )^{+}
        L\qty(
          \frac{1}{2} \dbG,
          \bSi(0),
          0
        )
      )
    ]
    +
    O\qty(
      \Lambda^{-2}
      ,
      \zeta^3,
      \nu_1^{-2}
    )
  \end{aligned}
\end{gathered}
\end{equation}
Now, choose a basis in which $\bw_\alpha^T \dbG \bw_\beta = \delta_{\alpha\beta}$ so that $\qty(\bPi_\bW \dbG \bPi_\bW)^+ = \bW \bW^T$.
Then, the cost function can be written as:
\begin{equation}
\begin{aligned}
  \dL_{\bW}
  &=
  \frac{1}{2}
  \tr\qty[
    \bSi(0)
    \dbG^{-1}
  ]
  -
  \frac{1}{2\Lambda}
  \tr\qty[
      \bW^T
    L\qty(
      \frac{1}{2} \dbG,
      \bSi(0),
      0
    )
    \bW
  ]
  +
  O\qty(
    \Lambda^{-2}
    ,
    \zeta^3,
    \nu_1^{-2}
  ),
  \qquad
  \qty(\bW^T \dbG \bW = \bm{I})
\end{aligned}
\end{equation}
which means that the optimal signaling modes maximize:
\begin{equation}
\begin{aligned}
  \max_{\bW}
  \;
  \tr\qty[
    \bW^T
    L\qty(
      \frac{1}{2} \dbG,
      \bSi(0),
      0
    )
    \bW
  ]
  \quad
  \text{s.t.}
  \quad
  \bW^T \dbG \bW = \bm{I}.
\end{aligned}
\end{equation}
This means the optimal signaling modes should satisfy the generalized eigenvalue problem:
\begin{equation}
\begin{aligned}
  L\qty(
    \frac{1}{2} \dbG,
    \bSi(0),
    0
  )
  \bw_\alpha
  =
  \ds_\alpha^2
  \dbG
  \bw_\alpha
\end{aligned}
\end{equation}
which can be expressed more elegantly using the integral representation of the Lyapunov equation solution:
\begin{equation}
\begin{aligned}
  \qty(
    \int_0^\infty
    \dd{\tau}
    e^{-\frac{1}{2} \dbG \tau}
    \bSi(0)
    e^{-\frac{1}{2} \dbG \tau}
  )
  \bw_\alpha
  =
  \ds_\alpha^2
  \dbG
  \bw_\alpha
  .
\end{aligned}
\end{equation}
In terms of the eigenvalue, the cost function can be written as:
\begin{equation}
\begin{aligned}
  \dL_{\bW}
  &=
  \frac{1}{2}
  \tr\qty[
    \bSi(0)
    \dbG^{-1}
  ]
  -
  \frac{1}{2\Lambda}
  \sum_{\alpha = 1}^M
  \ds_\alpha^2
  +
  O\qty(
    \Lambda^{-2}
    ,
    \zeta^3,
    \nu_1^{-2}
  ).
\end{aligned}
\end{equation}
Restoring units to the eigenvalue problem:
\begin{equation}
\begin{aligned}
  \qty(
    \int_0^\infty
    \dd{t}
    e^{-\frac{\varepsilon}{2} \bG t}
    \bSi(0)
    e^{-\frac{\varepsilon}{2} \bG t}
  )
  \bw_\alpha
  =
  \frac{\varepsilon}{\gamma^2}
  \ds_\alpha^2
  \bG
  \bw_\alpha
\end{aligned}
\end{equation}

\section{Asymptotic analysis of the two-peak Gaussian mixture example}
\label{si-sec:two_peak_gaussian}

Now, we will apply the above results to the two-peak Gaussian mixture example mentioned in the main text.
In particular, we will consider the one-dimensional distribution:
\begin{equation}
\begin{aligned}
  q(x)
  =
  \frac{1}{\sqrt{2 \pi \delta^2}}
  \qty(
    p
    e^{-\frac{(x-x_0)^2}{2\delta^2}}
    +
    (1-p)
    e^{-\frac{(x+x_0)^2}{2\delta^2}}
  )
\end{aligned}
\end{equation}
where we will assume $\delta \ll x_0$ so that the two peaks are well-separated and the distribution is bimodal.
We will typically be interested in the $p = 1/2$ case where the two peaks are equally weighted, but we will keep $p$ general for now.

\subsection{Transformation to the Schr\"odinger equation}

To begin our analysis, we are interested in solving the generalized eigenvalue problem:
\begin{equation}
\begin{aligned}
  -\nabla
  \cdot
  \qty(
    q(x)
    \nabla
    \chi_\mu(x)
  )
  =
  \kappa_\mu
  q(x)
  \chi_\mu(x)
\end{aligned}
\end{equation}
If we change variables $\psi_\mu(x) = \sqrt{q(x)}\, \chi_\mu(x)$, then we can write this as a Schr\"odinger equation:
\begin{equation}
\begin{aligned}
  -\nabla^2 \psi_\mu
  +
  \qty(
    \frac{1}{4}
    \abs{
      \nabla \log q
    }^2
    +
    \frac{1}{2}
    \nabla^2 \log q
  )
  \psi_\mu
  =
  \kappa_\mu
  \psi_\mu
  .
\end{aligned}
\end{equation}
As a warm-up, let us consider the case where $q(x)$ is a single Gaussian distribution with mean zero and variance $\sigma^2$:
\begin{equation}
\begin{aligned}
  q(x) = \frac{1}{\sqrt{2 \pi \sigma^2}} e^{-\frac{x^2}{2\sigma^2}}
  \implies
  \frac{1}{4}
  \abs{\nabla \log q}^2
  +
  \frac{1}{2}
  \nabla^2 \log q
  =
  \frac{x^2}{4\sigma^4}
  -
  \frac{1}{2\sigma^2}
\end{aligned}
\end{equation}
This means that the Schr\"odinger equation potential is a harmonic oscillator potential and the eigenvalues are $\kappa_\mu = \frac{\mu}{\sigma^2}$ where $\mu = 0,1,2,\ldots$.
The transformed modes are the Hermite polynomials: $\psi_\mu(x) = \sqrt{q(x)}\frac{1}{\sqrt{2^\mu \mu!}} H_\mu\qty(\frac{x}{\sqrt{2}\sigma})$, and the original modes are $\chi_\mu(x) = \psi_\mu(x)/\sqrt{q(x)}$.

\subsection{Asymptotic spectrum of the two-peak Gaussian mixture}

Now, we will consider the two-peak Gaussian mixture distribution.
We know that locally near each peak, the distribution looks like a Gaussian distribution with variance $\delta^2$, so we can expect the potential to look like a harmonic oscillator potential near each peak, and we expect each mode to be a ladder of symmetrized and anti-symmetrized Hermite polynomials near each peak.
However, the lowest mode will not be one of these Hermite polynomials and will have a nontrivial structure.
In particular, we expect $\psi_1$ to be two Gaussian peaks centered at each mode with a node somewhere between the two peaks.
After dividing by $\sqrt{q}$, we expect $\chi_1$ to look flat near the peaks and have some cross-over structure between the peaks, so we are primarily interested in this stitching region.
To make analysis easier, we will introduce the crossover point
\begin{equation}
\begin{aligned}
  x_c
  =
  \frac{\delta^2}{2 x_0}
  \log
  \frac{1-p}{p}
\end{aligned}
\end{equation}
which is the point at which the probability that a sample came from the left peak is equal to the probability that it came from the right peak.
We also define the centered and de-dimensionalized coordinate,
\begin{equation}
\begin{aligned}
  \xi
  =
  \frac{
    x_0
  }{
    \delta^2
  }
  \qty(
    x
    -
    x_c
  )
\end{aligned}
\end{equation}
With these definitions, we can write the distribution as
\begin{equation}
\begin{aligned}
  q(x)
  =
  \sqrt{
    \frac{
      2 p(1-p)
    }{
      \pi \delta^2
    }
  }
  e^{
    -\qty(
      x^2 + x_0^2
    )/(2\delta^2)
  }
  \cosh \xi
  .
\end{aligned}
\end{equation}
The effective Schr\"odinger potential is then given by:
\begin{equation}
\begin{aligned}
  V(x)
  =
  \frac{1}{4 x_0^2}
  \qty(
    \frac{
      x_0^2
    }{
      \delta^2
    }
    \tanh \xi
    -
    \xi
    -
    \frac{1}{2}
    \log
    \frac{1-p}{p}
  )^2
  +
  \frac{x_0^2}{2 \delta^4}
  \sech^2 \xi
  -
  \frac{1}{2 \delta^2}
  .
\end{aligned}
\end{equation}
We are interested in the asymptotic limit where the two peaks are well separated, so we can take the leading order term in the potential which is given by:
\begin{equation}
\begin{aligned}
  V(x)
  =
  \frac{
    x_0^2
  }{
    4 \delta^4
  }
  \qty(
    1 + \sech^2 \xi
  )
  +
  O\qty(
    \qty(
      \delta/x_0
    )^0
  )
\end{aligned}
\end{equation}
By keeping $\xi$ to be $O(1)$ but taking the limit outside of it, we are saying that we are interested in the structure of the modes near the crossover point where the two peaks are equally weighted, so this approximation is useful for the lowest mode.
Notice that this is exactly the (inverted) P\"oschl--Teller potential which is a well-known exactly solvable potential in quantum mechanics.
The eigenvalue equation is:
\begin{equation}
\begin{aligned}
  -
  \frac{
    x_0^2
  }{\delta^4}
  \dv[2]{\psi_\mu}{\xi}
  +
  \frac{
    x_0^2
  }{
    4 \delta^4
  }
  \qty(
    1 + \sech^2 \xi
  )
  \psi_\mu
  =
  \kappa_\mu
  \psi_\mu
  .
\end{aligned}
\end{equation}
To proceed, we will assume that the lowest mode has an eigenvalue $\kappa_1 \ll \frac{x_0^2}{\delta^4}$ so that we can ignore the eigenvalue term in the equation to leading order.
This ODE has the solution:
\begin{equation}
\begin{aligned}
  \psi_1(\xi)
  \propto
  \sqrt{\cosh \xi}
  \qty(
    A
    \arctan
    \tanh
    \frac{\xi}{2}
    +
    B
  )
  .
\end{aligned}
\end{equation}
To convert back to the original mode, we need the asymptotic form of the distribution in the crossover region:
\begin{equation}
\begin{aligned}
  q(x)
  \approx
  \sqrt{
    \frac{
      2 p (1-p)
    }{
      \pi \delta^2
    }
  }
  \cosh \xi
\end{aligned}
\end{equation}
The original mode is then given by:
\begin{equation}
\begin{aligned}
  \chi_1(x)
  =
  A
  \arctan
  \tanh
  \frac{\xi}{2}
  +
  B
\end{aligned}
\end{equation}
The constants $A$ and $B$ are fixed by the normalization conditions $\int \dd{x} q(x) \chi_1(x)^2 = 1$ and $\int \dd{x} q(x) \chi_1(x) = 0$.
The conditions are:
\begin{equation}
\begin{aligned}
  \int \dd{x}
  q(x) \chi_1(x)
  \approx
  p \cdot
  \chi_1(\infty)
  +
  (1-p) \cdot
  \chi_1(-\infty)
  =
  p
  \qty(
    A
    \frac{\pi}{4}
    +
    B
  )
  +
  (1 - p)
  \qty(
    - A
    \frac{\pi}{4}
    +
    B
  )
  =
  0
\end{aligned}
\end{equation}
\begin{equation}
\begin{aligned}
  \int \dd{x}
  q(x)
  \chi_1(x)^2
  \approx
  p \cdot
  \chi_1(\infty)^2
  +
  (1-p) \cdot
  \chi_1(-\infty)^2
  =
  p
  \qty(
    A
    \frac{\pi}{4}
    +
    B
  )^2
  +
  (1-p)
  \qty(- A
    \frac{\pi}{4}
    +
    B
  )^2
  =
  1
\end{aligned}
\end{equation}
This makes the solution:
\begin{equation}
\begin{aligned}
  \chi_1(x)
  =
  \frac{
    1
  }{
    \frac{\pi}{2}
    \sqrt{
      p (1-p)
    }
  }
  \qty[
    \arctan
    \tanh
    \frac{\xi}{2}
    -
    \frac{\pi}{2}
    \qty(p - \frac{1}{2})
  ]
\end{aligned}
\end{equation}
This means that $\chi_1$ is a switching mode that is approximately constant near each peak and has a crossover structure of lengthscale $\delta^2/x_0$ near the crossover point.
We can also get the eigenvalue $\kappa_1$ by computing $-\partial_x(q \partial_x \chi_1)/q$.
If we integrate by parts and use the normalization condition, we find:
\begin{equation}
\begin{aligned}
  \kappa_1
  &=
  \int \dd{x}
  q(x)
  \qty(
    \dv{\chi_1}{x}
  )^2
  \approx
  \eval{
    q
  }_{x = x_c}
  \frac{
    \delta^2
  }{
    x_0^2
  }
  \int_{-\infty}^\infty
  \dd{\xi}
  \qty(
    \dv{\chi_1}{\xi}
  )^2
  \\
  &=
  \frac{\sqrt{2}}{\pi^{3/2}}
  \frac{x_0}{\delta^3}
  \frac{1}{
    \sqrt{p(1-p)}
  }
  \exp\qty(
    -
    \frac{x_0^2}{2\delta^2}
  )
\end{aligned}
\end{equation}

For the higher modes, the two wells are exponentially weakly coupled in the limit $x_0/\delta \gg 1$, so each well locally supports a harmonic-oscillator ladder. Let
\begin{equation}
\begin{aligned}
  h_n(y)
  =
  \frac{1}{\sqrt{2^n n!}}
  H_n\qty(\frac{y}{\sqrt{2}}),
  \qquad
  n=0,1,2,\ldots
\end{aligned}
\end{equation}
so that
\begin{equation}
\begin{aligned}
  \int_{-\infty}^{\infty}
  \frac{\dd{y}}{\sqrt{2\pi}}
  e^{-y^2/2}
  h_n(y)h_m(y)
  =
  \delta_{nm}.
\end{aligned}
\end{equation}
For the symmetric mixture $p=1/2$, the eigenfunctions can be chosen to have definite parity.
For each local Hermite level $n=1,2,\ldots$, the two corresponding global modes are approximately
\begin{equation}
\begin{aligned}
  \chi_{2n}(x)
  &\approx
  \begin{cases}
    (-1)^n
    h_n\qty(\dfrac{x+x_0}{\delta}),
    & x \text{ near } -x_0,
    \\[1.2em]
    h_n\qty(\dfrac{x-x_0}{\delta}),
    & x \text{ near } +x_0,
  \end{cases}
  \\
  \chi_{2n+1}(x)
  &\approx
  \begin{cases}
    (-1)^{n+1}
    h_n\qty(\dfrac{x+x_0}{\delta}),
    & x \text{ near } -x_0,
    \\[1.2em]
    h_n\qty(\dfrac{x-x_0}{\delta}),
    & x \text{ near } +x_0 .
  \end{cases}
\end{aligned}
\end{equation}
The mode $\chi_{2n}$ is even under $x\mapsto -x$, while $\chi_{2n+1}$ is odd.
Their eigenvalues are
\begin{equation}
\begin{aligned}
  \kappa_{2n}
  \approx
  \kappa_{2n+1}
  \approx
  \frac{n}{\delta^2},
  \qquad
  n=1,2,\ldots,
\end{aligned}
\end{equation}
up to exponentially small splittings between the even and odd members of each pair.

\begin{figure*}[t]
  \includegraphics[width=1.0\linewidth]{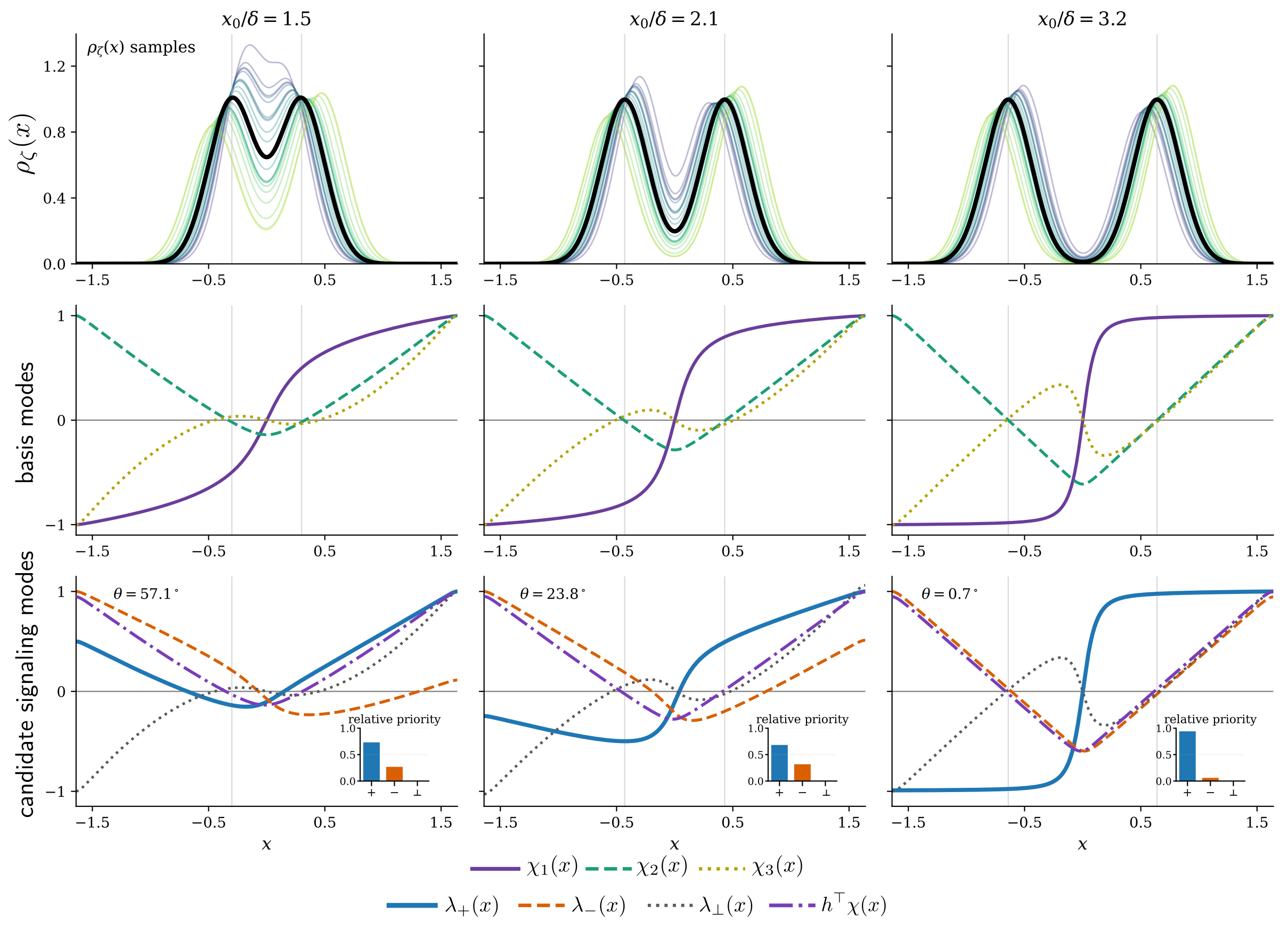}
  \caption{\label{si-fig:signaling-mixing}
    Signaling priority mixing angle example (Sec.~\ref{si-sec:signaling-priority-mixing-angle-example}).
    Top row: bimodal desired distributions $q$ with various peak separations (all with $\delta = 0.2$); samples from perturbation distribution ($\alpha = \pi/4, \beta = 0, \eta = 8$) are displayed as a cloud of curves around the desired distribution.
    With these perturbations, if the distribution is tilted right, then the modes shift outward, and if the distribution is tilted left, then the modes shift inward; note the magnitude of the peak means-shift perturbations is much larger than the magnitude of the de-balancing perturbations.
    Middle row: passive decay modes $\chi_1, \chi_2, \chi_3$.
    The $\chi_1$ mode is a switching mode that is approximately constant near each peak and has a crossover structure between the peaks, while the $\chi_2$ and $\chi_3$ modes are approximately degenerate and look like the first excited Hermite mode near each peak.
    Bottom row: candidate signaling modes and their signaling priorities.
    The signaling mode $\lambda_+$ (blue solid) is optimal, $\lambda_-$ (orange dashed) is the second choice, $\lambda_\perp$ (gray dotted) has zero signaling priority and should not be chosen.
    The relative signaling priority, $\ds_\alpha/(\sum_\beta \ds_\beta)$, is shown as an inset.
    The only eigenvector of $\bSi$ is $\bm{h}^T \bch$ (purple dash-dotted), the mode that perturbations excite.
    When $x_0 \gg \delta$, $\kappa_1$ becomes exponentially smaller than $\kappa_2$ and the optimal signaling mode $\lambda_+$ becomes very different from $\bm{h}^T \bch$, the direction the perturbations excite.
    This figure uses the same parameters as Fig.~\ref{fig:signaling-mixing} in the main text and is meant to be a companion figure that shows the same results in more detail.
  }
\end{figure*}

\subsection{Signaling priority mixing angle example}
\label{si-sec:signaling-priority-mixing-angle-example}

With this problem setup, we will work through the example in the main text (Sec.~\ref{sec:two-peak-example}) that demonstrates how signaling priority is influenced both by the passive decay structure and the perturbation structure.
The setup is a perturbation that has a low-rank structure that only excites the first three modes:
\begin{equation}
\begin{aligned}
  \bSi
  =
  \sigma^2\,
  \bm{h}\bm{h}^T,
  \qquad
  \bm{h}
  =
  \begin{bmatrix}
    \cos \alpha
    \\
    \eta \sin \alpha \cos \beta
    \\
    \eta \sin \alpha \sin \beta
    \\
    \vdots
  \end{bmatrix}
\end{aligned}
\end{equation}
where $\alpha$ is the mixing angle that controls how the perturbation is correlated between the first mode and the second and third modes; $\beta$ is the mixing angle that controls how the perturbation is correlated between the second and third modes; and $\eta$ controls the relative strength of the perturbation along the second and third modes compared to the first mode.
Recall that in the asymptotic limit, $\kappa_1 \ll \kappa_2 \approx \kappa_3$, so the first mode is much slower and more expensive to control than the second and third modes which are approximately degenerate.
For the signaling problem, the naive expectation is that the optimal signaling mode would simply be $\bm{h}$ because that is the mode that the perturbation excites, but as we will see, this is not necessarily the case because the signaling priority also depends on the passive decay structure.

We first calculate:
\begin{equation}
\begin{aligned}
  L_{\mu\nu}
  \equiv
  \qty(
    \int_0^\infty
    \dd{t}
    e^{-
    \frac{\varepsilon}{2} \bG t}
    \bSi
    e^{- \frac{\varepsilon}{2} \bG t}
  )_{\mu\nu}
  =
  \sigma^2
  h_\mu
  h_\nu
  \int_0^\infty
  \dd{t}
  e^{-\frac{\varepsilon}{2} (\kappa_\mu + \kappa_\nu) t}
  =
  \frac{2\sigma^2}{\varepsilon}
  \frac{h_\mu
  h_\nu}{\kappa_\mu + \kappa_\nu},
\end{aligned}
\end{equation}
so the signaling eigenvalue problem is:
\begin{equation}
\begin{aligned}
  2 \sigma^2
  \frac{
    h_\mu h_\nu
  }{\varepsilon(\kappa_\mu + \kappa_\nu)}
  W_{\alpha\nu}
  =
  \frac{\varepsilon}{\gamma^2}
  \ds_\alpha^2
  \kappa_\mu
  W_{\alpha\mu}.
\end{aligned}
\end{equation}
We will use the notation $\bm{e}_1, \bm{e}_2, \bm{e}_3$ to denote the basis vectors corresponding to the first three modes, so $\bm{h} = \cos \alpha \bm{e}_1 + \eta \sin \alpha \qty(\cos \beta \bm{e}_2 + \sin \beta \bm{e}_3)$.
Defining $\bm{e}_\parallel \equiv \cos \beta\, \bm{e}_2 + \sin \beta\, \bm{e}_3$ to be the unit vector aligned with the perturbation in the degenerate subspace, and $\bm{e}_\perp \equiv - \sin \beta \,\bm{e}_2 + \cos \beta \, \bm{e}_3$ to be the unit vector orthogonal to the perturbation in the degenerate subspace, we can write $\bm{h} = \cos \alpha \bm{e}_1 + \eta \sin \alpha \bm{e}_\parallel$.
Because $\bm{h}$ has no component along $\bm{e}_\perp$, $L \bm{e}_\perp = 0$, so $\bm{e}_\perp$ is an eigenvector of the signaling eigenvalue problem with eigenvalue zero $\ds_\perp = 0$.
This means, we need to solve the eigenvalue problem in the two-dimensional subspace spanned by $\bm{e}_1$ and $\bm{e}_\parallel$.
In this basis:
\begin{equation}
\begin{aligned}
  \bG
  =
  \begin{bmatrix}
    \kappa_1 & 0
    \\
    0 & \kappa_2
  \end{bmatrix}
  ,
  \qquad
  L
  =
  \frac{\sigma^2}{\varepsilon}
  \begin{bmatrix}
    \frac{1}{\kappa_1}
    \cos^2 \alpha
    &
    \frac{
      2\eta
    }{\kappa_1 + \kappa_2}
    \cos \alpha \sin \alpha
    \\
    \frac{
      2\eta
    }{\kappa_1 + \kappa_2}
    \cos \alpha \sin \alpha
    &
    \frac{\eta^2}{\kappa_2}
    \sin^2 \alpha
  \end{bmatrix}
\end{aligned}
\end{equation}
To solve this generalized eigenvalue problem, we can transform it into a regular eigenvalue problem:\begin{equation}
\begin{aligned}
  \bG^{-1/2}
  L
  \bG^{-1/2}
  \qty(
    \bG^{1/2}
    W_\alpha
  )
  =
  \frac{\varepsilon}{\gamma^2}
  \ds_\alpha^2
  \qty(
    \bG^{1/2}
    W_\alpha
  )
\end{aligned}
\end{equation}
where:
\begin{equation}
\begin{aligned}
  \bG^{-1/2}
  L
  \bG^{-1/2}
  =
  \sigma^2
  \begin{bmatrix}
    \frac{1}{\kappa_1^2}
    \cos^2 \alpha
    &
    \frac{
      2\eta
    }{\kappa_1^{1/2} \kappa_2^{1/2} (\kappa_1 + \kappa_2)}
    \cos \alpha \sin \alpha
    \\
    \frac{
      2\eta
    }{\kappa_1^{1/2} \kappa_2^{1/2} (\kappa_1 + \kappa_2)}
    \cos \alpha \sin \alpha
    &
    \frac{\eta^2}{\kappa_2^2}
    \sin^2 \alpha
  \end{bmatrix}
\end{aligned}
\end{equation}
Because this is a $2\times 2$ symmetric matrix, the eigenvectors can be written as:
\begin{equation}
\begin{aligned}
  \bm{w}_+
  \propto
  \frac{
    \cos \vartheta
  }{
    \sqrt{\kappa_1}
  }
  \bm{e}_1
  +
  \frac{\sin \vartheta
  }{\sqrt{\kappa_2}}
  \bm{e}_\parallel,
  \qquad
  \bm{w}_-
  \propto
  -
  \frac{
    \sin \vartheta
  }{
    \sqrt{\kappa_1}
  }
  \bm{e}_1
  +
  \frac{\cos \vartheta
  }{\sqrt{\kappa_2}}
  \bm{e}_\parallel
\end{aligned}
\end{equation}
with the mixing angle satisfying the relation:
\begin{equation}
\begin{aligned}
  \tan 2 \vartheta
  =
  \frac{
    \frac{
      4 \eta \tan \alpha
    }{
      (\kappa_1 + \kappa_2) \sqrt{\kappa_1 \kappa_2}
    }
  }{
    \frac{1}{\kappa_1^2}
    -
    \frac{\eta^2 \tan^2 \alpha}{\kappa_2^2}
  }
  =
  \frac{
    4\eta \tan \alpha
    \frac{
      \kappa_1^{3/2}
    }{
      \sqrt{\kappa_2}
      (\kappa_1 + \kappa_2)
    }
  }{
    1
    -
    \qty(
      \frac{\kappa_1}{\kappa_2}
      \eta
      \tan \alpha
    )^2
  }
\end{aligned}
\end{equation}
The signaling priorities are:
\begin{equation}
\begin{aligned}
  \ds_\perp = 0,
  \qquad
  \ds_\pm^2
  =
  \frac{\gamma^2\sigma^2}{2\varepsilon^2}
  \qty[
    \frac{
      \cos^2 \alpha
    }{
      \kappa_1^2
    }
    +
    \frac{
      \eta^2 \sin^2 \alpha
    }{
      \kappa_2^2
    }
    \pm
    \sqrt{
      \qty(
        \frac{
          \cos^2 \alpha
        }{
          \kappa_1^2
        }
        -
        \frac{
          \eta^2 \sin^2 \alpha
        }{
          \kappa_2^2
        }
      )^2
      +
      \frac{16 \eta^2 \cos^2 \alpha \sin^2 \alpha}{
        (\kappa_1 + \kappa_2)^2 \kappa_1 \kappa_2
      }
    }
  ]
  .
\end{aligned}
\end{equation}
This means that the optimal signaling mode $\bm{w}_+$ is a mixture of the first mode and the degenerate subspace, and the mixing angle depends on the perturbation structure (through $\alpha$ and $\eta$) and the passive decay structure (through $\kappa_1$ and $\kappa_2$).
In particular, there is a crossover when $\eta \abs{\tan \alpha} \sim \kappa_2/\kappa_1$ where the optimal signaling mode switches from being aligned with the first slowest mode to being aligned with the faster perturbed modes.
We know that $\kappa_1/\kappa_2 \approx (x_0/\delta) e^{-x_0^2/(2\delta^2)}$ is exponentially small in the limit of separated peaks.
In the $\kappa_1 \ll \kappa_2$ limit, the mixing angle vanishes $\vartheta \approx 0$ and the optimal signaling mode is aligned with the first slow mode $\bm{e}_1$ regardless of the perturbation structure.
This transition from a mixed signaling mode to a pure signaling mode aligned with the slowest mode is demonstrated in Fig.~\ref{si-fig:signaling-mixing}.

\clearpage
\section{Refresher on Wasserstein and Fisher geometry}
\label{si-sec:info-geom}

In this section, we will briefly review the Wasserstein and Fisher geometries of probability distributions in order to provide some geometric intuition for the results in the main text.
These concepts both belong to the field of information geometry, which studies the geometric structure of probability distributions and statistical models~\cite{Amari2016InformationGeometryApplications}.
To maintain generality, we will keep the same notation as the main text but not specialize to our particular parameterization and problem setup in this section.
We will use the notation $\rho_{\bze}(\bx)$ to denote the distribution parameterized by parameters $\zeta_\mu$.
In information geometry, we can think of the space of probability distributions as a manifold, and the parameters $\zeta_\mu$ as coordinates on this manifold.
There are many metrics that one can define on this manifold, each leading to distinct conceptual advantages and insights based on the resulting differential geometry.
For our purposes, the two most relevant metrics are the Wasserstein metric which captures the geometry of optimal transport~\cite{Ambrosio2008GradientFlows,Santambrogio2015OptimalTransportApplied} and the Fisher information metric which captures the geometry of statistical inference~\cite{Amari2016InformationGeometryApplications}.

\subsection{Fisher information geometry}

The Fisher information geometry captures the similarity between probability distributions in terms of their statistical distinguishability.
To first order, the KL divergence between two nearby distributions $\rho_{\bze}$ and $\rho_{\bze + \dd{\bze}}$ is given by:
\begin{equation}
\begin{aligned}
  D_{\rm KL}(\rho_{\bze} \Vert \rho_{\bze + \dd{\bze}})
  =
  \frac{1}{2}
  I_{\mu\nu}(\zeta)
  \dd{\zeta_\mu}
  \dd{\zeta_\nu}
  +
  O(\dd{\zeta}^3)
\end{aligned}
\end{equation}
where
\begin{equation}
\begin{aligned}
  I_{\mu\nu}(\zeta)
  =
  \int \dd{\bx}
  \rho_{\bze}(\bx)
  \qty(
    \pdv{\log \rho_{\bze}(\bx)}{\zeta_\mu}
    \pdv{\log \rho_{\bze}(\bx)}{\zeta_\nu}
  )
\end{aligned}
\end{equation}
is the Fisher information matrix.
It is symmetric and positive semi-definite, so it defines a Riemannian metric on the space of probability distributions.
In the parameterization we have been using, the Fisher information matrix is equal to the covariance matrix of the modes: $I_{\mu\nu} = C_{\mu\nu}$.
Let us say we re-parameterize the family of distributions with a change of coordinates $\zeta' = \zeta'(\zeta)$, then the Fisher information matrix transforms as a rank-2 tensor:
\begin{equation}
\begin{aligned}
  I'_{\mu\nu}(\bze')
  =
  \int \dd{\bx}
  \rho_{\bze'}(\bx)
  \qty(
    \pdv{\log \rho_{\bze'}(\bx)}{\zeta'_\mu}
    \pdv{\log \rho_{\bze'}(\bx)}{\zeta'_\nu}
  )
  =
  \pdv{\zeta_\omega}{\zeta'_\mu}
  \pdv{\zeta_\omega}{\zeta'_\nu}
  I_{\omega\pi}(\zeta)
\end{aligned}
\end{equation}
This all means that the Fisher information geometry is invariant to re-parameterization, so it captures the intrinsic geometry of the space of distributions independent of how we choose to parameterize it.
The Fisher information (like the KL divergence) strongly penalizes differences in the tails of the distribution (i.e., moving probability mass to a region where the original distribution has very low probability) and is relatively insensitive to differences in the bulk of the distribution (i.e., moving probability mass around within a region where the original distribution has high probability).
This is useful in contexts like statistical inference in which observing data that is very unlikely under the model is a strong signal that the model is wrong, while observing data that is likely under the model is not a strong signal that the model is correct.

The Fisher information geometry has no notion of actual spatial distance between probability distributions.
To see this, consider a change of variables $\bx' = \bx'(\bx)$ which is independent of $\zeta$, then the distribution transforms as $\rho'(\bx') = \rho(\bx) \abs{\pdv{\bx}{\bx'}}$, so the log-probability transforms as $\log \rho'(\bx') = \log \rho(\bx) + \log \abs{\pdv{\bx}{\bx'}}$.
The Fisher information matrix then transforms as:
\begin{equation}
\begin{aligned}
  I'_{\mu\nu}(\bze)
  &
  =
  \int \dd{\bx'}
  \rho_{\bze}'(\bx')
  \qty(
    \pdv{\log \rho_{\bze}'(\bx')}{\zeta_\mu}
    \pdv{\log \rho_{\bze}'(\bx')}{\zeta_\nu}
  )
  \\
  &
  =
  \int \dd{\bx}
  \abs{
    \pdv{\bx}{\bx'}
  }^{-1}
  \rho_{\bze}(\bx)
  \abs{
    \pdv{x}{x'}
  }
  \qty(
    \pdv{\log \rho_{\bze}(\bx)}{\zeta_\mu}
    +
    \cancel{\pdv{\log \abs{\pdv{\bx}{\bx'}}}{\zeta_\mu}}
  )
  \qty(
    \pdv{\log \rho_{\bze}(\bx)}{\zeta_\nu}
    +
    \cancel{\pdv{\log \abs{\pdv{\bx}{\bx'}}}{\zeta_\nu}}
  )
  \\
  &=
  I_{\mu\nu}(\bze),
\end{aligned}
\end{equation}
so the Fisher information geometry knows nothing about the actual spatial structure of the distributions and only cares about the actual probability values at each point in space.

\begin{figure}[t]
  \includegraphics[width=\linewidth]{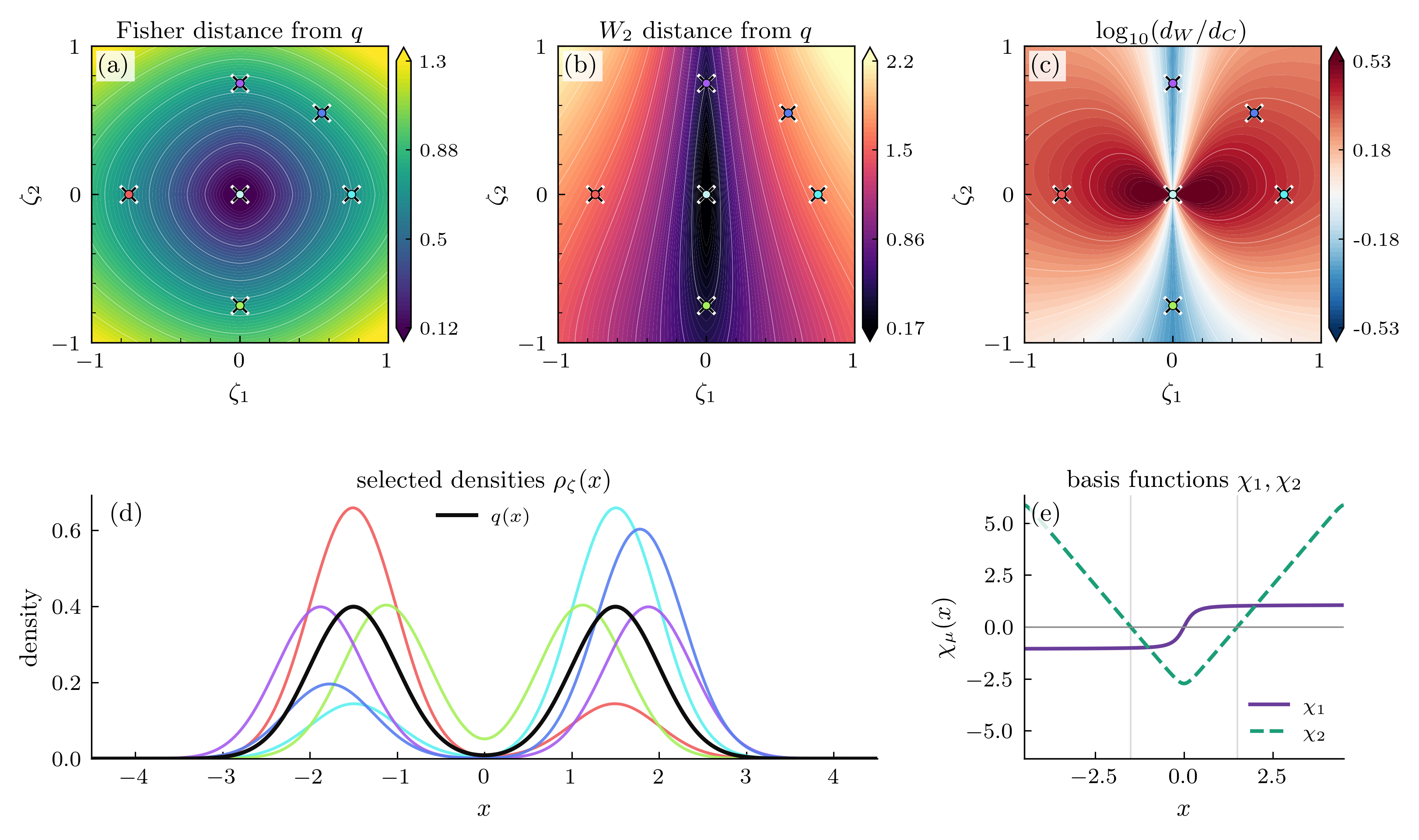}
  \caption{
    \label{si-fig:intrinsic-geodesics}
      Fisher and Wasserstein geometry of the two-dimensional exponential-family submanifold spanned by the modes $\chi_1$ and $\chi_2$.
      (a,b) Geodesic distance from the reference distribution $q=\rho_{\bze=0}$ under the Fisher metric $\bC$ and the Wasserstein metric $g^W = \bC \bG^{-1} \bC$.
      (c) Logarithmic distance ratio $\log_{10}(d_W/d_{\bC})$, showing where Wasserstein distances are large or small relative to Fisher distances.
      Colored markers indicate the same parameter values $\zeta=(\zeta_1,\zeta_2)$ in all panels.
      (d) The corresponding distributions $\rho_{\bze}(\bx)$, with $q(\bx)$ shown in black.
      (e) The basis functions $\chi_1(\bx)$ and $\chi_2(\bx)$ defining this submanifold.
      The Fisher geometry is close to isotropic near $q$, while the Wasserstein geometry is strongly anisotropic because $\chi_1$ is a slow inter-well mode with small $\kappa_1$, making transport along $\zeta_1$ much more costly than motion along the faster intra-well mode $\chi_2$.
  }
\end{figure}
\begin{figure}[t]
  \includegraphics[]{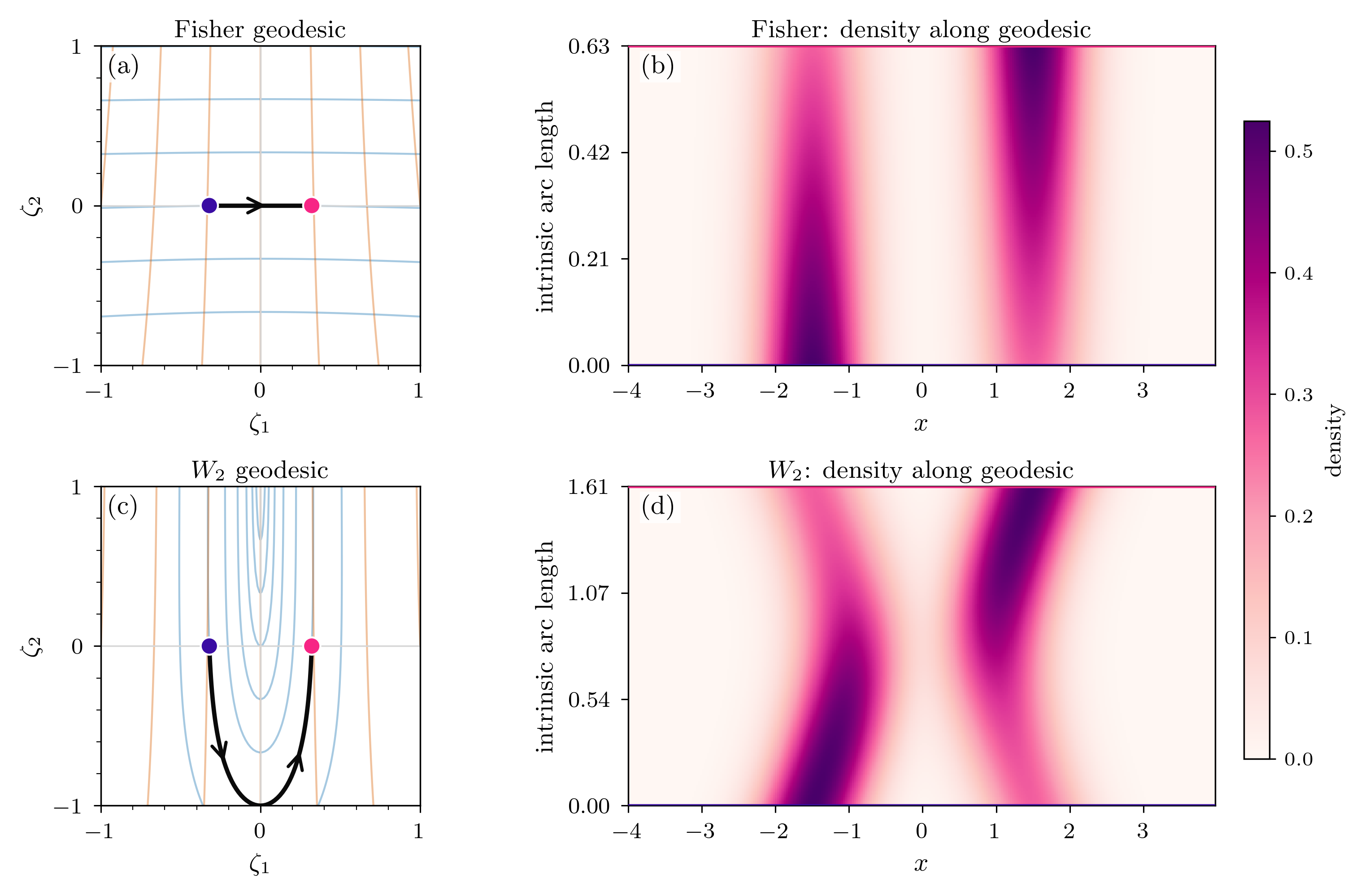}
  \caption{
    \label{si-fig:intrinsic-geodesic-grid}
    Geodesic grids for the Fisher and Wasserstein metrics on the same $\zeta_1,\zeta_2$ parameter plane.
    (a,c) Families of geodesics generated with initial velocities perpendicular to the coordinate axes, shown for the Fisher metric $\bC$ and the Wasserstein metric $g^W=\bC \bG^{-1}\bC$.
    The black curves highlight geodesics connecting the same pair of endpoint distributions.
    (b,d) Density evolution $\rho_{\bze}(\bx)$ along the highlighted geodesics, parameterized by intrinsic arc length.
    Under the Fisher metric, the highlighted geodesic follows an almost direct path along the inter-well coordinate.
    Under the Wasserstein metric, the geodesic bends strongly through negative $\zeta_2$, corresponding to intermediate density shapes that reduce the transport cost of changing the slow inter-well mode.
  }
\end{figure}

\begin{figure}[t]
  \includegraphics{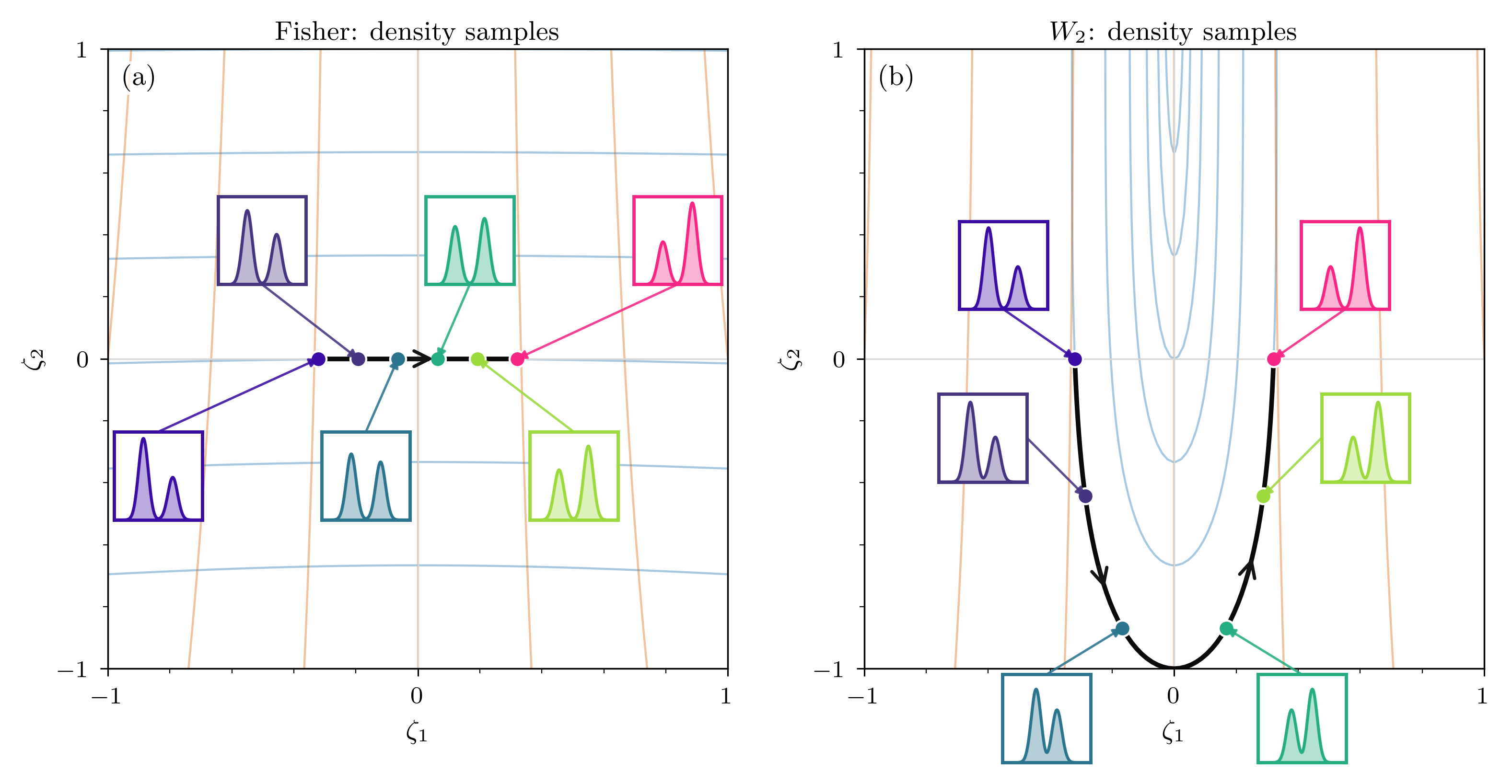}
  \caption{
    \label{si-fig:intrinsic-geodesic-grid-icons}
    Density snapshots along Fisher and Wasserstein geodesics.
    The background curves show the same geodesic grids as in Fig.~\ref{si-fig:intrinsic-geodesic-grid}, with black curves highlighting geodesics connecting the selected endpoint distributions.
    Insets show representative densities $\rho_{\bze}(\bx)$ sampled along each highlighted path.
    In Fisher geometry, the path mainly follows the direct coordinate displacement between the endpoints.
    In Wasserstein geometry, the path passes through intermediate density shapes before exchanging mass between wells, reflecting the fact that Wasserstein length measures physical transport cost rather than statistical distinguishability alone.
  }
\end{figure}

\subsection{Wasserstein geometry\label{si-sec:wasserstein-metric}}

Unlike the Fisher information geometry, the Wasserstein geometry captures the spatial structure of probability distributions and their similarity in terms of optimal transport.
The Wasserstein distance between two distributions $\rho$ and $\rho'$ is defined as the minimum cost of transporting the probability mass of $\rho$ to match $\rho'$:
\begin{equation}
\begin{aligned}
  W_2^2(\rho, \rho')
  =
  \inf_{\pi \in \Pi(\rho, \rho')}
  \int \dd{\bx} \dd{\bx'}
  \pi(\bx, \bx')
  \norm{\bx - \bx'}^2
\end{aligned}
\end{equation}
where $\Pi(\rho, \rho')$ is the set of all joint distributions $\pi(\bx, \bx')$ whose marginals are $\rho$ and $\rho'$ respectively.

Consider two nearby distributions $\rho_{\bze}$ and $\rho_{\bze + \dd{\bze}}$; let us see how the Wasserstein distance behaves between them.
First, we will want the relation:
\begin{equation}
\begin{aligned}
  \rho_{\bze + \dd{\bze}}(\bx)
  =
  \rho_{\bze}(\bx)
  +
  \dd{\rho(\bx)}
  =
  \rho_{\bze}(\bx)
  +
  \pdv{\rho_{\bze}(\bx)}{\zeta_\mu}
  \dd{\zeta_\mu}
\end{aligned}
\end{equation}
Let $T(\bx)$ be the optimal transport map (a vector field) that pushes $\rho_{\bze}$ to $\rho_{\bze + \dd{\bze}}$, so $\rho_{\bze + \dd{\bze}}(T(\bx)) \abs{\pdv{T}{x}} = \rho_{\bze}(\bx)$.
To first order, we can write $T(\bx) = \bx + \bu(\bx)$, where $\bu(\bx)$ is infinitesimal.
Let $f$ be an arbitrary test function; then we have:
\begin{equation}
\begin{aligned}
  \int \dd{\by}
  \rho_{\bze + \dd{\bze}}(\by)
  f(\by)
  &=
  \int \dd{\bx}
  f(T(\bx))
  \rho_{\bze}(\bx)
\end{aligned}
\end{equation}
and
\begin{equation}
\begin{aligned}
  f(T(\bx))
  =
  f(\bx + \bu(\bx))
  =
  f(\bx)
  +
  \nabla f(\bx) \cdot \bu(\bx)
\end{aligned}
\end{equation}
Putting these together and integrating by parts:
\begin{equation}
\begin{aligned}
  \int \dd{\bx}
  f(\bx) \, \dd{\rho(\bx)}
  &=
  \int \dd{\bx}
  \rho_{\bze}(\bx)
  \nabla f(\bx) \cdot \bu(\bx)
  =
  -
  \int \dd{\bx}
  f(\bx)
  \nabla \cdot
  \qty(
    \rho_{\bze}(\bx)
    \bu(\bx)
  )
\end{aligned}
\end{equation}
Therefore:
\begin{equation}
\begin{aligned}
  \dd{\rho(\bx)}
  =
  -
  \nabla \cdot
  \qty(
    \rho_{\bze}(\bx)
    \bu(\bx)
  )
\end{aligned}
\end{equation}
This is just the continuity equation for the probability distribution $\rho_{\bze}$ under the velocity field $\bu(\bx)$.

For this transport map, the quadratic cost is:
\begin{equation}
\begin{aligned}
  \int \dd{\bx}
  \rho_{\bze}(\bx)
  \norm{
    T(\bx) - \bx
  }^2
  =
  \int \dd{\bx}
  \rho_{\bze}(\bx)
  \norm{
    \bu(\bx)
  }^2
\end{aligned}
\end{equation}
Therefore, to find the Wasserstein distance between $\rho_{\bze}$ and $\rho_{\bze + \dd{\bze}}$, we need to find the velocity field $\bu(\bx)$ that satisfies the continuity equation and minimizes the quadratic cost:
\begin{equation}
\begin{aligned}
  \min_{\bu}
  \int \dd{\bx}
  \rho_{\bze}
  \norm{
    \bu
  }^2
  \qquad
  \text{s.t. }
  \dd{\rho}
  =
  -
  \nabla \cdot
  \qty(
    \rho_{\bze}
    \,
    \bu
  )
\end{aligned}
\end{equation}
We solve this using a Lagrange multiplier $\phi(\bx)$:
\begin{equation}
\begin{aligned}
  \mathcal{L}[\bu, \phi]
  &
  =
  \frac{1}{2}
  \int \dd{\bx}
  \rho_{\bze}
  \norm{
    \bu
  }^2
  -
  \int \dd{\bx}
  \phi
  \qty[
    \dd{\rho}
    +
    \nabla \cdot
    \qty(
      \rho_{\bze}
      \bu
    )
  ]
  \\
  &=
  \frac{1}{2}
  \int \dd{\bx}
  \rho_{\bze}
  \norm{
    \bu
  }^2
  +
  \int \dd{\bx}
  \rho_{\bze}
  \nabla \phi \cdot \bu
  -
  \int \dd{\bx}
  \phi
  \dd{\rho}
\end{aligned}
\end{equation}
where we have integrated the constraint term by parts. The variation with respect to $\bu$ gives:
\begin{equation}
\begin{aligned}
  \delta \mathcal{L}
  &=
  \int \dd{\bx}
  \rho_{\bze}
  \qty(
    \bu
    +
    \nabla \phi
  )
  \cdot
  \delta \bu
  =
  0
  \implies
  \bu
  =
  -
  \nabla \phi
\end{aligned}
\end{equation}
This means that all tangent vectors $\dd{\rho} = - \nabla \cdot (\rho \nabla \phi)$ are generated by potentials $\phi$.
For a parametric perturbation $\dd{\rho} = \pdv{\rho_{\bze}}{\zeta_\mu} \dd{\zeta_\mu}$, we can use linearity to assert the existence of potentials so that:
\begin{equation}
\begin{aligned}
  \dd{\rho}
  =
  -
  \nabla \cdot
  \qty(
    \rho_{\bze}
    \nabla \phi_\mu
  )
  \dd{\zeta_\mu}
\end{aligned}
\end{equation}
Substituting this back into the cost function, we find that the Wasserstein distance is given by:
\begin{equation}
\begin{aligned}
  W_2^2(\rho_{\bze}, \rho_{\bze + \dd{\bze}})
  &=
  \dd{\zeta_\mu}
  \dd{\zeta_\nu}
  g^W_{\mu\nu}(\zeta),
\end{aligned}
\end{equation}
where
\begin{equation}
\begin{aligned}
  g^W_{\mu\nu}(\zeta)
  =
  \int \dd{\bx}
  \rho_{\bze}
  \nabla \phi_\mu
  \cdot
  \nabla \phi_\nu
\end{aligned}
\end{equation}
is the Wasserstein metric.
If we define the weighted Laplacian operator $L_{\rho_{\bze}}$ as:
\begin{equation}
\begin{aligned}
  L_{\rho_{\bze}}[\phi]
  =
  -
  \nabla \cdot
  \qty(
    \rho_{\bze}
    \nabla \phi
  )
\end{aligned}
\end{equation}
Then the potentials $\phi_\mu$ are given by the inverse of this operator acting on the parametric perturbation:
\begin{equation}
\begin{aligned}
  \phi_\mu
  =
  L_{\rho_{\bze}}^{-1}
  \qty[
    \pdv{\rho_{\bze}}{\zeta_\mu}
  ]
\end{aligned}
\end{equation}
so the Wasserstein metric can be written as:
\begin{equation}
\begin{aligned}
  g^W_{\mu\nu}(\zeta)
  &=
  \int \dd{\bx}
  \pdv{\rho_{\bze}}{\zeta_\mu}
  L_{\rho_{\bze}}^{-1}
  \qty[
    \pdv{\rho_{\bze}}{\zeta_\nu}
  ]
\end{aligned}
\end{equation}
Because $L_{\rho_{\bze}}^{-1}$ is a non-local operator, the Wasserstein metric is a non-local metric that captures the spatial structure of the distributions and their similarity in terms of optimal transport.
The Wasserstein geometry is not invariant to re-mapping $x \to x'$ unlike the Fisher information geometry.

\subsection{Exponential family parameterization and co-diagonalizing the Wasserstein and Fisher metrics}

Now, we will return to our particular parameterization of the distributions and see how the Wasserstein and Fisher metrics look in this case.
We chose to parameterize the distributions as:
\begin{equation}
\begin{aligned}
  \rho_{\bze}(\bx)
  =
  q(\bx)
  \exp\qty(
    \sum_{\mu = 1}^\infty
    \zeta_\mu
    \chi_\mu(\bx)
    -
    F(\bze)
  )
\end{aligned}
\end{equation}
where $\chi_\mu$ are basis functions and $F(\bze) = \log \int \dd{\bx} q(\bx) \exp(\sum_\mu \zeta_\mu \chi_\mu(\bx))$ is the log-partition function that ensures normalization.
This is an exponential family parameterization of the distributions, and it has the nice property that the Fisher information matrix is equal to the covariance matrix of the modes: $I_{\mu\nu} = C_{\mu\nu}$.
It additionally has the property that:
\begin{equation}
\begin{aligned}
  \dd{\rho}
  =
  \rho_{\bze}
  \qty(
    \chi_\mu
    -
    \xi_\mu
  )
  \dd{\zeta_\mu}
\end{aligned}
\end{equation}
where $\xi_\mu = \int \dd{\bx} \rho_{\bze}(\bx) \chi_\mu(\bx)$ is the mean of the basis function $\chi_\mu$ under the distribution $\rho_{\bze}$.
We will assume that $\chi_\mu$ constitute a complete basis so we can expand the Wasserstein potentials as $\phi_\mu = A_{\mu\nu} \chi_\nu$.
The potential equation after integrating by parts becomes:
\begin{equation}
\begin{aligned}
  \int \dd{\bx}
  \rho_{\bze}
  \nabla \chi_\mu
  \cdot
  \nabla \phi_\nu
  =
  \int
  \dd{\bx}
  \rho_{\bze}
  \qty(
    \chi_\mu
    -
    \xi_\mu
  )
  \qty(
    \chi_\nu
    -
    \xi_\nu
  )
  \implies
  G_{\mu\pi}
  A_{\pi\nu}
  =
  C_{\mu\nu}
\end{aligned}
\end{equation}
where
\begin{equation}
\begin{aligned}
  G_{\mu\nu}(\bze)
  \equiv
  \int \dd{\bx}
  \rho_{\bze}
  \nabla \chi_\mu
  \cdot
  \nabla \chi_\nu
\end{aligned}
\end{equation}
With this, the Wasserstein metric can be written as:
\begin{equation}
\begin{aligned}
  g_{\mu\nu}^W
  &=
  A_{\mu\pi}
  G_{\pi\omega}
  A_{\omega\nu}
  =
  C_{\mu\pi}
  G_{\pi\omega}^{-1}
  C_{\omega\nu}
\end{aligned}
\end{equation}
In the main text we use the notation $J_{\mu\nu} = C_{\mu\pi}^{-1} G_{\pi\omega} C_{\omega\nu}^{-1}$, so the Wasserstein metric is the inverse of $J$: $g^W = J^{-1}$.
We can simultaneously diagonalize $\bC$ and $\bG$ about the origin by choosing the $\chi_\mu$ to solve the generalized eigenvalue problem:
\begin{equation}
\begin{aligned}
  -
  \nabla
  \cdot
  \qty(
    q
    \nabla \chi_\mu
  )
  =
  \kappa_\mu
  \chi_\mu
  q
\end{aligned}
\end{equation}
with the normalization
\begin{equation}
\begin{aligned}
  \int \dd{\bx}
  q
  \chi_\mu
  \chi_\nu
  =
  \delta_{\mu\nu}
\end{aligned}
\end{equation}
With this choice:
\begin{equation}
\begin{aligned}
  C_{\mu\nu}(\zeta = 0)
  =
  \delta_{\mu\nu},
  \qquad
  G_{\mu\nu}(\zeta = 0)
  =
  \kappa_\mu
  \delta_{\mu\nu},
  \qquad
  g_{\mu\nu}^W(\zeta = 0)
  =
  \frac{1}{\kappa_\mu}
  \delta_{\mu\nu}
   .
\end{aligned}
\end{equation}
This is an interesting result because it tells us that in the basis where the Fisher information is flat, the Wasserstein metric can be made diagonal but is heterogeneous with a nontrivial spectrum of eigenvalues.
The Wasserstein geometry can be highly anisotropic in this basis, with some directions being much more costly to move in than others.
An example of this is the two-peak Gaussian we have been considering in which $\chi_1$ is a slow inter-well balancing mode (see Fig.~\ref{fig:optimal-signaling-modes}(a)) with a very small $\kappa_1$ and $\chi_2$ is a fast intra-well mode with $\kappa_2 = O(1)$, so the Wasserstein metric is very anisotropic with $g_{11}^W \gg g_{22}^W$ despite the Fisher information being isotropic $C_{11} = C_{22} = 1$.
We explore this contrast between statistical distinguishability and transport geometry in Fig.~\ref{si-fig:intrinsic-geodesics}, where the two-dimensional exponential-family submanifold has nearly isotropic Fisher distances near $q$ but strongly anisotropic Wasserstein distances; in Figs.~\ref{si-fig:intrinsic-geodesic-grid} and~\ref{si-fig:intrinsic-geodesic-grid-icons}, we compare geodesics in the two geometries and show that the Wasserstein geodesics bend through intermediate density shapes to reduce transport cost, while the Fisher geodesics follow a more direct path along the inter-well coordinate.
It is also interesting to note that the Fisher and Wasserstein geometries can be severely not isometric, and the tangent vectors in the Wasserstein geometry are physical transport maps that move probability mass around in space, while the tangent vectors in the Fisher geometry are abstract statistical perturbations that can move probability mass around in any way as long as they satisfy the normalization constraint.
It is worth noting that this manipulation we performed is a general property of Riemannian geometry: any two metric $g,g'$ can be simultaneously diagonalized at a point by choosing coordinates that are normal for one of the metrics, and rotating the coordinates to diagonalize the other metric.

\clearpage
\section{Preconditioning, reservoir states, and a barycentric interpretation}
\label{si-sec:preconditioning}

\subsection{Motivating the passive drift}

We have been choosing the dynamics to be of the form:
\begin{equation}
\begin{aligned}
  \dot{\bx}_i
  =
  (
    \bu_0(\bx_i) + \bu(\bx_i;\rho_t)
  )
  +
  \sqrt{\varepsilon}\,
  \be_i(t)
\end{aligned}
\end{equation}
where
\begin{equation}
\begin{aligned}
  \bu_0(\bx)
  =
  -
  \nabla U_0(\bx),
  \qquad
  U_0(\bx)
  =
  -
  \frac{\varepsilon}{2}
  \log q(\bx)
\end{aligned}
\end{equation}
because this ensures that with $\bu = 0$, the system can maintain the desired distribution $q$.
To see this, look at the Fokker--Planck equation:
\begin{equation}
\begin{aligned}
  \partial_t
  \rho
  =
  -
  \frac{\varepsilon}{2}
  \nabla
  \cdot
  \qty(
    \rho
    \nabla
    \log q
  )
  +
  \frac{\varepsilon}{2}
  \nabla^2
  \rho
\end{aligned}
\end{equation}
One can immediately see that $\rho = q$ makes the RHS zero.
When the objective is undiscounted ($\gamma = 0$), this is a natural choice.
Otherwise, the optimal control policy as $t \to \infty$ would not hold $\rho$ at the desired state because the $\int \dd{\bx} \rho \norm{\bu}^2$ term in the objective would discourage the total drift from being $\bv = \bu_0 + \bu = \frac{\varepsilon}{2}\nabla \log q$.
This is all a property of having an infinite time horizon in which the controller has enough time to get $\rho$ all the way to $q$.
However, when $\gamma > 0$ and there is some finite time before a perturbation arises, the objective is no longer to get $\rho$ exactly to $q$ but as close as possible within the time available.
This can be accelerated by an alternative choice of the passive dynamics $\bu_0$ that is designed to help the system more effectively respond to perturbations.
This finite-recovery-time perspective is conceptually related to stochastic resetting and restart processes, where repeated restart events can change the optimal strategy for reaching a target state~\cite{evans2020stochastic}.
We will parameterize this in our basis as:
\begin{equation}
\begin{aligned}
  \bu_0^{\bph}(\bx)
  =
  -
  \nabla
  U_{\bph}(\bx),
  \qquad
  U_{\bph}(\bx)
  =
  -
  \frac{\varepsilon}{2}
  \log
  \qty[
    q(\bx)
    \exp\qty(
      \sum_{\mu = 1}^\infty
      \phi_\mu
      \chi_\mu(\bx)
    )
  ]
  .
\end{aligned}
\end{equation}
The coefficients $\bph$ specify the distribution which is maintained at steady-state by the passive dynamics (i.e., the distribution that makes the RHS of the Fokker--Planck equation zero when $\bu = 0$):
\begin{equation}
\begin{aligned}
  \rho_{\bph}(\bx)
  =
  q(\bx)
  \exp\qty(
    \sum_{\mu = 1}^\infty
    \phi_\mu
    \chi_\mu(\bx)
    -
    F(\bph)
  )
  .
\end{aligned}
\end{equation}
We call this distribution the auxiliary target distribution (as opposed to the desired distribution) and $\bu_0^{\bph}$ the preconditioned dynamics.
With this change in the passive dynamics, the $\bze$-dynamics become:
\begin{equation}
\begin{aligned}
  \dot \zeta_\mu
  &=
  -
  C_{\mu\nu}^{-1}
  G_{\nu\omega}
  \qty(
    \frac{\varepsilon}{2}
    (\zeta_\omega- \phi_\omega)
      +
      \frac{1}{r}
      C_{\omega \pi}^{-1}
    \pdv{S}{\zeta_\pi}
  )
\end{aligned}
\end{equation}
and the HJB equation becomes:
\begin{equation}
\begin{aligned}
  \gamma S
  &=
  D_{\rm KL}(\rho_{\bze} \Vert q)
  -
  \frac{1}{2r}
  \pdv{S}{\zeta_\mu}
  C_{\mu\nu}^{-1}
  G_{\nu\omega}
  C_{\omega\pi}^{-1}
  \pdv{S}{\zeta_\pi}
  -
  \frac{\varepsilon}{2}
  \pdv{S}{\zeta_\mu}
  C_{\mu\nu}^{-1}
  G_{\nu\omega}
  (\zeta_\omega - \phi_\omega)
  .
\end{aligned}
\end{equation}
Introducing this modified passive dynamics allows the system to implement controls that are more effective at responding to perturbations and allows it to incorporate prior knowledge about the perturbations it expects to encounter.
This means that including preconditioning makes the certainty equivalence assumption more accurate because any prior information that could be exploited by a non-feedback offset term can instead be exploited by the passive dynamics, so the optimal control policy can be more focused on responding to the perturbations that are not already being handled by the passive dynamics.

\subsection{Preconditioning as meta-optimization}

The goal of preconditioning is to choose the auxiliary target distribution (i.e., the passive dynamics) to help the system more effectively respond to perturbations.
This is essentially an outer-loop meta-optimization problem where the inner loop is the optimal control problem we have been solving so far, and the outer loop is optimizing over the auxiliary target distribution to minimize the cost of responding to perturbations.
We can write this as:
\begin{equation}
\begin{aligned}
  \min_{\bph}
  \min_{\bu}
  \mathcal{L}(\bu; \bph)
\end{aligned}
\end{equation}
where $\mathcal{L}(\bu; \bph)$ is the cost of responding to perturbations with control policy $\bu$ and preconditioning parameters $\bph$.
We will consider a case where the perturbation structure may depend on $\bph$ directly because we may conceive of $\bph$ as a homeostatic setpoint that the system is trying to maintain, and perturbations kick the system away from this setpoint, so the perturbation distribution should depend on the setpoint.
We determine $\bph$ numerically by performing gradient descent on $\mathcal{L}(\bph) = \min_{\bu} \mathcal{L}(\bu; \bph)$ with respect to $\bph$.
However, the gradient is nontrivial to compute because $\mathcal{L}(\bph)$ depends on the dynamics of the inner loop optimal control problem, and those dynamics depend on $\bph$.
We discuss how to compute this gradient as part of our numerical methods in SI~Sec.~\ref{si-sec:S-phi-gradients}.

\subsection{Refresher on (entropy-regularized) Wasserstein barycenters}

Wasserstein barycenters are a way to define a notion of a ``mean'' of a set of probability distributions under the Wasserstein metric~\cite{Santambrogio2015OptimalTransportApplied,Chewi2024StatisticalOptimalTransport}.
Here, we will briefly review the definition of Wasserstein barycenters and some of their properties.
If we consider a distribution over distributions $P(\rho)$, then the Wasserstein barycenter $\bar \rho$ is defined as the distribution that minimizes the expected squared Wasserstein distance to the distributions in $P$:
\begin{equation}
\begin{aligned}
  \bar \rho
  =
  \argmin_\rho
  \int \mathcal{D}\rho'\,
  P(\rho')
  W_2^2(\rho, \rho')
   .
\end{aligned}
\end{equation}
From an optimal transport perspective, the Wasserstein barycenter can be thought of as the distribution that minimizes the average cost of transporting probability mass from itself to the distributions in $P$.

We also will introduce the concept of entropy-regularized Wasserstein distances.
It is common to add an entropy regularization term to the optimal transport problem to make it more computationally tractable and define the $\varepsilon$-regularized Wasserstein distance as:
\begin{equation}
\begin{aligned}
  W^2_{2, \varepsilon}(\rho, \rho')
  =
  \inf_{\pi \in \Pi(\rho, \rho')}
  \qty{
    \frac{1}{2}
    \int \dd{\bx} \dd{x'}
    \pi(x, x')
    \norm{x - x'}^2
    +
    \varepsilon
    \int \dd{\bx} \dd{x'}
    \pi(x, x')
    \log \pi(x, x')
  }
\end{aligned}
\end{equation}
This can be conceived of as adding an entropy regularization term to the optimal transport problem which encourages the transport plan $\pi$ to be more diffuse and less sparse~\cite{Chewi2024StatisticalOptimalTransport}.
It can also be mapped to the Schr\"odinger bridge problem (SBP) which is a stochastic optimal control problem in which one finds stochastic dynamics that transport an initial distribution to a final distribution in a finite time interval $[0,1]$ while minimizing dissipation~\cite{Chen2014SchrodingerBridgesControl}.
In the SBP, the $\varepsilon$-regularized Wasserstein distance corresponds to the minimum cost of transporting probability mass from $\rho$ to $\rho'$ under the SBP dynamics with noise level $\varepsilon$.
With this, one can define an entropy-regularized Wasserstein barycenter, but it is common to add one more level of regularization on the barycenter itself to make it more computationally tractable~\cite{Chizat2023DoublyEntropicWassersteinBarycenters}:
\begin{equation}
\begin{aligned}
  \bar \rho_{\varepsilon, \tau}
  =
  \argmin_\rho
  \qty{
    \int \mathcal{D}\rho'\,
    P(\rho')
    W^2_{2, \varepsilon}(\rho, \rho')
    +
    \tau
    D_{\rm KL}(\rho \Vert \rho_0)
  }
\end{aligned}
\end{equation}
where $\rho_0$ is some reference distribution and $\tau$ is the regularization strength.
This can be thought of as finding a distribution that minimizes the average cost of transporting probability mass from itself to the distributions in $P$ under the $\varepsilon$-regularized Wasserstein distance while also being close to the reference distribution $\rho_0$ in terms of KL divergence.
This is all relevant because the optimal preconditioning problem can roughly be conceived of as finding an auxiliary target distribution $\rho_{\bph}$ that minimizes the average cost of transporting probability mass from itself to the perturbation distribution $p_{\rm pert}$ under the Wasserstein distance, so there is a close connection between optimal preconditioning and Wasserstein barycenters.

\subsection{Preconditioning as a barycenter problem}
\label{si-sec:preconditioning-wasserstein-barycenter}

The objective we are minimizing, $\ell$, takes the form of a stochastic optimal control problem in which we are trying to find a policy that minimizes dissipation ($\frac{r}{2}\int \dd{\bx} \rho \norm{\bu}^2$) while also bringing the distribution $\rho$ close to the desired distribution $q$ in terms of KL divergence.
This is conceptually similar to the Schr\"odinger bridge problem in which we are trying to find stochastic dynamics that transport an initial distribution to a final distribution in a finite time interval while minimizing dissipation~\cite{Chen2014SchrodingerBridgesControl}.
This means that there is a natural analogy between the optimal preconditioning problem and the doubly-regularized Wasserstein barycenter problem.

We now make more precise the sense in which the preconditioned stationary distribution $\rho_{\bph}$ can be interpreted as a barycenter-like object.
Recall that the perturbation ensemble $p_{\rm pert}$ is a probability distribution over initial collective states $\rho_0$.
For a fixed initial state $\rho_0$, let
\begin{equation}
\begin{aligned}
  S_\gamma(\rho_0; \rho_{\bph})
\end{aligned}
\end{equation}
denote the optimal discounted recovery cost starting from $\rho_0$ when the passive dynamics have stationary distribution $\rho_{\bph}$.
The preconditioning objective can then be written as
\begin{equation}
\begin{aligned}
  \mathcal{L}(\rho_{\bph})
  =
  \E_{\rho_0 \sim p_{\rm pert}}
  \qty[
    S_\gamma(\rho_0; \rho_{\bph})
  ],
  \qquad
  \rho_{\bph^\ast}
  =
  \argmin_{\rho_{\bph}}
  \mathcal{L}(\rho_{\bph}).
\end{aligned}
\end{equation}
This is not literally a Wasserstein barycenter problem because $\rho_{\bph}$ is not merely a point being optimized over in the space of distributions.
It also defines the passive drift,
\begin{equation}
\begin{aligned}
  \bu_0^{\bph}(\bx)
  =
  \frac{\varepsilon}{2}
  \nabla \log \rho_{\bph}(\bx),
\end{aligned}
\end{equation}
and therefore changes the dynamics by which the system recovers from perturbations.
Nevertheless, one can obtain a useful barycentric upper bound by considering a restricted class of recovery policies.

Consider any finite time $T>0$ and restrict to policies that transport the initial state $\rho_0$ to the auxiliary target $\rho_{\bph}$ during the interval $[0,T]$, and then turn off the control for $t\geq T$.
Let $\widetilde\rho_t$ denote the corresponding path in distribution space, so that
\begin{equation}
\begin{aligned}
  \widetilde\rho_{t=0}
  =
  \rho_0,
  \qquad
  \widetilde\rho_{t=T}
  =
  \rho_{\bph},
  \qquad
  \widetilde\rho_t
  =
  \rho_{\bph}
  \quad
  \text{for }t\geq T.
\end{aligned}
\end{equation}
Let $w_T(x,t)$ denote the velocity field that generates this path during $0\leq t\leq T$.
Since this is only one admissible class of policies, its cost upper-bounds the optimal recovery cost:
\begin{equation}
\begin{aligned}
  S_\gamma(\rho_0;\rho_{\bph})
  \leq
  &
  \int_0^T
  \dd{t}
  e^{-\gamma t}
  D_{\rm KL}(\widetilde\rho_t\Vert q)
  +
  \frac{r}{2}
  \int_0^T
  \dd{t}
  \int
  \dd{\bx}
  e^{-\gamma t}
  \widetilde\rho_t
  \norm{w_T}^2
  \\
  &+
  \int_T^\infty
  \dd{t}
  e^{-\gamma t}
  D_{\rm KL}(\rho_{\bph}\Vert q).
\end{aligned}
\label{eq:preconditioning-restricted-cost}
\end{equation}
The final term is the residual cost after the restricted policy has reached $\rho_{\bph}$ and turned off control.
Evaluating this tail gives
\begin{equation}
\begin{aligned}
  S_\gamma(\rho_0;\rho_{\bph})
  \leq
  &
  \int_0^T
  \dd{t}
  e^{-\gamma t}
  D_{\rm KL}(\widetilde\rho_t\Vert q)
  +
  \frac{r}{2}
  \int_0^T
  \dd{t}
  \int
  \dd{\bx}
  e^{-\gamma t}
  \widetilde\rho_t
  \norm{w_T}^2
  \\
  &+
  \frac{
    e^{-\gamma T}
  }{
    \gamma
  }
  D_{\rm KL}(\rho_{\bph}\Vert q).
\end{aligned}
\label{eq:preconditioning-upper-bound-before-convexity}
\end{equation}
This expression already displays the basic tradeoff: choosing $\rho_{\bph}$ close to typical perturbations can lower the transport cost, but choosing $\rho_{\bph}$ far from $q$ incurs a discounted long-time KL cost.

To make this tradeoff explicit, choose $\widetilde\rho_t$ to be the constant-speed Wasserstein geodesic from $\rho_0$ to $\rho_{\bph}$ over time $T$, with $w_T$ the corresponding Benamou--Brenier velocity field.
Then
\begin{equation}
\begin{aligned}
  \int_0^T
  \dd{t}
  \int
  \dd{\bx}
  \widetilde\rho_t
  \norm{w_T}^2
  =
  \frac{1}{T}
  W_2^2(\rho_0,\rho_{\bph}).
\end{aligned}
\end{equation}
Because this geodesic has constant speed,
\begin{equation}
\begin{aligned}
  \int
  \dd{\bx}
  \widetilde\rho_t
  \norm{w_T}^2
  =
  \frac{1}{T^2}
  W_2^2(\rho_0,\rho_{\bph}),
\end{aligned}
\end{equation}
so the discounted control cost is
\begin{equation}
\begin{aligned}
  \frac{r}{2}
  \int_0^T
  \dd{t}
  \int
  \dd{\bx}
  e^{-\gamma t}
  \widetilde\rho_t
  \norm{w_T}^2
  =
  \frac{r}{2}
  \frac{
    1-e^{-\gamma T}
  }{
    \gamma T^2
  }
  W_2^2(\rho_0,\rho_{\bph}).
\end{aligned}
\label{eq:discounted-benamou-brenier-action}
\end{equation}
Note that this uses a Wasserstein barycenter instead of an entropy-regularized Wasserstein barycenter.

Write $q(\bx)\propto e^{-U(\bx)}$, and assuming $U$ is semi-convex,
\begin{equation}
\begin{aligned}
  \nabla^2 U
  \succeq
  -\kappa \bm{I}.
\end{aligned}
\end{equation}
Then, by the notion of displacement convexity, for any $\widetilde \rho_t$ along the Wasserstein geodesic from $\rho_0$ to $\rho_{\bph}$ we have
\begin{equation}
\begin{aligned}
  D_{\rm KL}(\widetilde\rho_t\Vert q)
  \leq
  \qty(
    1-\frac{t}{T}
  )
  D_{\rm KL}(\rho_0\Vert q)
  +
  \frac{t}{T}
  D_{\rm KL}(\rho_{\bph}\Vert q)
  +
  \kappa
  \frac{
    t(T-t)
  }{
    T
  }
  W_2^2(\rho_0,\rho_{\bph}).
\end{aligned}
\label{eq:preconditioning-displacement-convexity-bound}
\end{equation}
Substituting Eq.~\eqref{eq:preconditioning-displacement-convexity-bound} into Eq.~\eqref{eq:preconditioning-upper-bound-before-convexity}, using Eq.~\eqref{eq:discounted-benamou-brenier-action}, and writing
\begin{equation}
\begin{aligned}
  \tau = \gamma T
\end{aligned}
\end{equation}
gives the finite-discount bound
\begin{equation}
\begin{aligned}
  S_\gamma(\rho_0;\rho_{\bph})
  \leq
  \min_{\tau>0}
  \Bigg\{
  &
  \frac{
    \tau - 1 + e^{-\tau}
  }{
    \gamma \tau
  }
  D_{\rm KL}(\rho_0\Vert q)
  +
  \frac{
    1-e^{-\tau}
  }{
    \gamma \tau
  }
  D_{\rm KL}(\rho_{\bph}\Vert q)
  \\
  &+
  \left[
    \frac{r}{2}
    \frac{
      \gamma(1-e^{-\tau})
    }{
      \tau^2
    }
    +
    \kappa
    \frac{
      (\tau - 2)
      +
      (\tau + 2)e^{-\tau}
    }{
      \gamma^2 \tau
    }
  \right]
  W_2^2(\rho_0,\rho_{\bph})
  \Bigg\}.
\end{aligned}
\label{eq:preconditioning-discounted-barycenter-bound}
\end{equation}
The minimization over $\tau$ reflects the freedom to choose the time $T=\tau/\gamma$ over which the restricted policy transports $\rho_0$ to $\rho_{\bph}$.
Equation~\eqref{eq:preconditioning-discounted-barycenter-bound} is an upper bound, not an equality for the true optimal recovery cost.
However, it makes the barycentric structure of the preconditioning problem explicit.

Averaging over the perturbation ensemble gives
\begin{equation}
\begin{aligned}
  \mathcal{L}(\rho_{\bph})
  \leq
  \E_{\rho_0\sim p_{\rm pert}}
  \Bigg[
    \min_{\tau>0}
    \Bigg\{
    &
    a_0(\tau)
    D_{\rm KL}(\rho_0\Vert q)
    +
    a_\phi(\tau)
    D_{\rm KL}(\rho_{\bph}\Vert q)
    +
    b(\tau)
    W_2^2(\rho_0,\rho_{\bph})
    \Bigg\}
  \Bigg],
\end{aligned}
\end{equation}
where
\begin{equation}
\begin{aligned}
  a_0(\tau)
  =
  \frac{
    \tau - 1 + e^{-\tau}
  }{
    \gamma \tau
  },
  \qquad
  a_\phi(\tau)
  =
  \frac{
    1-e^{-\tau}
  }{
    \gamma \tau
  },
\end{aligned}
\end{equation}
and
\begin{equation}
\begin{aligned}
  b(\tau)
  =
  \frac{r}{2}
  \frac{
    \gamma(1-e^{-\tau})
  }{
    \tau^2
  }
  +
  \kappa
  \frac{
    (\tau - 2)
    +
    (\tau + 2)e^{-\tau}
  }{
    \gamma^2 \tau
  }.
\end{aligned}
\end{equation}
For a fixed perturbation ensemble, $D_{\rm KL}(\rho_0\Vert q)$ is independent of $\rho_{\bph}$, although it can still affect the optimal transport time $\tau$ in the bound.
The dependence on the preconditioned stationary distribution is controlled by two competing contributions:
$
  D_{\rm KL}(\rho_{\bph}\Vert q),
$
and
$
  \E_{\rho_0\sim p_{\rm pert}}
  W_2^2(\rho_0,\rho_{\bph})
$.
The former penalizes choosing a passive stationary distribution far from the desired state.
The latter penalizes choosing a passive stationary distribution that is far, in transport distance, from the perturbations the system typically experiences.
This is the sense in which preconditioning is barycenter-like: $\rho_{\bph}$ behaves as a discounted Wasserstein barycenter of the perturbation ensemble, regularized by its divergence from the true target $q$.

This interpretation should not be read as saying that $\rho_{\bph^\ast}$ is exactly an ordinary Wasserstein barycenter.
An ordinary Wasserstein barycenter would minimize
\begin{equation}
\begin{aligned}
  \bar\rho
  =
  \argmin_{\rho}
  \E_{\rho_0\sim p_{\rm pert}}
  W_2^2(\rho_0,\rho).
\end{aligned}
\end{equation}
The preconditioned state $\rho_{\bph^\ast}$ differs from this object in two important ways.
First, it is regularized toward the desired state $q$ by the KL term.
Second, it is selected by a discounted control problem, so the relevant notion of centrality depends on the perturbation timescale $1/\gamma$.
Thus $\rho_{\bph^\ast}$ is better interpreted as a finite-horizon reservoir state: it is chosen to be accessible from typical perturbations on the discounted timescale, while remaining close enough to $q$ that the long-time bias is not too costly.
In metastable examples, this naturally leads to reservoir states between the modes of $q$, because placing probability mass in intermediate regions can reduce transport distances from perturbed states and accelerate short-time recovery.

\clearpage
\section{Numerical methods and simulation parameters}
\label{si-sec:numerics}

In this section, we discuss the numerical methods used throughout the paper.
In SI~Secs.~\ref{si-sec:numerics-chi-mu}, \ref{si-sec:numerics-S-optimization-overview}, \ref{si-sec:S-phi-gradients}, and \ref{si-sec:numerics-geometry-visualization} we discuss the high-level approaches of the methods used, and in SI~Sec.~\ref{si-sec:numerics-simulation-parameters}, we discuss specific parameters and implementation details used in the simulations.

\subsection{Solving for the basis functions $\chi_\mu$}
\label{si-sec:numerics-chi-mu}

To numerically solve for the basis functions, we chose a square stencil of evenly spaced points in the state space and discretized the generalized eigenvalue problem for $\chi_\mu$ using finite differences.
The stiffness matrix is computed by discretizing the operator $-\nabla \cdot (q \nabla)$ using finite differences, and the mass matrix is computed by discretizing the operator of multiplication by $q$ (a diagonal matrix with $q$ evaluated at the stencil points on the diagonal).
This gives a generalized eigenvalue problem of the form $\bm{A} \chi_\mu = \kappa_\mu \bm{B} \chi_\mu$ where $\bm{A}$ is the stiffness matrix and $\bm{B}$ is the mass matrix.
In one dimension, this is solved using \texttt{scipy.linalg.eigh} which is a dense solver.
In two dimensions, this is solved using the SciPy ARPACK setting in \texttt{scipy.sparse.linalg.eigsh}~\cite{Virtanen2020SciPy} which is a sparse solver that can efficiently compute the lowest few eigenvalues and corresponding eigenvectors; this is necessary because solving the full dense eigenvalue problem in two dimensions is computationally and memory intensive.
We use zero-flux boundary conditions by modifying the finite difference stencil at the boundaries to ensure that the flux of probability mass across the boundary is zero.
In these problems there is a trivial zero eigenvalue corresponding to the constant function, and we discard this mode and take the next few modes as our basis functions.
In order to evaluate quantities like $\rho_{\bze}$, $\bC$, and $\bG$ at arbitrary $\bze$, we used exact integration over $\bx$ and log-sum-exp tricks to compute these quantities in a numerically stable way.

\subsection{Solving for the value function $S$ using optimization}
\label{si-sec:numerics-S-optimization-overview}

Solving the HJB equation Eq.~\eqref{eq:hjb-limited-signaling} for $S(\bb)$ is highly nontrivial because it involves coefficient matrices that depend on $\bb$ and a quadratic term in the gradient of $S$.
By definition, $S$ is the value function of the optimal control problem, so we can solve for it by directly optimizing over $S$ to minimize the cost of responding to perturbations under the optimal policy.
By representing the state in the basis functions $\{\chi_\mu\}$, the dynamics of the system can be represented by a finite-dimensional system of ODEs for the coefficients $\zeta_\mu$ instead of a PDE (the Fokker--Planck equation) if we truncate the basis to a finite number of modes, closely related to projection methods for Fokker--Planck dynamics on finite-dimensional exponential families~\cite{brigo2009projectingfokkerplanckequationfinite}.
We can write the optimization problem over $S$ as:
\begin{align}
  \min_{S}
  &\quad
  \E_{\bze(0) \sim p_{\rm pert}}
  \qty[
    \int_0^\infty
    \dd{t}
    e^{-\gamma t}
    \qty(
      D_{\rm KL}(\rho_{\bze(t)}\Vert q)
      +
      \frac{r}{2}
      a_\mu
      G_{\mu\nu}
      a_\nu
    )
  ]
  \\
  \text{subject to }
  &\quad
  \dot \zeta_\mu
  =
  -
  C_{\mu\nu}^{-1}
  G_{\nu\omega}
  \qty(
    \frac{\varepsilon}{2}
    (\zeta_\omega- \phi_\omega)
    -
    a_\omega
  )
  \label{eq:numerics-zeta-dynamics}
  \\
  &\quad
  a_\mu(\bb(t))
  =
  -
  \frac{1}{r}
  W_{\alpha\mu}
  \widehat C_{\alpha\beta}^{-1}
  \pdv{S}{\beta_\beta}
  \label{eq:numerics-a-mu}
\end{align}
where $\bb$ and $\bze$ are related by the moment-matching condition as discussed before.
This means that we can solve for $S$ by directly optimizing this objective with respect to $S$.
This is done by rolling out trajectories under the optimal policy defined by $S$, computing the cost of those trajectories, and backpropagating through the dynamics to compute the gradient of the cost with respect to $S$.
In this implementation, the state cost that is being minimized is $D_{\rm KL}(\rho_{\bze(t)} \Vert q)$, not $D_{\rm KL}(\rho_{\hat{\bze}(t)} \Vert q)$; that is, the cost is evaluated at the true state, not the estimated state (but the policy only has access to the estimated state $\bb$).
In the above, matrices with hats are evaluated at $\hat{\zeta}_\mu(t) = W_{\alpha\mu}
  \beta_\alpha(t)
  +
  g_\mu
$ where $g_\mu$ are the prior offsets.
In order to determine the prior offset $g_\mu$, we optimize over $g_\mu$ to minimize the cost of responding to perturbations which makes the system select the prior that is most favorable for responding to perturbations and assume that the system has learned to exploit this prior in its policy.
This optimization of $S$ was performed by parameterizing $S$ as a neural network and performing gradient descent on the above objective with respect to the parameters of $S$.

A particularly useful trick for improving the stability of this optimization is to initialize $S$ as the weak-control approximate solution derived in SI~Sec.~\ref{si-sec:weak-control-solution} and then optimize from there:
\begin{equation}
\begin{aligned}
  S_\theta(\bb)
  =
  \frac{1}{\gamma}
  D_{\rm KL}(\rho_{\hat{\bze}} \Vert q)
  +
  \delta S_\theta(\bb)
\end{aligned}
\end{equation}
where $\delta S_\theta$ is a neural network with parameters $\theta$ that is initialized to zero and optimized.
It is additionally useful to parameterize $S_\theta$ in terms of $\bvp$ instead of $\bb$ because after using the change of variables $\pdv{\beta_\alpha}{\varphi_\beta} = \widehat C_{\alpha\beta}^{-1}$, the optimal policy becomes $a_\mu = -\frac{1}{r} W_{\alpha\mu} \pdv{S}{\varphi_\alpha}$ which avoids using $\widehat{\bC}^{-1}$ during the rollout of the dynamics which can be numerically unstable, especially early in training when $S$ may be inaccurate and have large gradients.

Throughout the simulation, it is necessary to solve for $\bb$ in terms of $\bze$ at each time step in order to compute the optimal policy and roll out the dynamics.
In the analytics we performed earlier, we approximated this by a linear projection, but in the numerics, we solve for $\bb$ using the moment-matching condition $W_{\alpha\mu} \xi_\mu(\hat{\bze}) = W_{\alpha\mu}\xi_\mu(\bze)$ where $\hat{\zeta}_\mu = W_{\alpha\mu} \beta_\alpha + g_\mu$ and $\xi_\mu(\bze) = \int \dd{\bx} \rho_{\bze}(\bx) \chi_\mu(\bx)$.
This is done by using Newton iteration to solve for $\bb$ at each time step, which is stable and efficient because the moment-matching condition is a convex optimization problem in $\bb$.

In Fig.~\ref{fig:dynamics-gallery}, we simulate the control of a swarm of particles under limited signaling and control.
The controller is trained (through learning the value function $S$) in $\bze$-space, not with a swarm of particles.
The trained controller is then rolled out using the empirical $\varphi_\alpha = \frac{1}{N} \sum_{i=1}^N \lambda_\alpha(\bx_i(t))$.
During these roll-outs, we solve for $\bb$ in terms of $\bvp$ by using the moment matching condition $W_{\alpha\mu} \xi_\mu(\hat{\bze}) = \varphi_\alpha$.

\begin{figure}
  \includegraphics{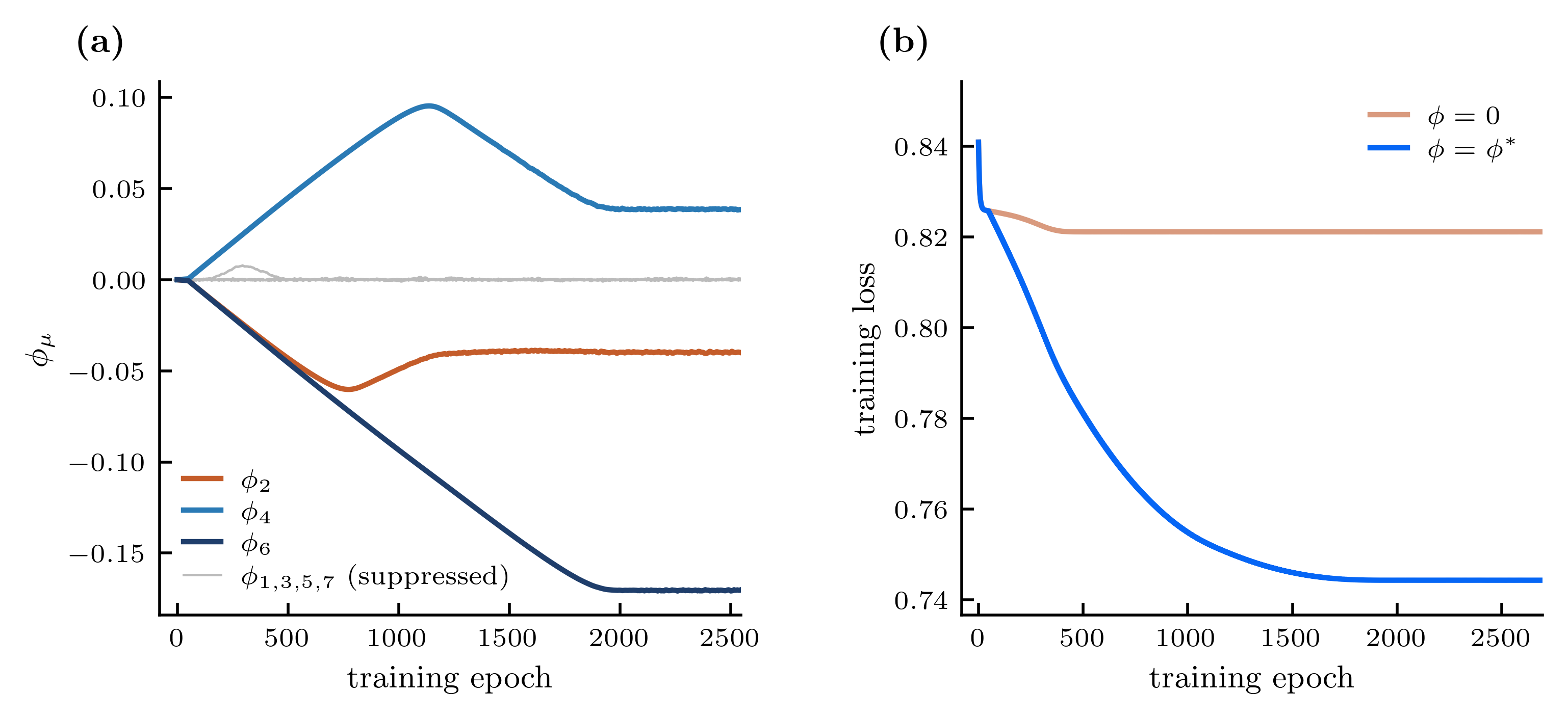}
  \caption{\label{fig:preconditioning-training-overview}
    (a) Evolution of the preconditioning parameters $\bph$ during training.
    The modes corresponding to odd $\mu$ are suppressed because the perturbation ensemble is symmetric and therefore the optimal preconditioning distribution should also be symmetric.
    (b) Training loss curve with and without preconditioning optimization.
    These are the training curves for the runs shown in Fig.~\ref{fig:preconditioning} in the main text.
  }
\end{figure}

\subsection{Differentiating through simulated dynamics}
\label{si-sec:S-phi-gradients}

Optimizing the value function $S$ and the preconditioning parameters $\bph$ requires differentiating trajectory losses through the simulated coefficient dynamics. 
This is nontrivial because $S$ and $\bph$ do not only enter the instantaneous loss; they also change the vector field for the ODE, and therefore change the entire trajectory $\bze(t)$ over which the loss is accumulated.
Thus, gradients must account for both the direct dependence of the loss on the parameters and the indirect dependence through the simulated dynamics.

Adjoint methods provide a standard way to compute such gradients for ODE-constrained objectives by propagating gradient information backward through the dynamics rather than storing all intermediate states explicitly~\cite{Sinha2023OptimalControlActiveParticles,Krishnan2026HamiltonianBridge}.
In practice, however, we did not implement an explicit continuous-adjoint equation. The trajectory losses were differentiated using automatic differentiation in JAX~\cite{Bradbury2018JAX}. 
We used the \texttt{RecursiveCheckpointAdjoint} setting in Diffrax~\cite{Kidger2021Diffrax}, which checkpointed intermediate states during the forward solve and recomputed trajectory segments during the backward pass to obtain gradients with reduced memory cost.

Forward and reverse integration during training was performed using the \texttt{Heun} method in Diffrax~\cite{Kidger2021Diffrax} for computational efficiency.
During evaluation of the trained controller, we used the more accurate \texttt{Tsit5} method. During training, we only simulate to a finite time $T$ which is chosen so that $e^{-\gamma T} < 0.01$ so that the tail cost is negligible.
In Fig.~\ref{fig:preconditioning-training-overview}, we show the evolution of the preconditioning parameters $\bph$ during training and the training loss curve with and without preconditioning optimization.

\subsection{Visualizing differential geometry of Wasserstein and Fisher information metrics}
\label{si-sec:numerics-geometry-visualization}

To visualize the Fisher and Wasserstein geometries in SI~Sec.~\ref{si-sec:info-geom}, we restricted attention to a two-dimensional exponential-family submanifold
\begin{equation}
\begin{aligned}
  \rho_{\bze}(\bx)
  =
  q(\bx)
  \exp\qty[
    \zeta_1 \chi_1(\bx)
    +
    \zeta_2 \chi_2(\bx)
    -
    F(\bze)
  ].
\end{aligned}
\end{equation}
where the basis functions $\chi_1$ and $\chi_2$ are calculated as described in SI~Sec.~\ref{si-sec:numerics-chi-mu}.
At each point $\zeta=(\zeta_1,\zeta_2)$ on a regular grid, we evaluated the Fisher metric
\begin{equation}
\begin{aligned}
  C_{\mu\nu}(\bze)
  =
  \int \dd{\bx}
  \rho_{\bze}(\bx)
  \qty(\chi_\mu(\bx)-\xi_\mu(\bze))
  \qty(\chi_\nu(\bx)-\xi_\nu(\bze)),
\end{aligned}
\end{equation}
where $\xi_\mu(\bze)=\int \dd{\bx}\rho_{\bze}(\bx)\chi_\mu(\bx)$.
We also evaluated
\begin{equation}
\begin{aligned}
  G_{\mu\nu}(\bze)
  =
  \int \dd{\bx}
  \rho_{\bze}(\bx)
  \nabla \chi_\mu(\bx)
  \cdot
  \nabla \chi_\nu(\bx),
\end{aligned}
\end{equation}
and to obtain the Wasserstein metric on the $\bze$-submanifold,
\begin{equation}
\begin{aligned}
  g^W_{\mu\nu}(\bze)
  =
  C_{\mu\omega}(\bze)
  G_{\omega\pi}^{-1}(\bze)
  C_{\pi\nu}(\bze).
\end{aligned}
\end{equation}

For the distance maps in Fig.~\ref{si-fig:intrinsic-geodesics}, we computed the intrinsic distance from the reference distribution $q=\rho_{\bze=0}$ under each metric.
The distance function $d(\zeta)$ is defined so that $d(\zeta)$ is the length of the shortest path, measured using the chosen metric, from the origin to $\zeta$.
Equivalently, infinitesimal displacements of the distance function satisfy
\begin{equation}
\begin{aligned}
  g^{\mu\nu}(\zeta)
  \pdv{d}{\zeta_\mu}
  \pdv{d}{\zeta_\nu}
  =
  1,
  \qquad
  d(0)=0,
\end{aligned}
\label{eq:intrinsic-distance-eikonal}
\end{equation}
where $g^{\mu\nu}$ is the inverse metric.
This equation says that the gradient of the distance function has unit norm with respect to the inverse metric, just as the ordinary Euclidean distance from the origin satisfies $\abs{\nabla d}=1$.
Eq.~\eqref{eq:intrinsic-distance-eikonal} is called the eikonal equation and is directly analogous to the Hamilton--Jacobi equation (notice the quadratic form in the gradient of $d$).
This equation is generally difficult to solve because one cannot directly obtain $\pdv*{d}{\zeta_\mu}$ and integrate along it to get $d$.
In fact, there are infinitely many gradient vectors that satisfy the above equation at each point (i.e., have unit norm).
Because of this, $d$ cannot be found by direct integration and must instead be solved self-consistently~\cite{Sethian1996FastMarching,Sethian2001OrderedUpwind,Zhao2005FastSweeping}.
To do so numerically, we discretize the parameter space $\zeta$ onto a regular grid with spacing $\Delta$, and let $d_{(i,j)}$ denote the (initially unknown) geodesic distance at the grid point $(\zeta_1, \zeta_2) = (i\Delta, j\Delta)$.
The physical constraint that determines a unique solution is that distance must grow monotonically outward from the origin.
A geodesic leaving the origin accumulates length as it travels, so $d_{(i,j)}$ must be larger than the distances at whichever neighboring grid points lie closer to the origin.
This suggests approximating the gradient $\pdv*{d}{\zeta_\mu}$ at each grid point using the difference between $d_{(i,j)}$ and the smaller of its two neighbors in the $\zeta_\mu$ direction (i.e., the neighbor that distance must have grown away from). Substituting these finite-difference approximations into the eikonal equation turns it into a quadratic equation for the single unknown $d_{(i,j)}$, in terms of the neighboring distances and the local metric $g^{\mu\nu}$ (including its off-diagonal components).
The two roots of this quadratic correspond to the gradient pointing inward or outward from the origin; we always take the root larger than both neighboring distances, since distance grows outward.
Starting from $d_{(0,0)}=0$ at the origin and $d_{(i,j)}=\infty$ at all other grid points, we sweep this local update repeatedly over the grid.
Each sweep propagates distance information outward by one grid step, and the process continues until the distance field stops changing, at which point every grid point holds a self-consistent estimate of the geodesic distance from the origin.

For the geodesic grids in Figs.~\ref{si-fig:intrinsic-geodesic-grid}
and~\ref{si-fig:intrinsic-geodesic-grid-icons}, we computed individual geodesic paths by directly
integrating the geodesic equation,
\begin{equation}
\begin{aligned}
  \ddot{\zeta}^\mu + \Gamma^\mu_{\nu\omega}(\zeta)\,\dot{\zeta}^\nu\dot{\zeta}^\omega = 0,
\end{aligned}
\label{eq:geodesic-ode}
\end{equation}
where overdots denote derivatives with respect to arc length and
\begin{equation}
\begin{aligned}
  \Gamma^\mu_{\nu \omega} = \frac{1}{2} g^{\mu\pi}\qty(\partial_\nu  g_{\pi \omega} + \partial_\omega g_{\pi \nu } -
  \partial_\pi g_{\nu \omega})
\end{aligned}
\end{equation}
are the Christoffel symbols of the metric, computed numerically from finite differences of $g_{\mu\nu}$ on the grid.
Given an initial position $\zeta(0)$ and velocity $\dot\zeta(0)$, we integrated this ODE forward and backward in arc length using a standard adaptive solver.

To produce a coordinate-like grid that reveals the global geometry, we generated two families of geodesics.
The first family is seeded at evenly spaced points along the $\zeta_1=0$ axis, with initial velocities chosen to be metric-orthogonal to the $\zeta_2$ direction at each seed point (i.e., $g_{\mu\nu}\dot\zeta^\mu\hat{e}_2^\nu=0$).
The second family is seeded analogously along the $\zeta_2=0$ axis, with metric-orthogonal initial velocities in the $\zeta_1$ direction.
At their seed points, the two families are perpendicular in the sense of the metric, so together they form a geodesic grid analogous to a coordinate grid, but adapted to the local geometry rather than to the flat parameter space.
Where the grid lines are nearly straight and evenly spaced, the metric is nearly flat; where they bend or compress, the geometry is curved. The black highlighted curves in each panel are geodesics connecting a specific pair of endpoint distributions, computed by the same ODE integration but with initial velocity tuned to reach the target point.

\clearpage
\subsection{Simulation parameters and figure details}
\label{si-sec:numerics-simulation-parameters}

\subsubsection*{Fig.~\ref{fig:dynamics-gallery}: finite-particle realization of the learned
signaling-mediated controller}

For Fig.~\ref{fig:dynamics-gallery}, we used a one-dimensional double-well target distribution
\begin{equation}
  q(x)
  =
  \frac{1}{2}\mathcal{N}(x;-x_0,\sigma^2)
  +
  \frac{1}{2}\mathcal{N}(x;x_0,\sigma^2),
  \qquad
  x_0=2,\quad \sigma=0.5,
\end{equation}
on the interval $x\in[-7,7]$ with reflecting boundaries.
The exponential-family basis was obtained by solving the weighted Neumann eigenproblem $-\partial_x(q\partial_x\chi_\mu)=\kappa_\mu q\chi_\mu$ on a grid of 801 points, discarding the constant mode, and retaining the first seven nonconstant modes to define the $\bze$-coordinates.
The physical parameters were $\varepsilon=1$, $r=0.1$, and $\gamma=1$.
The signaling-mediated controller was trained with a single signaling mode, $M=1$, chosen to be the slowest nonconstant relaxation mode, $\lambda_1(x)=\chi_1(x)$.
For comparison, the passive dynamics used the same static equilibrium landscape but no feedback control.
The learned value function used a two-hidden-layer neural-network correction with hidden widths $(16,16)$ added to the weak-control KL-reference initialization.
Training used perturbations
\begin{equation}
  \bze(0) = (\Delta,0,\ldots,0),
  \qquad
  \Delta \sim \mathrm{Unif}[-1.5,1.5],
\end{equation} with 2048 training initial conditions, batches of 512, and 1000 training epochs.
Both the value-function parameters and the prior offset $g_\mu$ were optimized.
For the deterministic curves in the top panel, we integrated the seven-dimensional truncated $\bze$-dynamics from
\begin{equation}
  \bze(0) = (1.5,0,\ldots,0),
\end{equation}
and plotted $D_{\rm KL}(\rho_{\bze(t)}\Vert q)$.
For the particle snapshots, we sampled $N=4096$ particles from $\rho_{\bze(0)}$ and evolved them with Euler--Maruyama time step $\Delta t=0.01$, saving every five time steps.
At each particle time step, the controller received only the empirical signal
\begin{equation}
  \varphi_1(t)=\frac{1}{N}\sum_{i=1}^N \lambda_1(x_i(t)).
\end{equation}
The corresponding belief coordinate $\beta$ was obtained by solving the moment-matching condition $W_{\alpha\mu}\xi_\mu(\hat{\bze})=\varphi_\alpha$, and the resulting feedback was applied to the particles.
The snapshot panels show particle histograms at $t=0,5,15,40$ together with the target density $q(x)$ and the instantaneous effective landscape.
In the plotted density panels, this potential is rescaled onto the density axis only for visual comparison with the histograms.

\subsubsection*{Fig.~\ref{fig:optimal-signaling-modes}: optimal signaling modes for mode-aligned
perturbations}

Fig.~\ref{fig:optimal-signaling-modes} was generated by solving the passive relaxation eigenproblem described in SI~Sec.~\ref{si-sec:numerics-chi-mu} for two example target distributions.
For both examples, we considered perturbations whose covariance is isotropic in the passive relaxation basis, $\Sigma_{\mu\nu}=\delta_{\mu\nu}$.
In this special case, the optimal signaling modes are the passive relaxation modes themselves, ordered by the signaling priority derived in the main text.
Since the perturbation variance is the same for all modes in this example, the ordering shown in the figure is equivalent to ordering by the inverse priority scale $\ds_\alpha^{-1}$, which is proportional to the passive relaxation eigenvalue $\kappa_\alpha$ for the parameters used in the plot.
In particular, we work in the $\gamma \to 0$ limit, so $s_\alpha \sim \gamma/(\varepsilon \kappa_\alpha)$ and the values of $s_\alpha^{-1}$ displayed are measured in units of $\varepsilon/\gamma$.
We choose to plot $s_\alpha^{-1}$ rather than $s_\alpha$ because in the presented examples, the signaling priorities are separated by several orders of magnitude, so the inverse scale is more visually informative.

For the one-dimensional example in Fig.~\ref{fig:optimal-signaling-modes}(a), the target density
was
\begin{equation}
  q(x)
  =
  \frac{1}{2}\mathcal{N}(x;-2,0.5^2)
  +
  \frac{1}{2}\mathcal{N}(x;2,0.5^2).
\end{equation}
The eigenproblem was solved on $x\in[-8,8]$ using Mathematica's \texttt{NDEigensystem} with finite-element discretization and maximum cell measure $10^{-3}$.
We computed the lowest ten eigenfunctions, discarded the constant mode, and plotted the first nine nonconstant modes.
The corresponding nonzero eigenvalues were approximately
\begin{equation}
\begin{aligned}
  2.48\times 10^{-3},\,
  4.00,\,
  4.03,\,
  8.01,\,
  8.17,\,
  12.1,\,
  12.5,\,
  16.3,\,
  17.2 .
\end{aligned}
\end{equation}
The first mode is the slow inter-well population-imbalance mode, while the higher modes describe increasingly localized intra-well features.

For the two-dimensional example in Fig.~\ref{fig:optimal-signaling-modes}(b), the target density was an equal mixture of four isotropic Gaussians,
\begin{equation}
  q(x,y)
  =
  \frac{1}{4}
  \sum_{s_x=\pm 1}
  \sum_{s_y=\pm 1}
  \mathcal{N}\!\left((x,y);(3s_x,3s_y),0.8^2 I\right).
\end{equation}
The eigenproblem was solved on $(x,y)\in[-8,8]^2$ using Mathematica's \texttt{NDEigensystem} with finite-element discretization and maximum cell measure $10^{-2}$.
We computed the lowest forty eigenfunctions, discarded the constant mode, and displayed the low-priority modes as smaller thumbnails.
The first nonzero eigenvalues were approximately
\begin{equation}
\begin{aligned}
  2.36\times10^{-3},\,
  2.36\times10^{-3},\,
  4.71\times10^{-3},\,
  1.56,\,
  1.56,\,
  1.57,\ldots
\end{aligned}
\end{equation}
reflecting that the two slowest modes are degenerate diagonal-crossing modes, followed by a slightly faster non-degenerate mode describing population imbalance between neighboring quadrants (checkerboard), and then by a cluster of modes describing intra-quadrant features.

\subsubsection*{Figs.~\ref{fig:signaling-mixing} and~\ref{si-fig:signaling-mixing}: crossover in the optimal signaling direction for the symmetric two-peak mixture}

Figs.~\ref{fig:signaling-mixing} and~\ref{si-fig:signaling-mixing} were generated from the same symmetric two-peak Gaussian-mixture example discussed in SI~Sec.~\ref{si-sec:two_peak_gaussian}, with
\begin{equation}
  q(x)
  =
  \frac{1}{2}\mathcal{N}(x;-x_0,\delta^2)
  +
  \frac{1}{2}\mathcal{N}(x;x_0,\delta^2),
  \qquad
  \delta = 0.2.
\end{equation}
In both figures, the perturbation covariance was taken to be rank one in the passive relaxation basis,
\begin{equation}
  \Sigma
  =
  \sigma^2 \bm{h}\bm{h}^T,
  \qquad
  \bm{h}
  =
  \cos\alpha\, \bm{e}_1
  +
  \eta\sin\alpha
  \qty(
    \cos\beta\, \bm{e}_2
    +
    \sin\beta\, \bm{e}_3
  ),
\end{equation}
with
\begin{equation}
  \alpha = \frac{\pi}{4},
  \qquad
  \beta = 0,
  \qquad
  \eta = 8,
  \qquad
  \sigma = 0.06.
\end{equation}
Thus, for these plots the perturbation direction lies in the span of the slow switching mode $\chi_1$ and the even member of the nearly degenerate $(\chi_2,\chi_3)$ pair.

For each value of $x_0$, we solved the weighted Sturm--Liouville eigenproblem
\begin{equation}
  -\partial_x\qty(q\,\partial_x \chi_\mu)
  =
  \kappa_\mu q \chi_\mu
\end{equation}
using the finite-volume discretization with natural no-flux boundary conditions described in SI~Sec.~\ref{si-sec:numerics-chi-mu}.
We used a uniform grid of $2001$ points and discarded the constant mode, retaining the first three nonconstant modes $\chi_1,\chi_2,\chi_3$.
For the main-text sweep in Fig.~\ref{fig:signaling-mixing}, the computational domain was chosen using the largest peak separation in that sweep, corresponding to
\begin{equation}
  x \in [-1.68,\,1.68],
\end{equation}
while for the detailed SI figure, Fig.~\ref{si-fig:signaling-mixing}, the three displayed cases were computed on the common interval
\begin{equation}
  x \in [-1.64,\,1.64].
\end{equation}
Fig.~\ref{fig:signaling-mixing} is the compact summary of the crossover.
The top panel was generated by sweeping
\begin{equation}
  x_0/\delta \in [1.2,\,3.4]
\end{equation}
using $29$ evenly spaced values.
At each value of $x_0/\delta$, we computed $\kappa_1,\kappa_2,\kappa_3$, defined
\begin{equation}
  \kappa_\parallel = \frac{\kappa_2+\kappa_3}{2},
\end{equation}
and evaluated the mixing angle
\begin{equation}
  \vartheta
  =
  \frac{1}{2}
  \operatorname{atan2}\qty(
    4\qty(\frac{\kappa_1}{\kappa_\parallel})^{3/2}\eta\tan\alpha,\,
    1-\qty(\frac{\kappa_1}{\kappa_\parallel}\eta\tan\alpha)^2
  ).
\end{equation}
The lower panel of Fig.~\ref{fig:signaling-mixing} shows the ratio $\kappa_2/\kappa_1$ extracted from the same numerical solutions.

The inset comparisons in Fig.~\ref{fig:signaling-mixing}, and the three columns of Fig.~\ref{si-fig:signaling-mixing}, correspond to the same three representative peak separations,
\begin{equation}
  x_0 = 0.30,\qquad 0.43,\qquad 0.64,
\end{equation}
i.e.
\begin{equation}
  x_0/\delta = 1.5,\qquad 2.15,\qquad 3.2.
\end{equation}
In the panel titles these were rounded to one decimal place, so the middle case is labeled as $x_0/\delta=2.1$ in the figure.
For these three cases, the numerically obtained mixing angles were
\begin{equation}
  \vartheta \approx 57.1^\circ,\qquad 23.8^\circ,\qquad 0.7^\circ.
\end{equation}
In the main-text figure insets, we plotted the real-space optimal signaling mode
\begin{equation}
  \lambda_+(x) = \bm{w}_+^T\bch(x)
\end{equation}
together with the perturbation direction
\begin{equation}
  \bm{h}^T\bch(x),
\end{equation}
normalizing each curve by its maximum absolute value for visual comparison.

Fig.~\ref{si-fig:signaling-mixing} shows the same three cases in more detail.
The top row displays the target density $q(x)$ together with sampled perturbed distributions
\begin{equation}
  \rho_{\bze}(x)
  \propto
  q(x)\exp\qty(\zeta_\mu \chi_\mu(x)),
  \qquad
  \bze = a\,\bm{h}.
\end{equation}
For each column, we plotted $20$ such samples.
To make the displayed perturbation magnitudes comparable across the three values of $x_0$, we used the same set of shared quantile levels across all columns: we drew $20$ quantiles uniformly from the interval $[0.08,0.92]$, transformed them to standard-normal quantiles, and then centered and rescaled them to unit variance.
These common amplitudes were then multiplied by a column-dependent scale chosen so that the median KL divergence of the sampled perturbations satisfied
\begin{equation}
  \operatorname{median}\qty[D_{\rm KL}(\rho_{\bze}\Vert q)] = 0.04.
\end{equation}
For numerical stability, the exponent field $\zeta_\mu\chi_\mu(x)$ was clipped to the interval $[-30,30]$ before exponentiation.

The middle row of Fig.~\ref{si-fig:signaling-mixing} shows the first three passive relaxation modes $\chi_1,\chi_2,\chi_3$, each normalized by its own maximum absolute value for display.
The bottom row shows the candidate signaling modes
\begin{equation}
  \lambda_+(x),\qquad
  \lambda_-(x),\qquad
  \lambda_\perp(x),
\end{equation}
together with the perturbation direction $\bm{h}^T\bch(x)$.
Here
\begin{equation}
  \lambda_\pm(x)=\bm{w}_\pm^T\bch(x),
  \qquad
  \lambda_\perp(x)=\bm{w}_\perp^T\bch(x),
\end{equation}
with
\begin{equation}
  \bm{w}_+
  \propto
  \frac{\cos\vartheta}{\sqrt{\kappa_1}}\,\bm{e}_1
  +
  \frac{\sin\vartheta}{\sqrt{\kappa_\parallel}}\,\bm{e}_\parallel,
  \qquad
  \bm{w}_-
  \propto
  -\frac{\sin\vartheta}{\sqrt{\kappa_1}}\,\bm{e}_1
  +
  \frac{\cos\vartheta}{\sqrt{\kappa_\parallel}}\,\bm{e}_\parallel,
\end{equation}
and
\begin{equation}
  \bm{e}_\parallel = \bm{e}_2,
  \qquad
  \bm{e}_\perp = \bm{e}_3
\end{equation}
for the present choice $\beta=0$.
All plotted modes in the bottom row were again normalized by their maximum absolute value for visual comparison.
The priority insets show the relative signaling priorities
\begin{equation}
  \ds_\alpha\Big/\sum_\beta \ds_\beta,
\end{equation}
for $\alpha\in\{+,-,\perp\}$.
Because the perturbation covariance is rank one and has no support along $\bm{e}_\perp$, we obtain $\ds_\perp=0$ in this example.

\subsubsection*{Figs.~\ref{fig:preconditioning} and~\ref{fig:preconditioning-training-overview}:
preconditioning with limited signaling}

Figs.~\ref{fig:preconditioning} and~\ref{fig:preconditioning-training-overview} use the same one-dimensional two-peak target distribution and seven-mode basis as Fig.~\ref{fig:dynamics-gallery}.
The controller was restricted to a single signaling mode,
\begin{equation}
  \bW = (1,0,\ldots,0),
\end{equation}
so that the feedback policy had access only to the moment of the slowest inter-peak relaxation mode.
The physical and optimization parameters were
\begin{equation}
  \varepsilon = 1,\qquad
  r = 1,\qquad
  \gamma = \frac{1}{3}.
\end{equation}

The preconditioned run optimized both the preconditioning vector $\bph$ and the limited-signaling value function
$S$, while the matched baseline fixed $\bph=\bm{0}$ and optimized only the corresponding controller.
Both runs used the same fixed perturbation ensemble
\begin{equation}
  \bze(0)=(\Delta,0,\ldots,0),
  \qquad
  \Delta\sim \mathrm{Unif}[-2,2],
\end{equation}
represented by 4096 antithetic (i.e., symmetric under $\Delta\mapsto -\Delta$) initial conditions.
The perturbation pool was sampled once and reused throughout training.
Since the perturbation distribution was fixed rather than centered at the current value of $\bph$, no score-function correction was used; gradients were taken only through the simulated dynamics and trajectory loss.

For the preconditioned run, $\bph$ was updated for 500 outer steps.
During each outer step, the value function and prior offset $g$ were trained by inner-loop gradient descent, with 50 initial $S/g$ epochs, 5 additional $S/g$ epochs per outer $\bph$ step, and 150 final refinement epochs after the last $\bph$ update.
The total number of inner-loop training epochs was 2695.
The matched $\bph=\bm{0}$ controller was trained for the same total number of epochs using the same perturbation pool, batch size 1024, and learning rates.
Gradient descent was performed using the ADAM optimizer and the rates were $10^{-4}$ for the value-function and $g$ updates and $5\times10^{-4}$ for the $\bph$ updates.
The KL-reference anchor in the value function was fixed at $\zeta_{\rm ref}=0$ throughout.

Because the perturbation ensemble is symmetric under $x\mapsto -x$, the optimal preconditioned stationary distribution should also preserve this symmetry.
To prevent a near-degeneracy between odd components of $\bph$ and the learned prior offset $g$, the final preconditioned run included a small quadratic penalty, with coefficient $10^{-2}$, on the odd-parity components of $\bph$.
The final optimized preconditioning vector was approximately
\begin{equation}
  \bph^\ast
  =
  (5.8\times10^{-4},\,
  -4.0\times10^{-2},\,
  1.4\times10^{-4},\,
  3.8\times10^{-2},\,
  -1.6\times10^{-5},\,
  -1.7\times10^{-1},\,
  -2.5\times10^{-4}),
\end{equation}
and the resulting stationary distribution satisfied
\begin{equation}
  D_{\rm KL}(\rho_{\bph^\ast}\Vert q) \simeq 3.24\times10^{-2}.
\end{equation}

For Fig.~\ref{fig:preconditioning}, all four dynamical curves report $D_{\rm KL}(\rho(t)\Vert q)$, including the preconditioned passive and controlled trajectories.
The thin horizontal reference line is $D_{\rm KL}(\rho_{\bph^\ast}\Vert q)$.
The time axis is displayed in units of the perturbation timescale, $t/(1/\gamma)$.
Fig.~\ref{fig:preconditioning-training-overview} uses the same training artifacts: panel (a) shows the evolution of $\bph$ during the preconditioned run, and panel (b) compares the final training losses of the matched $\bph=\bm{0}$ and $\bph=\bph^\ast$ controllers.

\subsubsection*{Fig.~\ref{fig:trajecs-schem}: attractor schematic}

This figure is a schematic illustration of flows through distribution space and was not generated from numerical simulations.
If it were to be taken literally, the space could be interpreted as a Wasserstein manifold of distributions, and the representation would be in normal coordinates around the target distribution $q$.
In Riemannian geometry, normal coordinates are a special coordinate system defined in a neighborhood of a point (in this case, the target distribution $q$) such that geodesics through that point are represented as straight lines and the metric at that point is the identity matrix.
In these normal coordinates, more efficient transport paths look like straighter lines, while less efficient paths look more curved; this is reflected in the schematic in which the passive relaxation path is more curved/indirect than the feedback-controlled path.

\subsubsection*{Figs.~\ref{si-fig:intrinsic-geodesics}--\ref{si-fig:intrinsic-geodesic-grid-icons}: Fisher and Wasserstein geometry on a two-mode exponential-family submanifold}

Figs.~\ref{si-fig:intrinsic-geodesics}, \ref{si-fig:intrinsic-geodesic-grid}, and \ref{si-fig:intrinsic-geodesic-grid-icons} were generated from a one-dimensional symmetric two-peak Gaussian-mixture reference density
\begin{equation}
  q(x)
  =
  \frac{1}{2}\mathcal{N}(x;-x_0,\delta^2)
  +
  \frac{1}{2}\mathcal{N}(x;x_0,\delta^2),
  \qquad
  x_0 = 1.5,
  \qquad
  \delta = 0.5,
\end{equation}
on the interval
\begin{equation}
  x \in [-4.5,\,4.5].
\end{equation}
The weighted Neumann eigenproblem for the passive relaxation modes was solved on a uniform grid with spacing $\Delta x=0.025$ (equivalently, $361$ grid points), and the constant mode was discarded.  The figures use the two-dimensional exponential-family submanifold spanned by the first two nonconstant modes, $\chi_1$ and $\chi_2$.

The Fisher metric $C_{\mu\nu}(\zeta)$ and Wasserstein metric $g^W_{\mu\nu}(\zeta)$ were evaluated on a regular
\begin{equation}
  (\zeta_1,\zeta_2)\in[-1,1]^2
\end{equation}
grid with $151\times 151$ sample points.  In Fig.~\ref{si-fig:intrinsic-geodesics}, the colored markers correspond to the parameter values
\begin{equation}
  (0,0),\quad
  (\pm 0.75,0),\quad
  (0,\pm 0.75),\quad
  (0.55,0.55).
\end{equation}
The accompanying density and mode panels use these same points.

For Figs.~\ref{si-fig:intrinsic-geodesic-grid} and~\ref{si-fig:intrinsic-geodesic-grid-icons}, the geodesic grids were generated from two families of metric-orthogonal geodesics seeded at seven evenly spaced points along each coordinate axis.  The highlighted comparison paths were chosen to connect the same pair of endpoint distributions under the two metrics: in the Wasserstein panel the path is the selected curved geodesic, while in the Fisher panel symmetry makes the corresponding comparison geodesic lie along $\zeta_2=0$.  The density heatmaps in Fig.~\ref{si-fig:intrinsic-geodesic-grid} show $\rho_{\bze(s)}(x)$ along these highlighted paths as a function of intrinsic arc length $s$, and Fig.~\ref{si-fig:intrinsic-geodesic-grid-icons} shows six representative density samples along each path, including the two endpoints.

\section{Miscellaneous optimal signaling mode figures}
\label{si-sec:additional-optimal-signaling-modes}

Here, we show some additional figures showing the optimal signaling modes and signaling priority spectra for various choices of $q(\bx)$ with $p_{\rm pert} \propto \exp(- \frac{1}{2 \sigma} \zeta_\mu \delta_{\mu\nu} \zeta_\nu)$.

\begin{figure}[h]
  \includegraphics{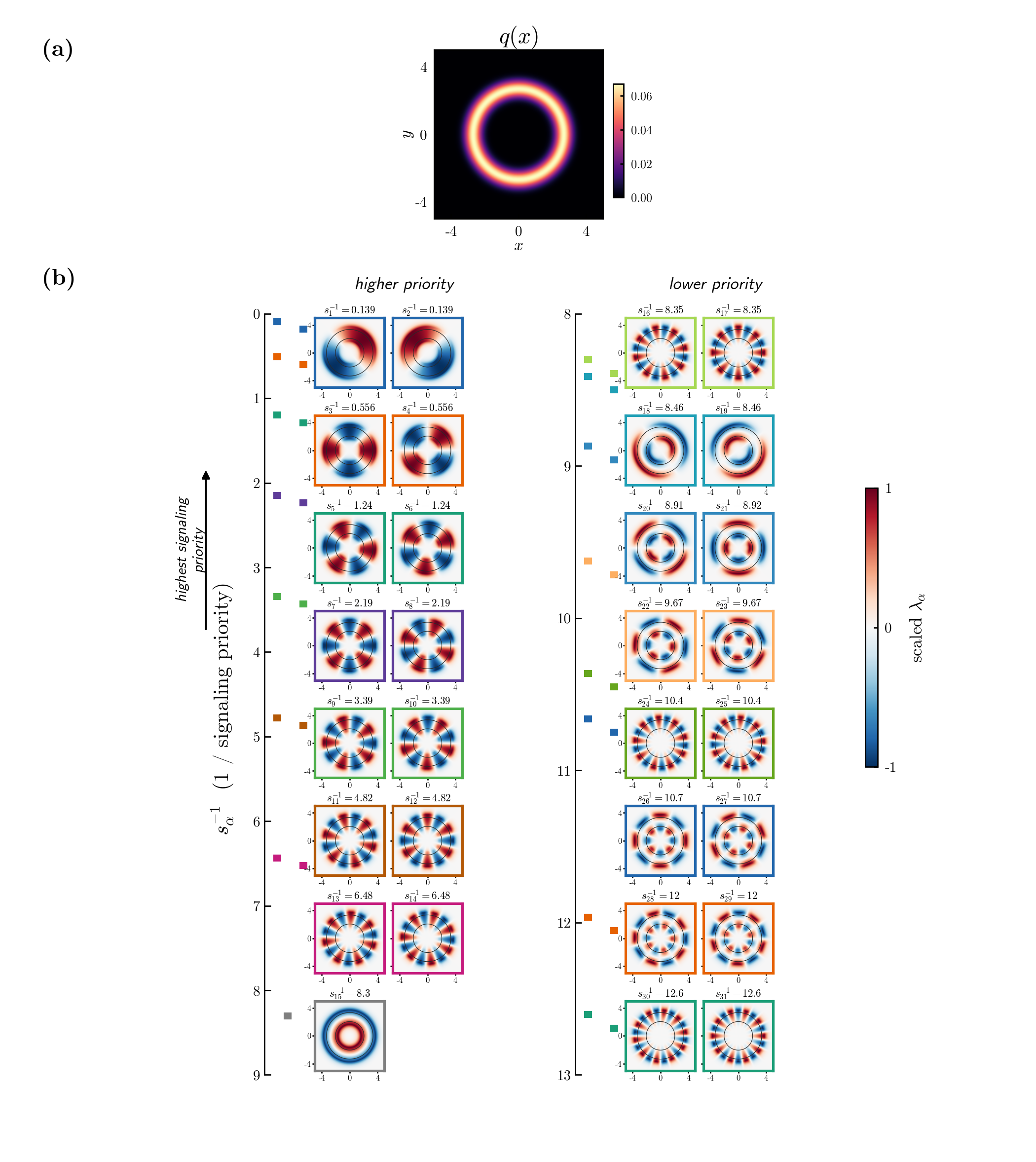}
  \caption{Optimal signaling modes for a circular target distribution.}
\end{figure}

\begin{figure}[h]
  \includegraphics{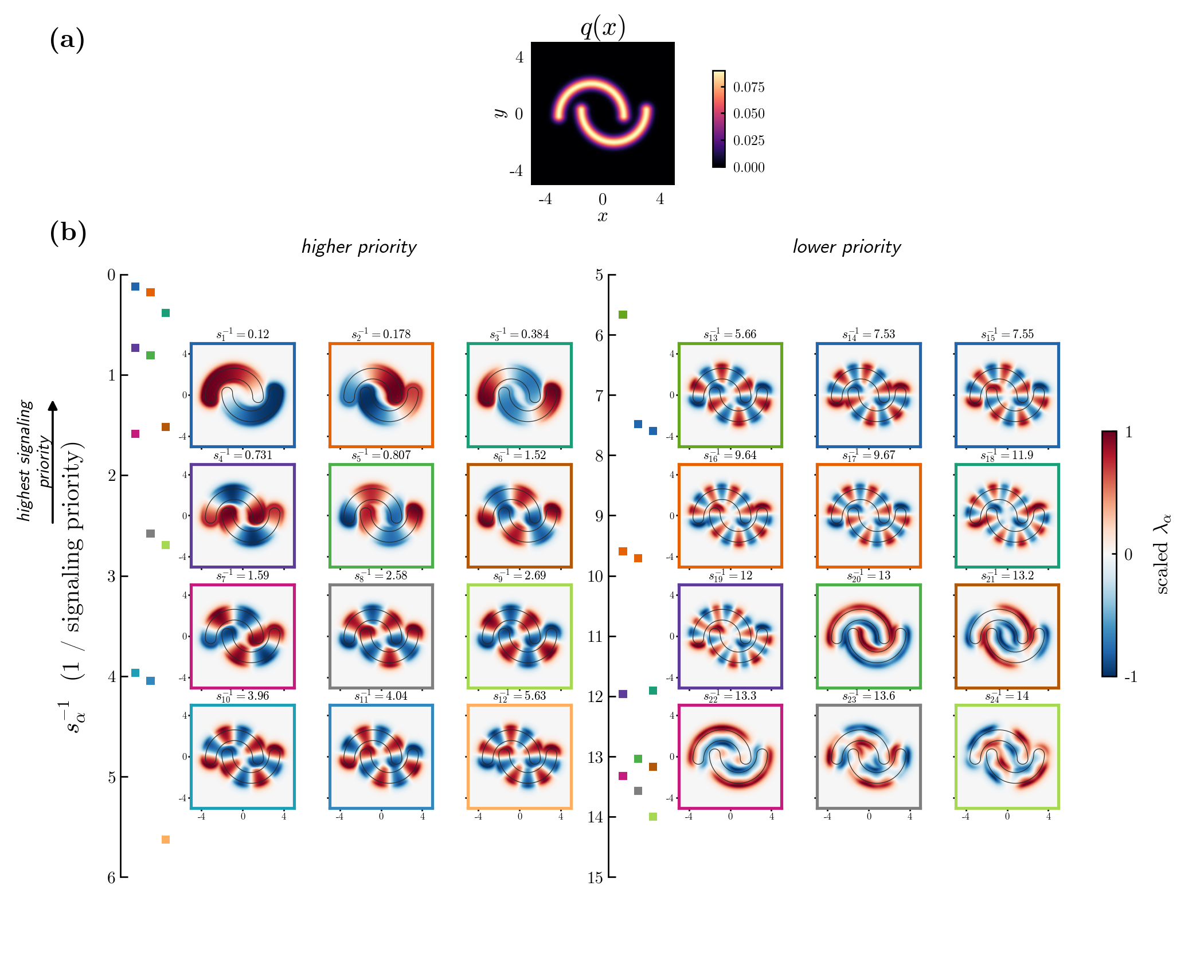}
  \caption{Optimal signaling modes for a double crescent target distribution.}
\end{figure}

\begin{figure}
  \includegraphics{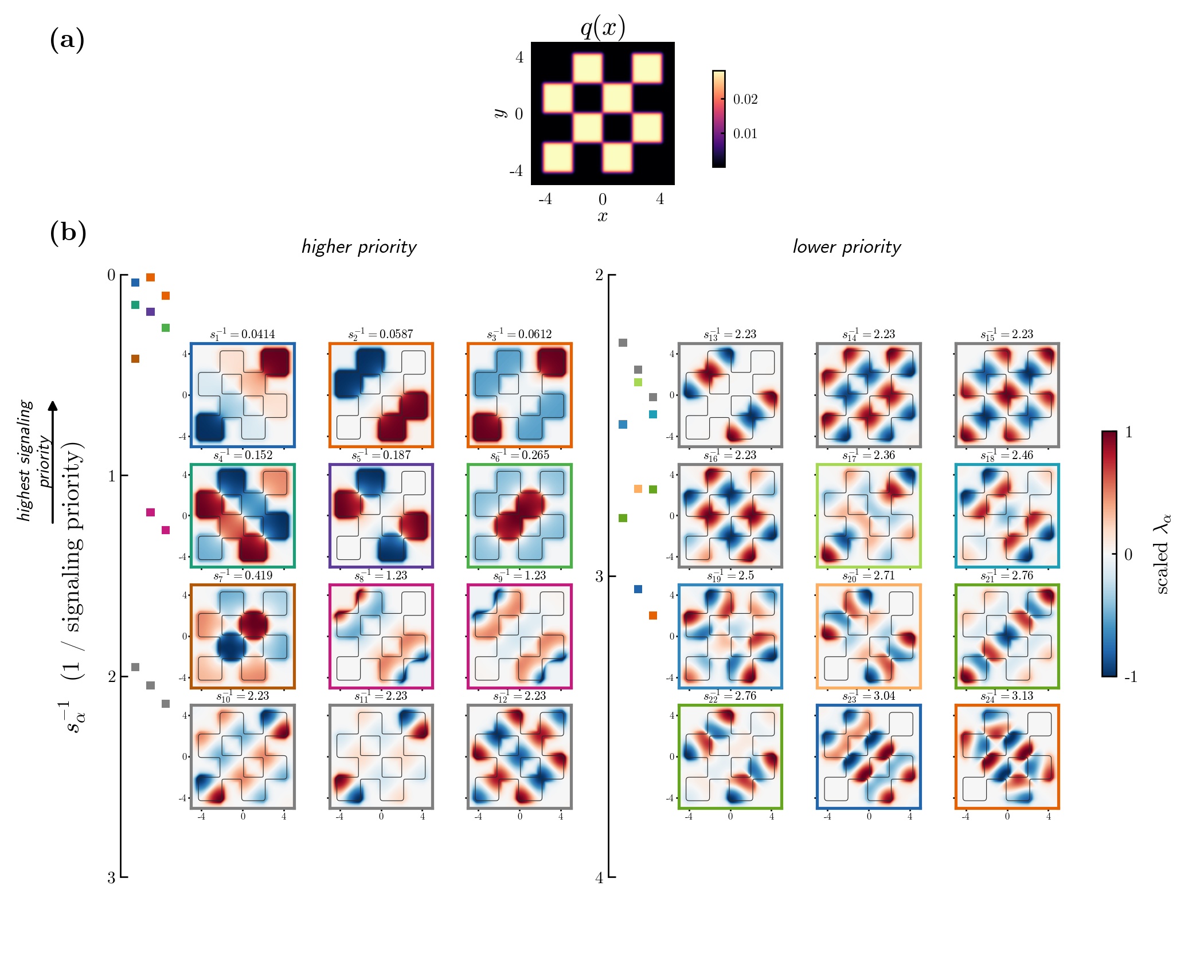}
  \caption{Optimal signaling modes for a checkerboard target distribution.}
\end{figure}

\begin{figure}
  \includegraphics{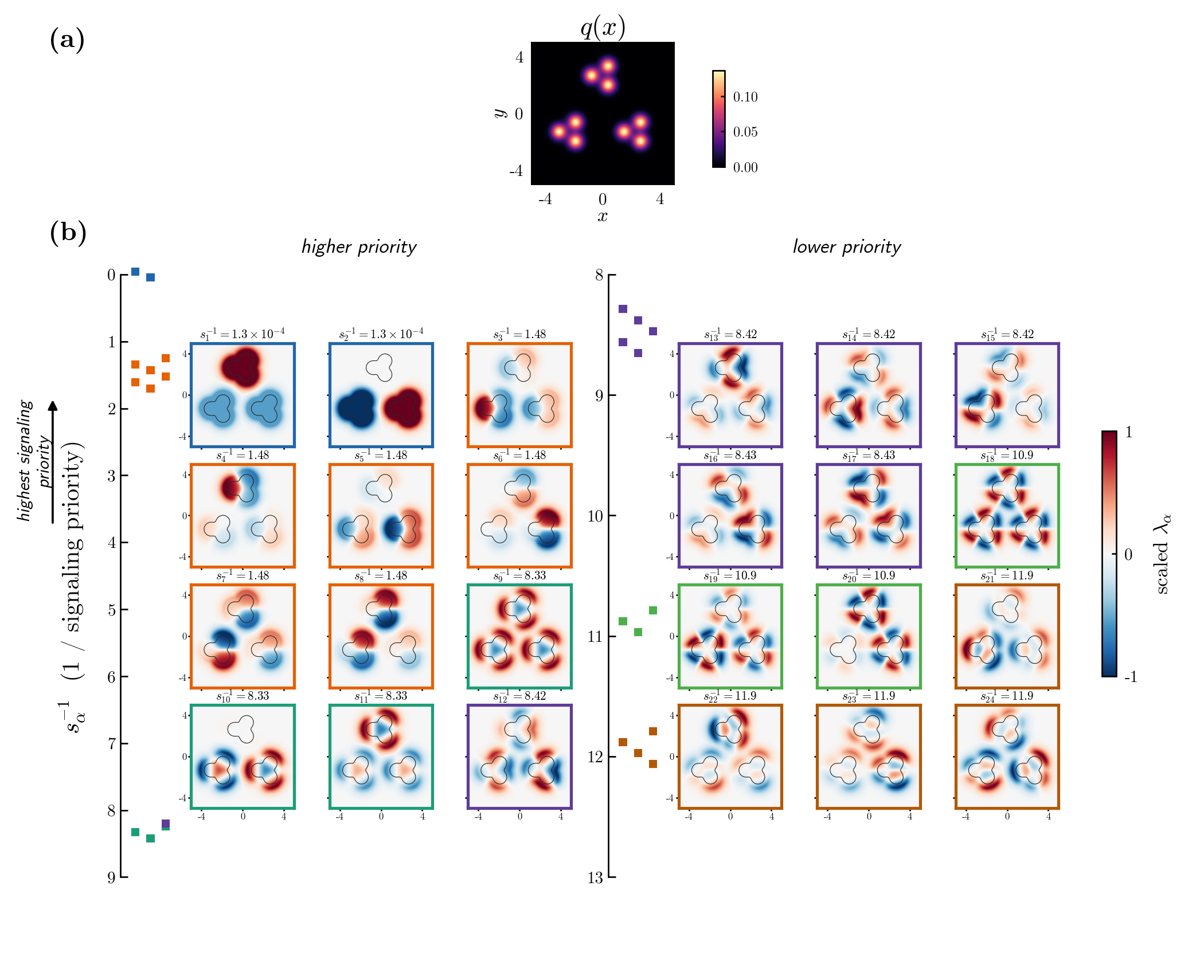}
  \caption{Optimal signaling modes for a hierarchical clusters target distribution.}
\end{figure}

\end{document}